\documentclass[
 reprint, amsmath,amssymb,
 aps]{revtex4-2}

\usepackage{graphicx}
\usepackage{dcolumn}
\usepackage{bm}
\usepackage{hyperref}
\usepackage{ulem}

\newcommand{\dd}{\mathrm{d}}
\newcommand{\data}{\boldsymbol{d}}

\newcommand{\Nobs}{N_\mathrm{obs}}
\newcommand{\trigger}{\mathrm{trig}}

\newcommand{\ma}{m_a}
\newcommand{\mb}{m_b}

\newcommand{\mmin}{m_\mathrm{min}}
\newcommand{\mmax}{m_\mathrm{max}}

\newcommand{\lambdapp}{\psi_\mathrm{PP}}
\newcommand{\lambdapg}{\psi_\mathrm{PG}}
\newcommand{\lambdagg}{\psi_\mathrm{GG}}

\newcommand{\classa}{\mathrm{class}_a}
\newcommand{\lambdavec}{\boldsymbol{\lambda}}
\newcommand{\Lambdavec}{\boldsymbol{\Lambda}}
\newcommand{\thetavec}{\boldsymbol{\theta}}
\newcommand{\classpp}{\mathrm{class}_\mathrm{PP}}
\newcommand{\classpg}{\mathrm{class}_\mathrm{PG}}
\newcommand{\classgg}{\mathrm{class}_\mathrm{GG}}

\newcommand{\Lobs}{\mathcal{L}^\mathrm{obs}}

\newcommand{\powerpeakmodel}{Power-law+Peak}

\usepackage[dvipsnames]{xcolor}
\definecolor{flatirons}{HTML}{8B2131}
\definecolor{sunshine}{HTML}{CA9500}
\definecolor{skyline}{HTML}{1D428A}
\definecolor{midnight}{HTML}{0E2240}

\newcommand\apjl{Astrophys. J. Letters} 
\newcommand\mnras{Monthly Notices of the Royal Astronomical Society} 
\newcommand\aap{Astronomy and Astrophysics}                
\newcommand\apjs{Astrophysical Journal, Supplement}               
\newcommand{\pasp}{Publications of the Astronomical Society of the Pacific}  

\usepackage{orcidlink}
\newcommand{\IMRELEASENO}{LLNL-JRNL-862034-DRAFT}
\begin{document}

\title{Investigating the mixing between two black hole populations in LIGO-Virgo-KAGRA GWTC-3}

\author{Ming-Feng Ho \orcidlink{0000-0002-4457-890X}$^1$}
\email{mho026@ucr.edu}
\author{Scott E. Perkins \orcidlink{0000-0002-5910-3114}$^2$}
\email{perkins35@llnl.gov }
\author{Simeon Bird \orcidlink{0000-0001-5803-5490}$^1$}
\email{sbird@ucr.edu}
\author{William A. Dawson \orcidlink{0000-0003-0248-6123}$^2$}
\email{dawson29@llnl.gov}
\author{Nathan Golovich \orcidlink{0000-0003-2632-572X}$^2$}
\author{Jessica R. Lu \orcidlink{0000-0001-9611-0009}$^3$}
\author{Peter McGill \orcidlink{0000-0002-1052-6749}$^2$}

\affiliation{$^1$Department of Physics and Astronomy, University of California at Riverside, 900 University Ave., Riverside, CA 92521, USA}

\affiliation{$^2$Space Science Institute, Lawrence Livermore National Laboratory, 7000 East Ave., Livermore, CA 94550, USA}

\affiliation{$^3$Department of Astronomy, University of California at Berkeley, 501 Campbell Hall \#3411, Berkeley, CA 94720, USA}


\date{\today}

\begin{abstract}
We introduce a population model to analyze the mixing between hypothesised power-law and $\sim 35 M_\odot$ Gaussian bump black hole populations in the latest gravitational wave catalog, GWTC-3, estimating their co-location and separation. We find a relatively low level of mixing, $3.1^{+5.0}_{-3.1}\%$, between the power-law and Gaussian populations, compared to the percentage of mergers containing two Gaussian bump black holes, $5.0^{+3.2}_{-1.7}\%$. Our analysis indicates that black holes within the Gaussian bump are generally separate from the power-law population, with only a minor fraction engaging in mixing and contributing to the $\mathcal{M} \sim 14 M_\odot$ peak in the chirp mass. This leads us to identify a distinct population of Binary Gaussian Black Holes (BGBHs) that arise from mergers within the Gaussian bump. We suggest that current theories for the formation of the massive $35 M_\odot$ Gaussian bump population may need to reevaluate the underlying mechanisms that drive the preference for BGBHs.\end{abstract}

\maketitle

\section{\label{sec:intro}Introduction}

Gravitational wave astronomy is shifting focus from in-depth analysis of single events to population inference that addresses key questions in astrophysics \cite{2023PhRvX..13a1048A,Nitz:2023}, fundamental physics \cite{2021arXiv211206861T}, and cosmology \cite{2021arXiv211103604T}.
Research using the third Gravitational-Wave Transient Catalog (GWTC-3) \cite{gwtc3}, published by the LIGO Scientific Collaboration \cite{2021arXiv211103604T}, Virgo Collaboration \citep{2015CQGra..32b4001A}, and KAGRA Collaboration \cite{kagra}, demonstrated this \citep{2019ApJ...882L..24A,Venumadhav:2019,2021ApJ...913L...7A,Wadekar:2023}.
GWTC-3 has become a vital tool for understanding binary black hole (BBH) formation physics \citep[e.g.,][]{Zevin:2017,Zevin:2020gbd,2021ApJ...913L...7A,Edelman:2022,2023PhRvX..13a1048A}.
Investigating the formation history of BBHs through a single gravitational wave (GW) event is a difficult task.
However, the population analysis of numerous merging BBH events can provide insights into their formation channels (e.g., \cite{Stevenson:2015,Zevin:2017}).
For example, the lack of black holes with masses $\sim 2 - 5 M_\odot$ \citep{Ozel:2010,Farr:2011,Fishbach:2020,Farah:2021} may indicate maximum neutron star masses \citep{Li:2021,Patton:2022,Siegel:2023},
and also the timescale related to supernova explosions (such as \citep{Fryer:2012,Zevin:2020,Mandel:2020}) and mass transfer~\citep[e.g.,][]{vanSon:2022}.

LIGO-Virgo-KAGRA's (LVK, hereafter) population analysis of the GWTC-3 catalog indicates distinct substructures within the primary black hole mass spectrum~\citep{2023PhRvX..13a1048A}.
In the primary mass distribution, two prominent peaks are observed at $m_1 \sim 10 M_\odot$ and $m_1 \sim 35 M_\odot$ with high significance. Another peak, at $m_1 \sim 20 M_\odot$, appears to be less certain.
The peak at approximately $10 M_\odot$ is postulated to exist above the black hole-neutron star (BH-NS) low-mass gap and can arise from the stellar initial mass function (IMF). The corresponding peak in the binary black hole (BBH) mass function could be attributed to the evolution of binary star systems \citep{2023PhRvX..13a1048A,vanSon:2022,Schneider:2023}.
Several options have been suggested to explain the peaks at $m_1 \sim 20 M_\odot$ and $m_1 \sim35 M_\odot$.
A popular explanation for the peak at around $m_1 \sim 35 M_\odot$ is that it results from pulsational pair-instability supernovae (PPSNe) originating from stars initially ranging between $100 M_\odot$ and $150 M_\odot$ \citep{Fowler:1964,Barkat:1967,Heger:2002,Heger:2003,Woosley:2015,Belczynski:2016,Talbot:2018,Marchant:2019,Woosley:2019,Renzo:2020,Farmer:2019},
though some recent studies suggest the $35 M_\odot$ peak is unlikely to be due to the PPSNe \citep{Hendriks:2023,Briel:2023}. 
The pair-instability mechanism also predicts a sharp cutoff at masses greater than $40 M_\odot$, attributed to the absence of remnants from pair-instability supernovae occurring in stars with initial masses ranging from $150 M_\odot$ to $250 M_\odot$.
Other exotic formation channels that may explain the $35 M_\odot$ peak include primordial black holes \citep{George:1975,Carr:1975,Bird:2016,Sasaki:2018}, massive triple stars \citep{Antonini:2017,Silsbee:2017}, low-metallicity star progenitors~\citep{Inayoshi:2016}, and hierarchical mergers~\citep{Tiwari:2021b}.

As the number of BBH detections from gravitational wave observations increases, it becomes feasible to test BBH formation channels by examining the statistical properties of the population of secondary black holes.
In situations where black holes merge within dense environments (e.g.~star clusters), following a dynamical channel, the underlying mass distribution would likely appear similar as the comparable component masses have a higher binding energy \citep{Rodriguez:2016PhRvD..93h4029R,Amaro-Seoane:2016,Farah:2023b}, though it is possible for a dynamical channel to produce unequal component mass binaries through ultra-wide binaries \citep{Michaely:2019}.
While the isolation formation channel (``field binaries'') prefers BBHs with comparable masses \cite{Dominik:2015}, some isolation channels can produce unequal component masses \cite{Giacobbo:2018,Spera:2019}.
This variation could be influenced by a range of uncertain physical processes, such as the binary IMF \cite{Grudic:2023},
the evolution of binary star systems \cite{Dominik:2012,Stevenson:2017,Giacobbo:2018,Spera:2019}, and possible mechanisms like mass transfer or inversion \cite{Laplace:2021,Olejak:2021,Broekgaarden:2022,Zevin:2022}.

A widely used parameterization for the BBH mass spectrum is the {\powerpeakmodel} model \citep{Talbot:2018}, which involves modeling a combination of the primary mass ($m_1$), the heavier black hole in the BBH, and the mass ratio ($q = m_2 / m_1 < 1$), representing the ratio between the secondary and primary masses. The {\powerpeakmodel} model posits a power-law function with a Gaussian peak for the primary mass distribution, while the physical distinction between primary and secondary masses is modeled using a power-law model on the mass ratio. This approach has been utilized in several studies (e.g., \cite{Kovetz:2017,Fishbach:2017,Tiwari:2021,Edelman:2022,Callister:2023,Godfrey:2023}). Beyond the {\powerpeakmodel} model, Ref.~\cite{Fishbach:2020b,Farah:2023b} explore various models for the secondary mass spectrum in BBHs.

In this paper, we construct a population model to estimate the mixing fractions between the populations of black holes from the power-law distribution (corresponding to the peak at $m_1 \sim 10 M_\odot$) and those from the Gaussian bump (at $m_1 \sim 35 M_\odot$).
We want to understand how likely it is for black holes originating from different peaks to mix in the Universe, forming the BBHs we observe. Inferring the mixing fraction between the power-law and Gaussian peak populations can provide new perspective into BBH formation mechanisms. For instance, low mixing between the two populations might indicate that black holes produced by different formation mechanisms remain separated, with binaries likely forming within their respective populations.

Our population model begins by forward-sampling black hole masses from either a power-law or Gaussian mass spectrum. Each pair in a BBH is then sampled from a mixture of these mass populations. We vary the relative abundance in the mixture model, thus controlling the mixing fraction between the power-law and Gaussian mass populations. Using a mixture of power-law and Gaussian populations ensures our primary mass function aligns well with the {\powerpeakmodel} model from Ref.~\cite{2023PhRvX..13a1048A}.

Our inference suggests that a significant portion (approximately $5.0^{+3.1}_{-1.7}\%$ of the total population) of the BBHs consist of Binary Gaussian Black Holes (BGBHs), where both black holes originate from the $\sim 35 M_\odot$ Gaussian bump. 
We also observe a low mixing fraction between the power-law and the Gaussian bump, $3.1^{+5.0}_{-3.1}\%$, indicating that the Gaussian bump black holes are primarily separate from the power-law population.
Another interesting aspect of our model is the alignment of the second chirp mass peak at $\mathcal{M} \sim 14 M_\odot$ with the mixing between the power-law distribution peak ($ \sim 10 M_\odot$) and the Gaussian distribution peak ($ \sim 35 M_\odot$), calculated as $\mathcal{M} \sim \frac{(10 M_\odot \times 35 M_\odot)^{3/5}}{(10 M_\odot + 35 M_\odot)^{1/5}} \simeq 15 M_\odot$. Among notable features in the black hole mass spectrum, the second peak around $m_1 \sim 20 M_\odot$ has been identified as marginally significant in primary mass \citep{Wong:2022, Callister:2023, Farah:2023a, Tiwari:2024}. However, its nature remains debated. Some argue it may be a result of Poisson fluctuations within the power-law function \citep{Farah:2023a}, while others suggest that the corresponding second peak in the chirp mass spectrum ($\mathcal{M} \sim 14 M_\odot$) is more pronounced than the primary mass substructure \citep{Tiwari:2024}.
Our inference suggests a way to interpret the $\mathcal{M} \sim 14 M_\odot$ chirp mass peak as a result of mixing between two populations with different formation mechanisms.

This paper is structured as follows:
Section~\ref{sec:population_models} introduces our population model for BBHs.
Section~\ref{sec:selection_effect} outlines the Bayesian inference approach we employ, taking into account detection efficiency.
Section~\ref{sec:results} presents our inference results and the predicted black hole mass functions.
Section~\ref{sec:conclusion} offers concluding remarks.

\section{\label{sec:population_models} Population Model}

In this section, we discuss our BBH population model designed to understand the mixing between different black hole populations.
Our population model is detailed in Section~\ref{subsec:population_model_3_shapes}, where we discuss the three different subpopulations of BBHs and the forward model for generating samples of BBHs.
Next, in Section~\ref{subsec:viz_model}, we show the exploratory models of the predicted chirp mass and mass ratios according to different parameters of the population model.
In Section~\ref{subsec:average_spectrum}, we discuss a method for acquiring fiducial parameters for our population mixture model.

\subsection{\label{subsec:population_model_3_shapes} Population model: Subpopulations}

In this section, we discuss the reasoning behind our population model, which is motivated by the Gaussian bump in the primary mass function.
LVK population analysis identified a notable Gaussian peak mixed into the primary mass function \citep{2023PhRvX..13a1048A}, with the mixing fraction represented by $\lambda_\mathrm{peak} \sim 3.8^{+5.8}_{-2.6}\%$, suggesting roughly 3.8\% of the primary mass black holes are from the Gaussian distribution.

\begin{figure}
  \includegraphics[width=\columnwidth]{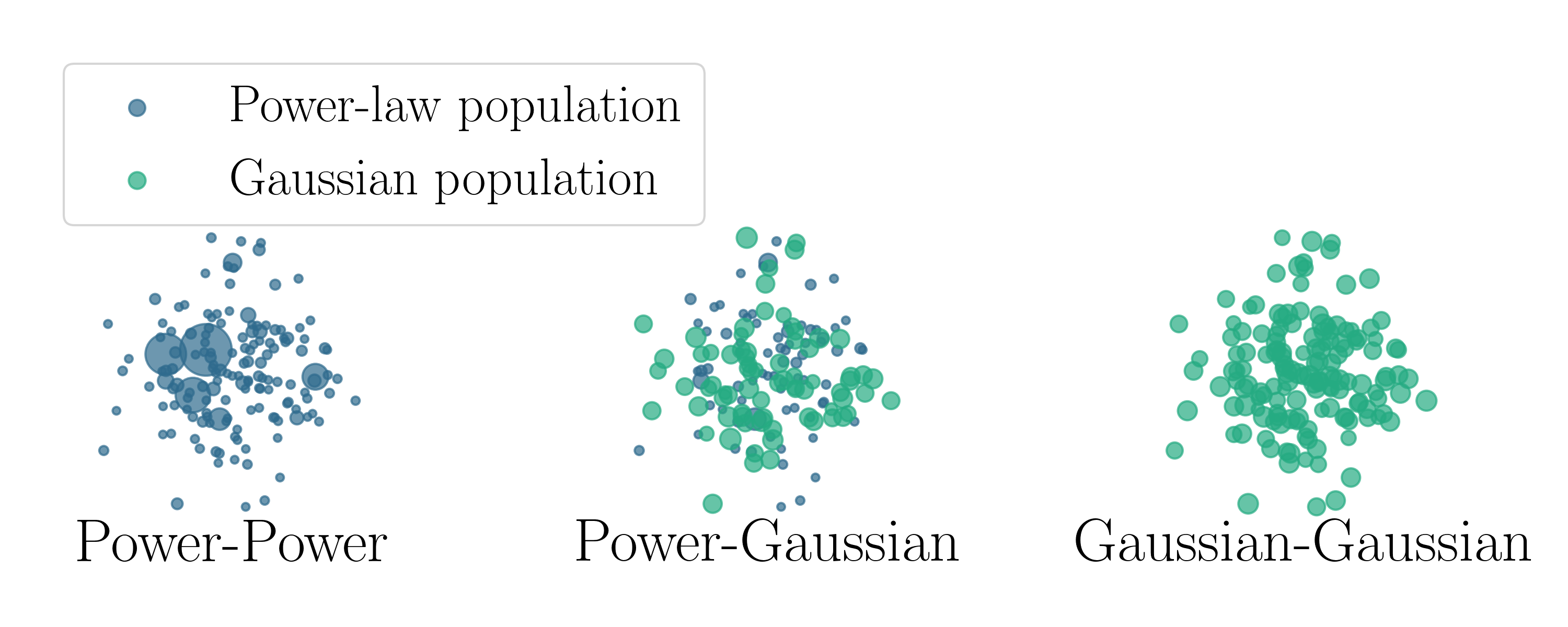}
  \caption{
  A cartoon illustrates the mixing scenarios used in this work.
  The size of the circles represents the masses of the black holes, while the color indicates the underlying population. If the power-law and Gaussian bump BHs are mixed, as in the PG model, the resulting two-dimensional probability density of chirp mass and mass ratio $(\mathcal{M}, q)$ will exhibit a distinct morphology, as shown in Figure~\ref{fig:hist2d}.
  }
  \label{fig:cover}
\end{figure}

To begin, our population model assumes the black holes in BBHs are drawn from either a power-law distribution or a Gaussian distribution.
This points to three distinct subpopulations of BBHs in our model: Power-Power (PP), Power-Gaussian (PG), and Gaussian-Gaussian (GG).
\begin{itemize}
  \item Power-Power (PP) model: A black hole from the Power-law population merging another power-law population black hole.
  \item Gaussian-Gaussian (GG) model: A black hole from the Gaussian population merging with another Gaussian population black hole. If the fraction of GG events is high relative to PG events, then it is likely Gaussian bump black holes are separate from the rest of the black holes. We name this population of BBHs as BGBHs.
  \item Power-Gaussian (PG) model: A black hole from the Power-law population merging with a black hole from the Gaussian population. If the fraction of PG events is high relative to GG events, then it is likely that the Gaussian bump black holes are mixed with the rest of the black holes.
\end{itemize}
A cartoon version representing these three scenarios can be found in Fig~\ref{fig:cover}.
Measuring a high fraction of GG events alongside a low fraction of PG events would suggest that the Gaussian bump is separate from the power-law population. Conversely, a significantly high fraction of PG events suggests that they are part of a singular, co-located population containing a mixture of black holes. The co-location (and separation) here refers to a broader concept of co-locating (and separation) in the phase space of space or time.


  We chose to use a broad definition of co-location (and separation) because the GWTC data only provides measurements of BBH mergers. 
  Therefore, the separation of the Gaussian peak population in the measured mass spectrum is not necessarily due to spatial separation.
  There are some degeneracies, such as these black holes being temporally separated (formed at different redshifts).
Therefore, in this work, when we say that BH populations are separate, this could imply that they are distributed separately in space or time.

Following Ref.~\cite{Talbot:2018}, we define the power-law population as
\begin{equation}
  \begin{split}
    \mathcal{B}(m \mid -\alpha, \delta_m, \mmin, \mmax) &= \\
    \frac{m^{- \alpha}}{Z_{m}(\mmin, \mmax)} &S(m \mid \mmin, \delta_m).
  \end{split}
\end{equation}
Here, $-\alpha$ represents the spectral index of the power-law.
The $Z_{m}(\mmin, \mmax)$ is the normalization factor for the power-law
\begin{equation}
  Z_{m}(\mmin, \mmax) = \frac{\mmax^{-\alpha + 1} - \mmin^{-\alpha + 1}}{-\alpha + 1}
\end{equation}
with the smoothing function at the low-mass end
\begin{equation}
  \begin{split}
    &S(m \mid \mmin, \delta_m) = \\&\left(\exp{\left(\frac{\delta_m}{m - \mmin} + \frac{\delta_m}{m - \mmin - \delta_m}\right)  + 1}\right)^{-1}    
  \end{split}
  \label{eq:smoothing_kernel}
\end{equation}
where $S(m \mid \mmin, \delta_m) = 0$ for $m < \mmin$ and $S(m \mid \mmin, \delta_m) = 1$ for $m \geq \mmin + \delta_m$.
That is, a smoothing kernel in the range $\mmin \leq m < \mmin + \delta_m$.
Additionally, we incorporate a parameter for the maximum mass cutoff, $\mmax$.
The same smoothing kernel is also applied to the Gaussian distribution 
\begin{equation}
  \begin{split}
    \mathcal{G}(m \mid \mu, \sigma, \delta_m, \mmin) &=\\ 
    &\frac{1}{\sigma \sqrt{2 \pi}} e^{-\frac{1}{2}\left( \frac{m - \mu}{\sigma} \right)^2} S(m \mid \mmin, \delta_m).
  \end{split}
\end{equation}
Here, $\mu$ is the mean and $\sigma$ is the standard deviation. 
Our power-law and Gaussian models are the same as the ones in the {\powerpeakmodel} model \citep{Talbot:2018}.
We do not explicitly model the primary and secondary mass spectra, instead we draw black hole masses from one of the population models. Explicitly modelling the primary and secondary requires us to ensure that the primary is the more massive object, which makes our model more complex.


Next, we build a forward model that draws samples of BBHs.
We do not directly model the primary and secondary masses, but instead, we model the mass function of the component black holes in binaries.
For clarity, when referencing arbitrary masses in a BBH system, we will use $\ma$ and $\mb$.
We will use the chirp mass and mass ratio as observables, which are derived from $\ma$ and $\mb$ samples we draw from the subpopulation model.
For PP subpopulation, the two-dimensional $(m_a, m_b)$ probability density is given by:
\begin{equation}
  \begin{split}
    &p_\mathrm{PP}(\ma, \mb \mid -\alpha, \delta_m, \mmin, \mmax) \propto \\
    &\mathcal{B}(\ma \mid -\alpha, \delta_m, \mmin, \mmax)\mathcal{B}(\mb \mid -\alpha, \delta_m, \mmin, \mmax),
  \end{split}
  \label{eq:power_power}
\end{equation}
For GG subpopulation, the probability density is
\begin{equation}
  \begin{split}
    p_\mathrm{GG}(\ma, \mb &\mid \mu, \sigma, \delta_m, \mmin) \propto \\
    &\mathcal{G}(\ma \mid \mu, \sigma, \delta_m, \mmin)
    \mathcal{G}(\mb \mid \mu, \sigma, \delta_m, \mmin).
  \end{split}
  \label{eq:peak_peak}
\end{equation}
And for the PG subpopulation, the probability density is
\begin{equation}
  \begin{split}
    &p_\mathrm{PG}(\ma, \mb \mid -\alpha, \mu, \sigma, \delta_m, \mmin,\mmax) \propto \\
    &\mathcal{B}(\ma \mid -\alpha, \delta_m, \mmin, \mmax)
    \mathcal{G}(\mb \mid \mu, \sigma, \delta_m, \mmin).
  \end{split}
  \label{eq:power_peak}
\end{equation}
Here, we do not repeat the sampling for $\ma \sim \mathcal{G}$ and $\mb \sim \mathcal{B}$ because we assume the shape parameters are the same for $\ma$ and $\mb$, so the $(\ma, \mb)$ labels are interchangeable in this sampling.

Since we are not modeling the primary and secondary mass directly, we cannot directly use the probability density of $(\ma, \mb)$ as a likelihood function and apply it on the data.
Instead, we convert the $(\ma, \mb)$ parameters into quantities that we observe in gravitational events, i.e., chirp mass and mass ratio, $(\mathcal{M}, q)$, with $\mathcal{M} = (\ma \mb)^{3/5}/(\ma + \mb)^{1/5}$ and $ q = \min(\ma, \mb)/\max(\ma, \mb)$.
They have the same statistical information as $(\ma, \mb)$ and are some of the more directly measured parameters in gravitational wave events.
In addition, in this paper, we focus on the binary-centric properties, i.e., the mixing fractions of BBH subpopulations; it is thus reasonable to use chirp mass and mass ratio (the properties specific to BBHs) over component masses.

After defining the three different subpopulation models, we can now construct a mixture model to infer the relative abundances of the three subpopulations:
\begin{equation}
  \begin{split}
    p(\mathcal{M}, q &\mid \lambdapp, \lambdapg, \lambdagg, -\alpha, \mu, \sigma, \delta_m, \mmin) \propto \\
    &\lambdapp~p_\mathrm{PP}(\mathcal{M}, q \mid -\alpha, \delta_m, \mmin, \mmax)+\\
    &\lambdapg~p_\mathrm{PG}(\mathcal{M}, q \mid -\alpha, \mu,\sigma, \delta_m, \mmin, \mmax)+\\
    &\lambdagg~p_\mathrm{GG}(\mathcal{M}, q \mid \mu,\sigma, \delta_m, \mmin),
  \end{split}
  \label{eq:inference}
\end{equation}
where $(\lambdapp, \lambdapg, \lambdagg)$ are the relative abundances for PP, PG, and GG subpopulations, with $\lambdapp + \lambdapg + \lambdagg = 1$ and $\lambdapp, \lambdapg, \lambdagg \in [0, 1]$.

  The PG subpopulation is the key to measuring the mixing between the power-law and Gaussian bump populations.
  The {\powerpeakmodel} models the mass ratios from all BBHs as a single power-law, which does not allow us to separate the contributions of PG, PP, and GG to the mass ratio distribution.
  Even if the mass ratio from {\powerpeakmodel} prefers equal-mass binaries, this preference could be driven by the majority of power-law black holes.
  Our model allows us to separate the mass ratio contribution of the PG from the rest of the BBHs, providing a more direct measurement of the separation of the Gaussian bump.

\subsection{Visualizations of the population model\label{subsec:viz_model}}

\begin{figure}
  \includegraphics[width=\columnwidth]{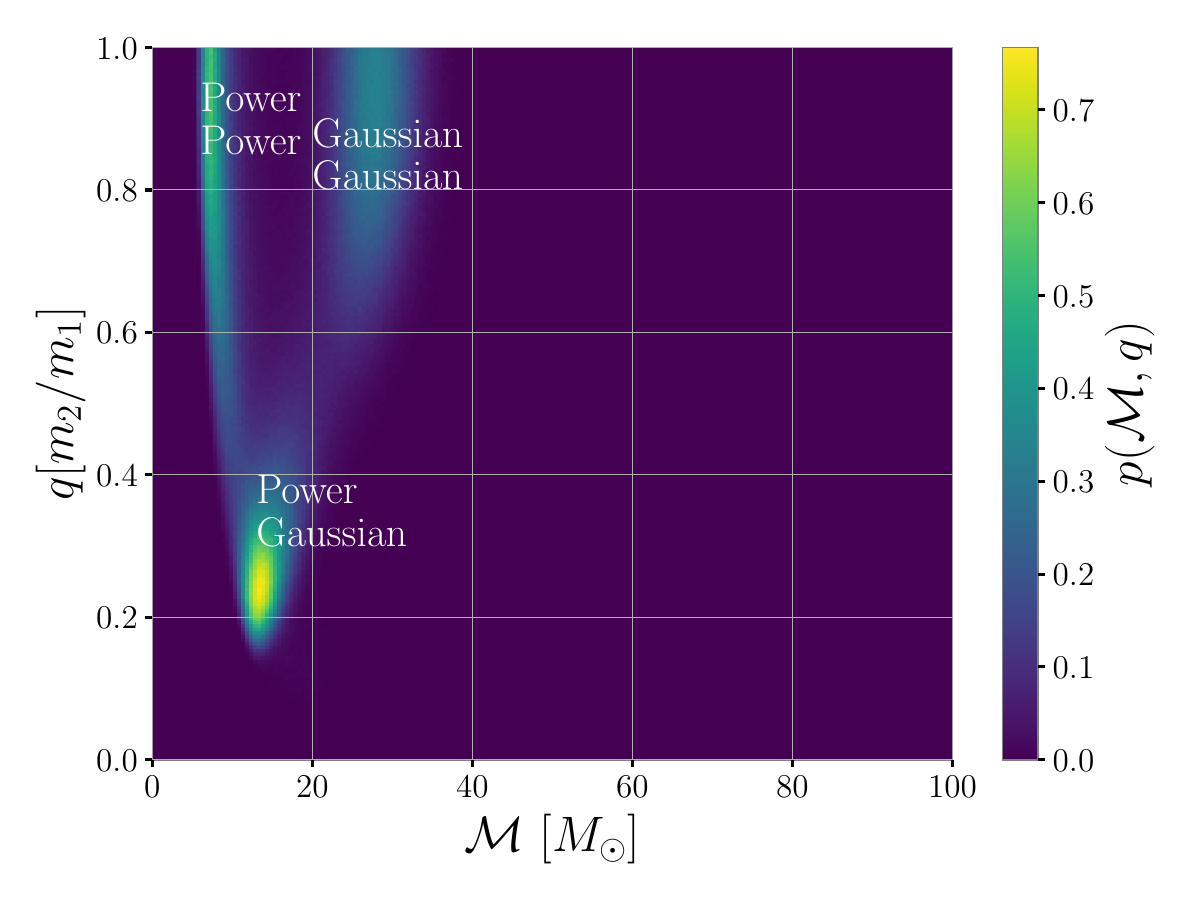}
  \caption{
  Map of likelihood density in chirp mass ($\mathcal{M}$) versus mass ratio ($q$) space for three subpopulation models, PP, PG, and GG. The shape parameters, $\lambdavec = (-\alpha, \mu, \sigma) = (-3.66, 31.59, 5.51)$ used to generate the map come from the average mass spectrum of the GWTC-3's {\powerpeakmodel} model as derived in Section~\ref{subsec:average_spectrum}.
  }
  \label{fig:hist2d}
\end{figure}

To help gain intuition for the population model, we generate histograms from the Monte Carlo samples of the three subpopulations across the parameter space of $(\mathcal{M}, q)$, shown in Figure~\ref{fig:hist2d}.
Each subpopulation covers a unique region within this $(\mathcal{M}, q)$ space.
These distinct areas will aid in determining the mixing fraction of each subpopulation in the gravitational wave data.

\begin{figure}
  \includegraphics[width=\columnwidth]{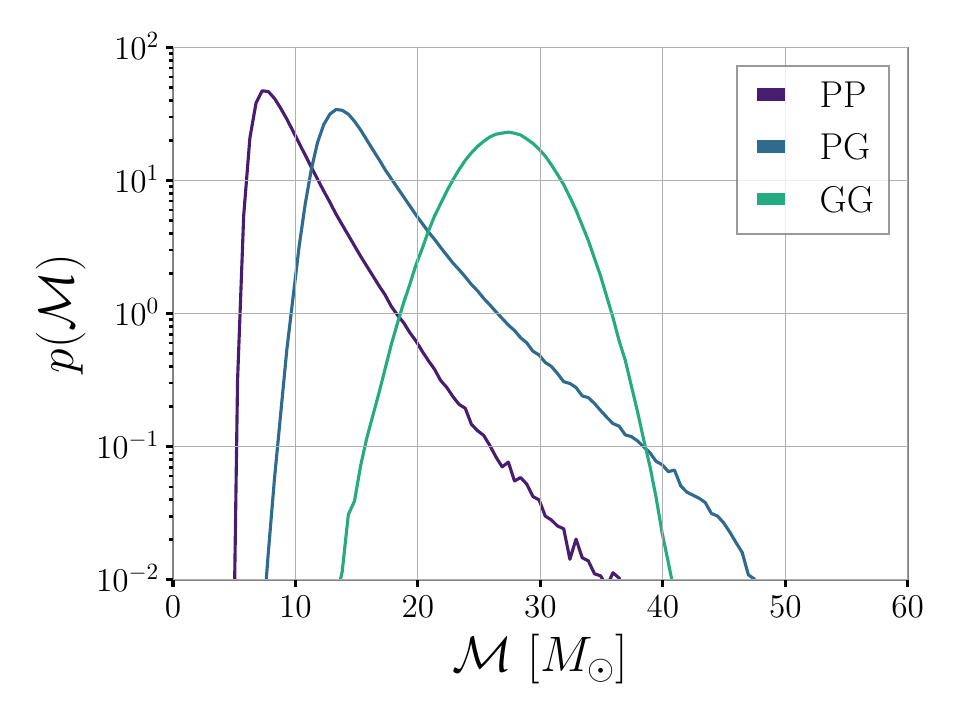}
  \includegraphics[width=\columnwidth]{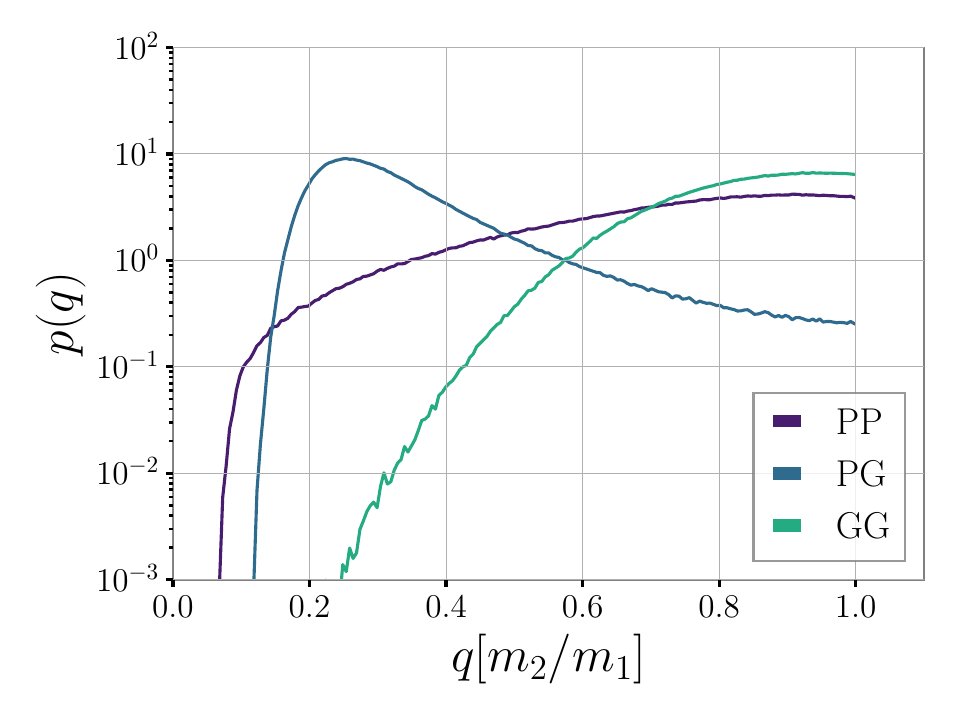}
  \caption{The one-dimensional marginal distribution of the two-dimensional density shown in Figure~\ref{fig:hist2d}. The chirp mass spectrum, as shown in the upper panel, features three peaks at $\mathcal{M} \sim 8 M_\odot$, $14 M_\odot$, and $28 M_\odot$.
  The mass ratio spectrum reveals a bump at $q \sim 0.2$ for the PG population,  highly equal-mass binaries in the GG population, and a smooth mass ratio distribution for the PP population.
  }
  \label{fig:hist1d}
\end{figure}

Figure~\ref{fig:hist1d} shows the 1-D marginal distributions derived from each subpopulation model.
The chirp mass distributions align with the three peak structures observed in the GWTC-3 chirp mass spectrum, i.e., $(8 M_\odot, 14 M_\odot, 28 M_\odot)$. 
The mass ratio distributions exhibit a bump at $q \sim 0.2$ for the PG population, highly equal-mass binaries in the GG population, and a smooth mass ratio distribution for the PP population.

\begin{figure*}
  \includegraphics[width=2\columnwidth]{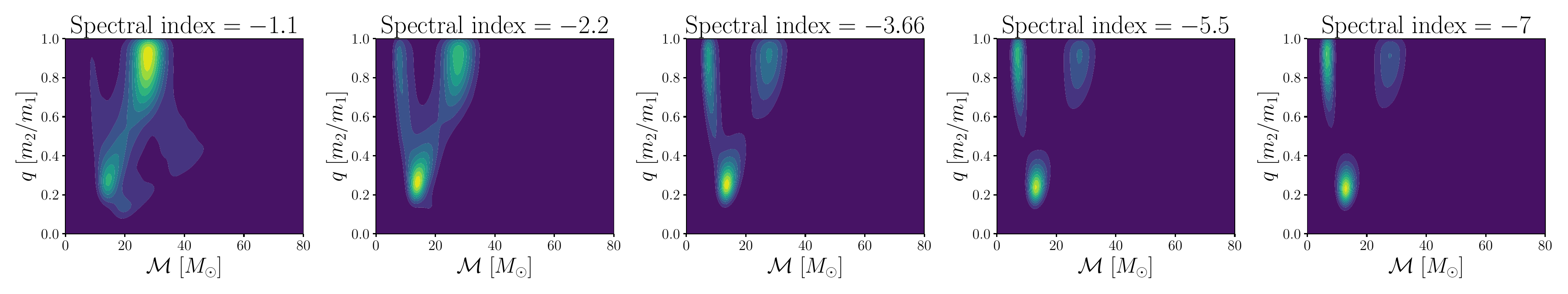}
  \caption{
  The exploratory models with varying spectral indices, $-\alpha$, in the space of chirp mass and mass ratio, $(\mathcal{M}, q)$. It ranges from a flat spectral index (left panel) to a sharp spectral index (right panel). In these exploratory plots, each subpopulation model has the same relative abundance, $1/3$.
  }
  \label{fig:demo_mchirp_q}
\end{figure*}

Figure~\ref{fig:demo_mchirp_q} illustrates various potential outcomes of our population model with different values for the spectral index.
We have fixed the relative abundances for each subpopulation model at equal weights, $(\lambdapp, \lambdapg, \lambdagg) = (1/3, 1/3, 1/3)$.
The spectral index, $-\alpha$, is varied from $-7$ to $-2$.
A flatter spectral index results in a more diffuse (less peaked) density of the PP and PG models in the $(\mathcal{M}, q)$ space.
Conversely, a steeper spectral index (e.g., $-\alpha = 7$) results in a more distinct separation of the density of each subpopulation model in the $(\mathcal{M}, q)$ space.

\begin{figure}
  \includegraphics[width=\columnwidth]{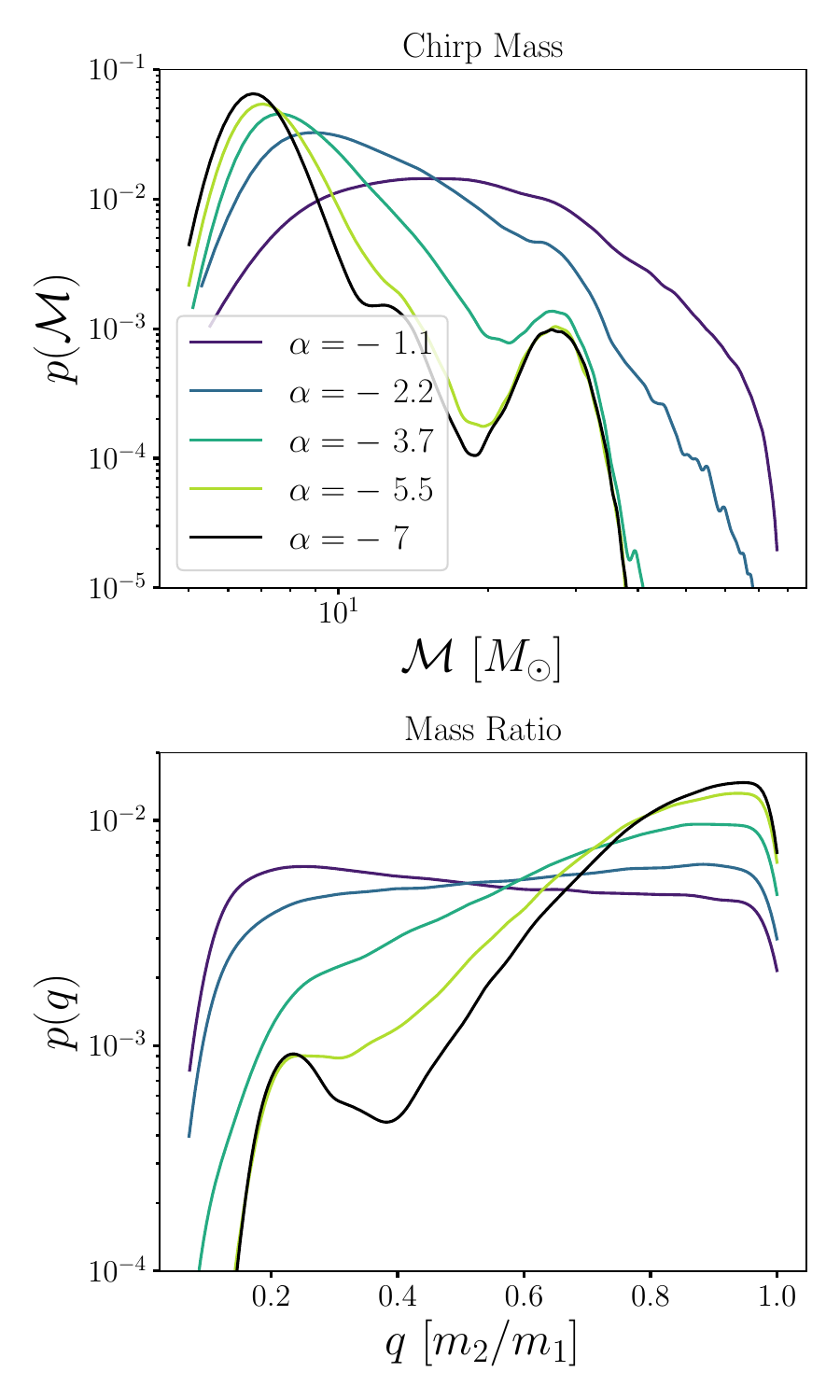}
  \caption{
  The exploratory models with varying spectral indices, $-\alpha$, on the chirp mass ($p(\mathcal{M})$) and mass ratio ($p(q)$) marginal distributions.
  The relative abundance is fixed to $(\lambdapp, \lambdapg, \lambdagg) = (0.92, 0.03, 0.05)$, matching the maximum a posteriori (MAP) of the model averaging results in Section~\ref{subsec:model_averaging_inference}.
  }
  \label{fig:demo_1d_mchirp_q}
\end{figure}

Figure~\ref{fig:demo_1d_mchirp_q} presents the 1D marginal  distributions with varying spectral indices.
The relative abundances are set to $(\lambdapp = 0.92, \lambdapg = 0.03, \lambdagg = 0.05)$, which are close to the inferred relative abundance from the model averaging results in Section~\ref{subsec:model_averaging_inference}.
The chirp mass spectrum exhibits three peaks at $\mathcal{M} \sim 8 M_\odot$, $14 M_\odot$, and $28 M_\odot$ for $-\alpha \lesssim -3.7$.
For $-\alpha \gtrsim -3.7$, the chirp mass spectrum shows a relatively uniform density across the mass ratio spectrum.

\subsection{\label{subsec:average_spectrum} Average mass spectrum}

Our model operates on the average mass spectrum for individual black holes, treating black hole masses without distinguishing them as primary or secondary.
Thus, the fiducial values for our model's shape parameters, $\lambdavec = (-\alpha, \mu, \sigma)$, will be different from those defined by the {\powerpeakmodel} published in GWTC-3.
We transform the primary and secondary masses, $(m_1, m_2)$, from the {\powerpeakmodel} into a single black hole mass spectrum.
We then fit a combination of the power-law and Gaussian model to this average mass spectrum, obtaining the fiducial values for our population model.
This single mass spectrum reflects what we aim to represent by $(\ma, \mb)$. Throughout the paper, this will be referred to it as the ``average mass spectrum'' for black holes.

\begin{figure}
  \includegraphics[width=\columnwidth]{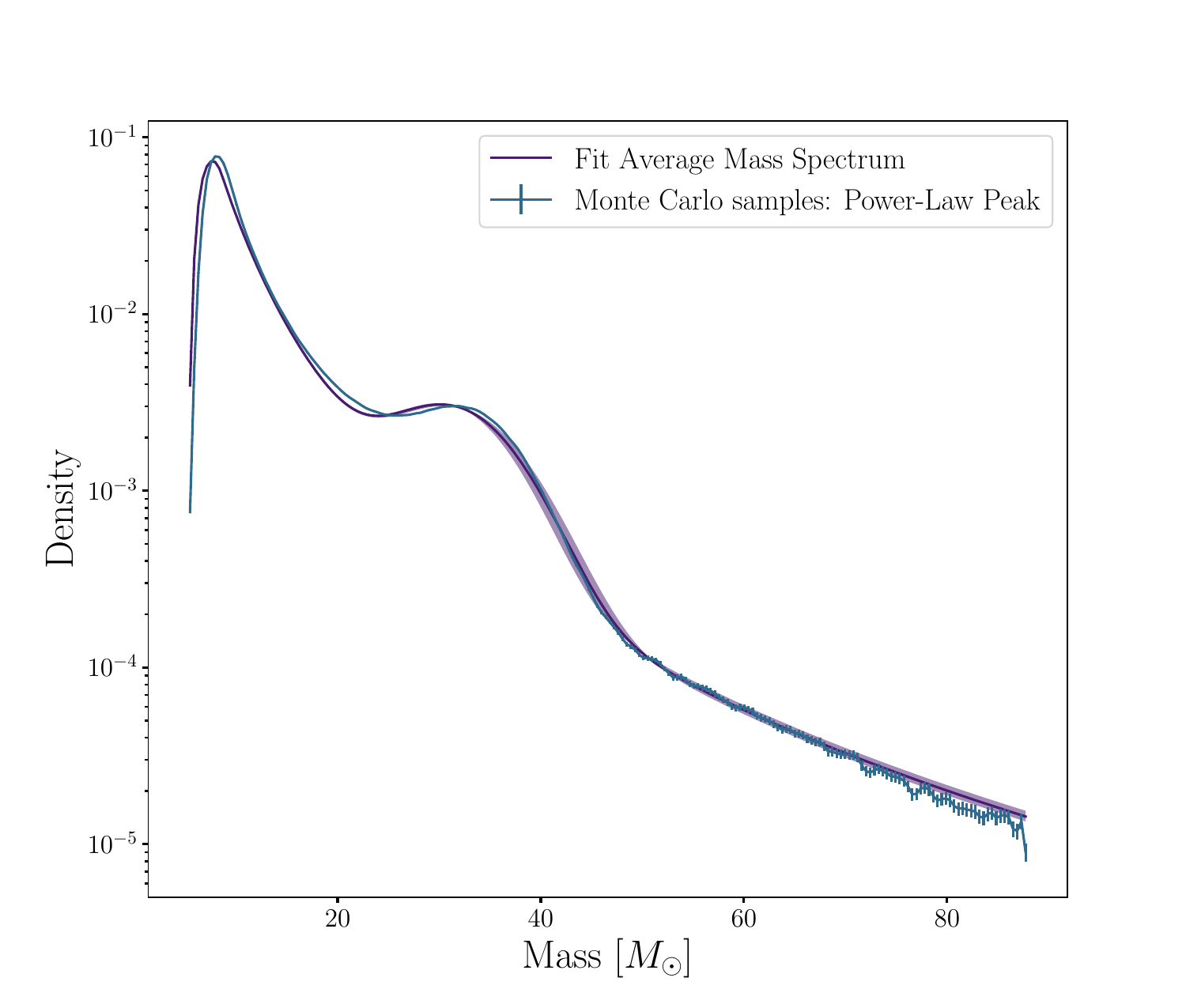}
  \caption{The average mass spectrum sampled from GWTC-3 {\powerpeakmodel} model and the best-fit mass spectrum with the fiducial values of our population model.
  The data points represent the Monte Carlo samples from the {\powerpeakmodel} with best-fit parameters from \citep{2023PhRvX..13a1048A}, with the Poisson uncertainty of the Monte Carlo samples.
  The purple line represents the best-fit power-law function with a Gaussian peak to the sampled average mass spectrum.}
  \label{fig:average_mass_spectrum}
\end{figure}

The primary mass distribution in the {\powerpeakmodel} is parameterized as a combination of a power-law and a Gaussian distribution:
\begin{equation}
  \begin{split}
    p(m_1 &\mid -\alpha, \delta_m, \mmax, \mmin, \mu, \sigma, \lambda_p)
    =\\
    &(1 - \lambda_p) \mathcal{B}(m_1 \mid -\alpha, \mmax, \mmin, \delta_m)
    \\
    &+\lambda_p \mathcal{G}(m_1 \mid \mu, \sigma, \mmin, \delta_m).
  \end{split}
  \label{eq:power_law_peak}
\end{equation}
Note that we have integrated the smoothing kernel directly into the functions of $\mathcal{G}$ and $\mathcal{B}$.
Besides the primary mass model, the {\powerpeakmodel} model also includes a mass ratio model conditioned on $m_1$, which is parameterized as follows:
\begin{equation}
  p(q \mid \beta, m_1, \mmin, \delta_m) \propto q^\beta S( q m_1 \mid \mmin, \delta_m).
  \label{eq:power_law_peak_mass_ratio}
\end{equation}
Here, the $S$ function refers to the smoothing kernel, as previously presented in Eq~\ref{eq:smoothing_kernel}. This function is a conditional probability for the secondary mass, represented as $q \, m_1$. The desired average mass spectrum combines both $(m_1, m_2)$.

Unfortunately, there is no straightforward analytical method to transform the probability density function of {\powerpeakmodel} model to an average mass spectrum.
This difficulty is primarily because of the smoothing kernel at the low-mass end in both primary mass and mass ratio parameterizations.
Even if we were to ignore this smoothing kernel, the combination of the probability density functions between $m_1$ and $q$ via $m_2 = q m_1$ would still lead to the $m_2$ integral that is not computable analytically.
Therefore, we have opted for a numerical strategy to acquire the average mass spectrum from {\powerpeakmodel}.

We generate Monte Carlo samples for $(m_1, m_2)$ using the fiducial values from Ref.~\cite{2023PhRvX..13a1048A} (see Table~\ref{table:fiducial}), then merge these samples to create a unified average mass spectrum.
We then apply a Kernel Density Estimation (KDE) on the combined values of $(m_1, m_2)$, thereby deriving the average mass spectrum.
This spectrum, shown in Figure~\ref{fig:average_mass_spectrum}, generally preserves the original shape of the {\powerpeakmodel} model.
We fit a mix of the power-law and Gaussian model to this numerically-derived average mass spectrum using the same parameterization in Eq~\ref{eq:power_law_peak}.
From this fitting, we obtain the new shape parameters as fiducial values for our population model.

Table~\ref{table:fiducial} shows the best-fit parameters from fitting the average mass spectrum.
The power-law spectral index, $\alpha$, rises from around 3.5 to roughly 3.7. This shows a steeper power-law shape in the average mass spectrum than in the primary mass.
This is expected, given that the secondary mass is lighter than the primary one, leading to a steeper average power-law.
The Gaussian bump shifts from around 33.5 $M_\odot$ to approximately 31.5 $M_\odot$, with its standard deviation expanding from around 4.6 $M_\odot$ to roughly 5.6 $M_\odot$.
We outline the details of this average mass spectrum fitting procedure in Appendix.~\ref{sec:appendix_average_mass_spectrum}.

\begin{table*}
	\centering
	\caption{The fiducial shape parameters for our population model, transforming the fiducial values of {\powerpeakmodel} to our average mass spectrum parameterization.
  The uncertainty in converting the shape parameters from one population model to another can be arbitrarily small, depending on the number of Monte Carlo samples used to construct the KDE for the average mass spectrum.
  Therefore, we do not include this uncertainty in the table.
  We do not vary $\mmin$ or $\mmax$.
  }
	\label{table:fiducial}
	\begin{tabular}{lcccc}
		\hline
		Parameter & Description & Best-fit values & {\powerpeakmodel} values\\
		\hline
		$\delta_m$   & The $\delta m$ for the low-mass mass spectrum smoothing & $4.62$  &  $4.95$   \\
		$\mmax$   & Maximum mass bound for the power-law model & $87.73$   &  $87.73$ \\
    $\mmin$   & Minimum mass bound for both power-law and Gaussian models & $5.06$   &  $5.06$ \\
		$-\alpha$ & Spectral index of the power-law & $3.66$ &  $3.51$ \\
    $\mu$ & Mean of the Gaussian model & $31.59$ &  $33.56$ \\
    $\sigma$ & Standard deviation of the Gaussian model & $5.51$ &  $4.61$ \\
    $\lambda_p$ & Mixing fraction of the Gaussian model & $0.034$ &  $0.038$ \\    
		\hline
	\end{tabular}
\end{table*}





\section{\label{sec:selection_effect} Population Inference}

In this section, we explain the population inference framework implemented in this work. Section~\ref{subsec:inference} specifies the posterior of the mixing fractions that we aim to infer, taking into account the detection efficiency of the gravitational wave detectors. Section~\ref{subsec:model_averaging} demonstrates the model averaging approach we have adopted to marginalize over the shape parameters.

\subsection{\label{subsec:inference}Inference Framework}
We now discuss how we infer the hyperparameters of the population model.
The population inference is determined using a set of $\Nobs$ gravitational wave events with data $\data_i$ for the $i^{\text{th}}$ event. The set of data for the entire catalog will be denoted as $\{\data_i\}$.
In this work, we use GWTC Releases 1, 2, and 3 with a selection criteria specified by a False Alarm Rate (FAR) $< 1\,\mathrm{yr}^{-1}$, which includes 73 BBH events.
For GWTC-1, we use the re-analysis of the events in GWTC-2.1 \footnote{\url{https://zenodo.org/records/6513631}} \citep{gwtc2_1:2021}.
Compared to Ref.~\cite{2023PhRvX..13a1048A}, we do not include \texttt{GW170817}, \texttt{GW200105\_162426}, and \texttt{GW190426\_152155}, as their chirp masses are below the minimum chirp mass of our population model.
For this work, we use only the event posteriors of chirp mass and mass ratio with a combined analysis of \texttt{C01:IMRPhenomXPHM} \citep{Pratten:2021} and \texttt{C01:SEOBNRv4PHM} \citep{Ossokine:2020} waveforms.


We define some notation below to align with the notation in the literature.
We differentiate between the event parameters, $\thetavec = \{\mathcal{M}, q\}$, and the population hyperparameters, $\Lambdavec$. 
The population hyperparameters encompass the mixing fractions, or relative abundances, which are given by ${\boldsymbol{\psi}} = \{\lambdapp, \lambdapg, \lambdagg\}$, as well as the shape parameters, symbolized by $\lambdavec \equiv (-\alpha, \mu, \sigma)$. We use the subscript $a$ denotes which hyperparameter set comes from which subpopulation model, namely $\classa \in \{\classpp, \classpg, \classgg\}$. For clarity, we use $a = \{\mathrm{PP}, \mathrm{PG}, \mathrm{GG}\}$ to represent each model.
The mixing fractions and shape parameters jointly describe the entire population model: $\Lambdavec \equiv \lambdavec \cup \boldsymbol{\psi}$.

Our primary goal is to infer the relative abundance of each subpopulation model, and the mixing fraction posterior is defined as follows~\cite{Loredo:2004,Vitale:2020aaz,Taylor:2018iat,Mandel:2018mve}
\begin{equation}
  \begin{split}
    p(\boldsymbol{\psi} &\mid \{\data_i\}, \{\trigger\}, \Nobs, \lambdavec)
    \propto\\
    &\frac{
      p(\boldsymbol{\psi}) p(\trigger \mid \Lambdavec)^{-\Nobs}
    }{
      p(\{\data_i\}, \{\trigger\}, \Nobs)
    }
    \prod_{i=0}^{\Nobs} \Lobs_i\,.
  \end{split}
  \label{eq:psi_posterior}
\end{equation}
Here, we use a Dirichlet prior over the mixing fractions, $\boldsymbol{\psi}$:
\begin{equation}
  \begin{split}
    p(\lambdapp, \lambdapg, \lambdagg&\mid  \alpha_1, \alpha_2, \alpha_3) = \\
    &\frac{1}{\mathbb{B}(\alpha_1, \alpha_2, \alpha_3)}
    \left(
      \lambdapp^{\alpha_1 - 1} + \lambdapg^{\alpha_2 - 1} + \lambdagg^{\alpha_3 - 1}
    \right).
  \end{split}
\end{equation}
Here, the normalization factor, $\mathbb{B}(\alpha_1, \alpha_2, \alpha_3)$, is a multivariate beta function
\begin{equation}
  \mathbb{B}(\alpha_1, \alpha_2, \alpha_3) = \frac{\Gamma(\alpha_1)\Gamma(\alpha_2)\Gamma(\alpha_3)}{\Gamma(\alpha_1 + \alpha_2 + \alpha_3)}.
\end{equation}
We use a Dirichlet prior to ensure that the mixing fractions sum up to one, $\lambdapp + \lambdapg + \lambdagg = 1$.
This reduces the number of parameters we need to infer to two.
We opt for a non-informative prior, setting $(\alpha_1, \alpha_2, \alpha_3) = (1, 1, 1)$.
This ensures we do not initially favor any specific mixing fraction.
The relative abundances, $(\lambdapp, \lambdapg, \lambdagg)$, physically represent the fraction of BBHs coming from each mixing scenario.

Our population model adopts the average mass spectrum approach (see Section~\ref{subsec:average_spectrum}), so it starts with the shape parameters that are close to the best-fit fiducial values from GWTC-3's {\powerpeakmodel}.
This work primarily focuses on estimating the mixing fraction between the $35 M_\odot$ Gaussian bump and the power-law population, not accurately estimating the shape parameters.
To simplify the computation, the uncertainty in the shape parameters is taken into account through model averaging.
We use a set of pre-computed models within a Latin hypercube of shape parameters, which will be detailed in Section~\ref{subsec:model_averaging}.

In Eq.~\ref{eq:psi_posterior}, we have implicitly marginalized over the total rate of BBH mergers~\cite{Fishbach:2018edt}.
We have incorporated the concept of ``detection'' in the formalism by introducing the trigger term, $\{\trigger\}$, into our notation.
This term represents the criteria that determines if an individual event is selected, typically based on a specific signal-to-noise threshold of the observational instrument. 
Mathematically, the probability of detection given an actual realization of data is defined as 
\begin{equation}
p(\trigger| \data_i) = 
\begin{cases}
0 & \rho(\data_i) < \rho_{\rm threshold} \,, \\
1 & \rho(\data_i) \geq \rho_{\rm threshold} \,, \\
\end{cases}
\label{eq:tigger}
\end{equation}
where $\rho(\data_i)$ defines some deterministic calculation on the data (the signal-to-noise ratio, for example) which classifies data as containing an event or not, based on some threshold value $\rho_{\text{threshold}}$.
This notation is inherited from Ref.~\cite{Perkins:2023} which gives a detailed explanation of work originally derived in past literature~\cite{Loredo:2004,Vitale:2020aaz,Taylor:2018iat,Mandel:2018mve}.
We use $p(\trigger \mid \Lambdavec)$ to represent the detection efficiency, which quantifies the proportion of detectable sources based on the population model represented by population parameters, $\Lambdavec$.
The detection efficiency, $p(\trigger \mid \Lambdavec)$, can be explicitly expressed as follows:
\begin{equation}
  \begin{split}
    p(\trigger \mid \Lambdavec)
    &= \int \dd \data \int \dd \thetavec \,p(\trigger \mid \data) p(\data \mid \thetavec) p(\thetavec \mid \Lambdavec)\\
    &= \int \dd \thetavec \,p(\trigger \mid \thetavec) p(\thetavec \mid \Lambdavec).
  \end{split} 
  \label{eq:alpha} 
\end{equation}
Here, $p(\trigger \mid \thetavec)$ is known as the detection probability and depends on the event parameters, $\thetavec$, not population parameters, $\Lambdavec$.
Note that the concept of detection fundamentally relies exclusively on the data itself, $p(\trigger \mid \data)$ defined above, and is only connected to the event parameters $\thetavec$ through the event likelihood $p(\data \mid \thetavec) $. The detection probability implicitly marginalizes over this hierarchical relationship.
Mathematically, this quantity is defined as 
\begin{equation}
    p(\trigger \mid \thetavec) \equiv \int p(\trigger \mid \data) p(\data \mid \thetavec) \dd \data \,.
\end{equation}
As an approximation to this integral, we use the calculation for detection probability graphically shown in Fig.~(3) of Ref.~\cite{Perkins:2021} denoted as $p_{\text{det}}(\thetavec)$ in that work, which is pre-marginalized over extrinsic parameters (sky location and orientation) using standard distributions (uniform on the sky and uniform in orientation)~\cite{Finn:1995ah,Finn:1992xs,Dominik:2014yma}.
The details of that calculation can be found in Ref.~\cite{Perkins:2021}, for example, but essentially amounts to evaluating this sky-location-averaged detection statistic as a function of the SNR for an optimally oriented binary.
We neglect the BH spin in this calculation, which should have a small effect on the population-averaged detection rate or the sensitive volume. While spins can drastically increase the detectability of individual events~\citep[e.g.][]{Tiwari:2018,PhysRevD.74.041501}, the quantity of interest in hierarchical inference is Eq.~\ref{eq:alpha}, which includes information about the distribution of spins coming from a population model. As the latest inference on the spin distributions of BBH indicate a distribution clustered around $\chi_{\rm eff} \sim 0$, this factor should be negligible. See Ref.~\cite{Tiwari:2018} and their calculation of the impact on the detectable volume for isotropically distributed spins (the distribution most consistent with LVK's results), which indeed shows negligible impact. 
This means $p(\trigger \mid \thetavec)$ simply involves an integral over the mass parameters $\mathcal{M}$ and $q$.

We evaluate Eq.~\ref{eq:alpha} with a fixed Power Spectral Density (PSD) function for all events, where we use the analytic \texttt{AdVMidHighSensitivityP1200087}~\citep{lalsuite:2018,Biwer:2019} PSD throughout this work, but we also discuss the impact of using different PSDs in Appendix~\ref{subsec:appendix_psd}.
The pre-marginalized approach in Ref.~\cite{Perkins:2021} factors out the detector dependent quantities from SNR on $p_{\text{det}}(\thetavec)$ is explained in Ref.~\cite{Dominik:2015,Finn:1993,Finn:1996}.
In this work, injections with expected SNR less than 8 ($\rho_\mathrm{threshold} = 8$) are not considered detected. To calculate the SNR, we employed the \texttt{IMRPhenomD}~\cite{Khan:2016,Husa:2016,Purrer:2023}.
We discuss the priors used in calculating $p_{\text{det}}(\thetavec)$ in Appendix~\ref{subsec:appendix_psd}.
While this semi-analytic method is an approximation of the more accurate method (which involves injecting signals from known distributions into the entire detection pipeline ontop of real data), it has been utilized extensively in the literature~\citep[e.g.,][]{Farah:2021,Gerosa:2019dbe,Fishbach:2017,Tiwari:2018} and shown to accurately capture the salient features of selection bias~\citep{2019ApJ...882L..24A,Dominik:2014yma,LIGOScientific:2016ebi,LIGOScientific:2016kwr}. The main approximations of this method relevant to this study relate to the non-Gaussian and non-stationarity of gravitational wave detectors, which are reasonable approximations in current detector networks.

In practice, we numerically estimate the detection efficiency by Monte Carlo sampling the event parameters, $\thetavec$, under a given set of population parameters, $\Lambdavec$:
\begin{equation}
  \begin{split}
    p(\trigger \mid \Lambdavec) &\approx \frac{1}{S} \sum_{i=0}^{S} p(\trigger \mid \thetavec_i);\\
    \thetavec_i &\sim p(\thetavec \mid \Lambdavec),
  \end{split}
\end{equation}
where we generate $S = 500\,000$ samples of event parameters according to  $p(\thetavec \mid \Lambdavec)$.
For the event parameters $\thetavec$,
we use primary/secondary masses and luminosity distance, $L$.
For the primary and secondary masses, we sample from the population model.
For the luminosity distance, we sample with a prior of $p(L) \propto L^{2}$, which is uniform in volume.

The mixture model construction allows us to simplify the estimation of detection efficiency to the sum of detection efficiency of each subpopulation model:
\begin{equation}
  \begin{split}
    p(\trigger \mid \Lambdavec) &= \sum_{a=\{\mathrm{PP}, \mathrm{PG}, \mathrm{GG}\}} \psi_a p(\trigger \mid \lambdavec, \classa);\\
    p(\trigger \mid \lambdavec, \classa) &= \int \dd \thetavec p(\trigger \mid \thetavec) p(\thetavec \mid \lambdavec, \classa).
  \end{split}
\end{equation}
This substantially simplifies computing the detection efficiency.
In practice, we pre-compute Monte Carlo samples from subpopulation models with a fixed set of shape parameters, $\lambdavec$.
Therefore, we can quickly compute the $p(\trigger \mid \Lambdavec)$ via a weighted sum by varying mixing fractions, $\boldsymbol{\psi}$.

For the likelihood of each data set $\data_i $, $\Lobs_i \equiv p(\data_i | \Lambdavec)$, we ultimately compare the distribution of the event parameters to the predictions from our population model.
For each event, $i$, the likelihood is
\begin{equation}
  \begin{split}
    \Lobs_i &= \sum_{a=\{\mathrm{PP}, \mathrm{PG}, \mathrm{GG}\}} p(\classa \mid \Lambdavec ) p(\data_i \mid \classa, \Lambdavec) , \\
    &= \sum_{a=\{\mathrm{PP}, \mathrm{PG}, \mathrm{GG}\}} \psi_a \, p(\data_i \mid \classa, \lambdavec) , 
  \end{split}
  \label{eq:class_likelihood}
\end{equation}
where $p(\classa \mid \Lambdavec)$ is the prior probability that an event belongs to each subpopulation model.
Eq~\ref{eq:class_likelihood} describes the likelihood of the data marginalized over the class of the event.
We can explicitly write it as
\begin{equation}
    p(\data_i \mid \classa, \lambdavec) = \int \dd \thetavec \, p(\data_i \mid \thetavec) p(\thetavec \mid \classa, \lambdavec) ,
\end{equation}
where the integral can be approximated via importance sampling \cite{Hogg:2010}
\begin{equation}
  \begin{split}
    \int \dd \thetavec \, p(\data_i \mid \thetavec) &p(\thetavec \mid \classa, \lambdavec) \approx\\
    &\frac{1}{S} \sum_{c = 0}^{S} \frac{p(\thetavec_c \mid \classa, \lambdavec)}{\pi(\thetavec_c)}
  \end{split}
  \label{eq:likelihood_importance_sampling}
\end{equation}
with
\begin{equation}
  \thetavec_c \sim p(\thetavec \mid \data_i).
\end{equation}
Here, $p(\thetavec \mid \data_i)$ is the event posterior provided by LVK's template fitting.
We need to divide the posterior by the event parameter prior, $\pi(\thetavec)$, to get the likelihood for each event.
The event parameter priors used by LVK are
\begin{equation}
  \begin{split}
    \pi({\cal M}) &\propto {\cal M}\\
    \pi(q)        &\propto \frac{(1 + q)^{2/5}}{q^{6/5}}.
  \end{split}
\end{equation}

Finally, $p(\thetavec_c \mid \classa, \lambdavec)$ represents evaluating the event posterior samples on the histogram likelihoods shown in Figure~\ref{fig:hist2d}.
Thus, Eq~\ref{eq:likelihood_importance_sampling} simply states that we evaluate the Monte Carlo samples of event likelihoods on the numerical probability density derived from the histograms.

\subsection{\label{subsec:model_averaging} Model averaging}

We describe our numerical method to integrate the population posterior over $\boldsymbol{\psi}$ with a fixed set of shape parameters $\lambdavec$ in Eq~\ref{eq:psi_posterior}.
Our primary goal is to infer $\boldsymbol{\psi}$, not accurately estimate the shape parameters, $\lambdavec$.
One way to marginalize over the uncertainty of $\lambdavec$ in this case is through model averaging.
We run a fixed set of 1,000 choices for the shape parameters, $\lambdavec$, and obtain the MCMC posterior of $\boldsymbol{\psi}$ for each mixture model.
For the model averaging approach, we approximate the marginalization by treating each $\lambdavec$ as a model, and we use the posterior of $\lambdavec$ to weight the contribution of each model to the population posterior through a Monte Carlo sum with a discrete set of $\lambdavec$:
\begin{equation}
  \begin{split}
    p(\boldsymbol{\psi} \mid \{\data_i\}, \{\trigger\}, &\Nobs)
    \simeq\\
    \frac{1}{S}\sum_{j=1}^{S} p(\boldsymbol{\psi} &\mid \{\data_i\}, \{\trigger\}, \Nobs, \lambdavec^j) \times\\
    &p(\lambdavec^j \mid \{\data_i\}, \{\trigger\}, \Nobs)
  \end{split}
  \label{eq:model_averaging}
\end{equation}
with
\begin{equation}
  \lambdavec^j \sim p(\lambdavec).
\end{equation}
The shape parameters in the $\lambdavec$ space have a prior volume $\alpha \sim \mathcal{U}(1, 6)$, $\mu \sim \mathcal{U}(25, 40)$, and $\sigma \sim \mathcal{U}(3, 8)$. We sample $\lambdavec$ using a Latin hypercube,
maximizing coverage of the parameter space.
This effectively searches the hyperspace of $\lambdavec$ and marginalizes out uncertainty in the shape parameters.
The posterior $p(\lambdavec \mid \{\data_i\}, \{\trigger\}, \Nobs)$, is obtained by evaluating the model evidence for all events, where we assume a uniform, model prior for each choice of $(-\alpha, \mu, \sigma)$.

\section{\label{sec:results} Results}

In this section, we present the inference results of our population model.
First, in Section~\ref{subsec:fiducial_inference}, we discuss the inference results of the population model with fixed shape parameters, as obtained from the average mass spectrum approach.
Subsequently, we present the results of model averaging in Section~\ref{subsec:model_averaging_inference}, where we marginalize over the shape parameters.

\subsection{\label{subsec:fiducial_inference} Fiducial model}

\begin{figure}
  \includegraphics[width=\columnwidth]{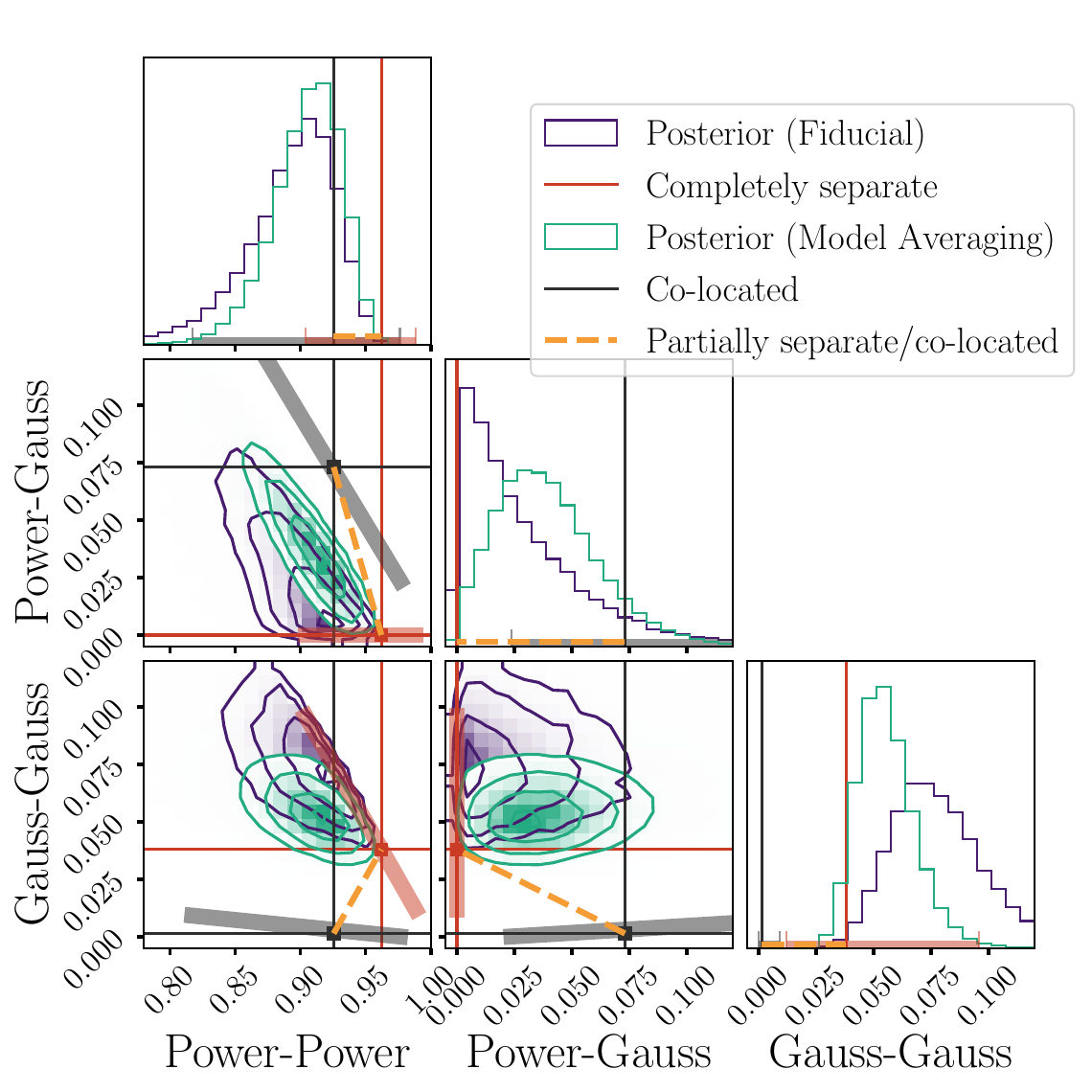}
  \caption{
  Posterior probability for mixing fraction parameters, $\boldsymbol{\psi}$ under different model assumptions (``Fiducial'' model in purple and the ``Model Averaging'' in green).
  Two extreme hypothetical scenarios are also shown: (1). Red error bars show the case where the Gaussian bump is completely separate from the power-law population. (2). Black error bars show the case where the Gaussian bump and power-law populations are co-located.
  These hypothetical scenarios are defined using $\lambda_\mathrm{peak} = 3.8^{+5.8}_{-2.6}\%$ for the Gaussian bump reported in LVK~\cite{2023PhRvX..13a1048A}.
  The orange dashed lines represent the ``Partially Separate/Co-located'' scenario, showing a situation in which a portion of the Gaussian bump black holes is co-located with the power-law distribution, while the remainder is separate.
  }
  \label{fig:MCMC_model_average}
\end{figure}

We have obtained the fiducial shape parameters, $(-\alpha, \mu, \sigma, \delta_m, \mmax, \mmin)$, from the average mass spectrum as detailed in Section~\ref{subsec:average_spectrum}.
The posterior distribution for the mixing fraction is shown in Figure~\ref{fig:MCMC_model_average}.
The $95\%$ posterior confidence intervals for the mixing fractions are $(\lambdapp = 92.9^{+2.2}_{-11.1}\%, \lambdapg < 8.7\%, \lambdagg = 7.1^{+4.8}_{-3.0}\%)$.
Interestingly, the mode of the relative abundance of the PG mixing, $\lambdapg$, is consistent with zero. This indicates some evidence for the separation of the two populations, based on the fiducial shape parameters.

To illustrate two extreme situations, we define two hypothetical population model scenarios: ``Completely separate'' and ``Co-located.'' In the ``Completely separate'' scenario, we assume that the power-law and Gaussian populations are entirely separate, resulting in no PG mixing. The relative abundances of the PP and GG populations reflect the fractions of power-law and Gaussian populations in the single-mass distribution, respectively. For the completely separate scenario, we adopt $(\lambdapp, \lambdapg, \lambdagg) = (96.2^{+2.6}_{-5.8}\%, 0.0\%, 3.8^{+5.8}_{-2.6}\%)$. The choice of $\lambdagg = 3.8^{+5.8}_{-2.6}\%$ is based on the relative abundance of the Gaussian bump from the {\powerpeakmodel} model, with the $90\%$ credible intervals reported in Ref.~\cite{2023PhRvX..13a1048A} indicating a relative abundance of the Gaussian bump, $\lambda_\mathrm{peak} = 3.8^{+5.8}_{-2.6}\%$. Assuming the Gaussian bump population is separate from the power-law, this implies a GG mixing with a relative abundance of approximately $3.8^{+5.8}_{-2.6}\%$ and a zero mixing abundance, $\lambdapg \approx 0\%$.
In the ``Co-located'' scenario, we assume the power-law and Gaussian populations are completely mixed together, resulting in the mixing fraction of PG equal to $2 \times \lambda_\mathrm{peak} (1 - \lambda_\mathrm{peak})$, giving $(\lambdapp, \lambdapg, \lambdagg) = ((1 - \lambda_\mathrm{peak})^2, 2 \times \lambda_\mathrm{peak} (1-\lambda_\mathrm{peak}), \lambda_\mathrm{peak}^2) \approx (92.5^{+5.1}_{-10.8}\%, 7.3^{+11.7}_{-5.1}\%, 0.1^{+0.8}_{-0.1}\%)$. 
Interestingly, Figure~\ref{fig:MCMC_model_average} suggests that the posterior distribution from the fiducial model prefers the ``Completely separate'' scenario over the ``Co-located'' scenario, although the error bars from each scenario remain substantial.

\begin{figure*}
  \includegraphics[width=2\columnwidth]{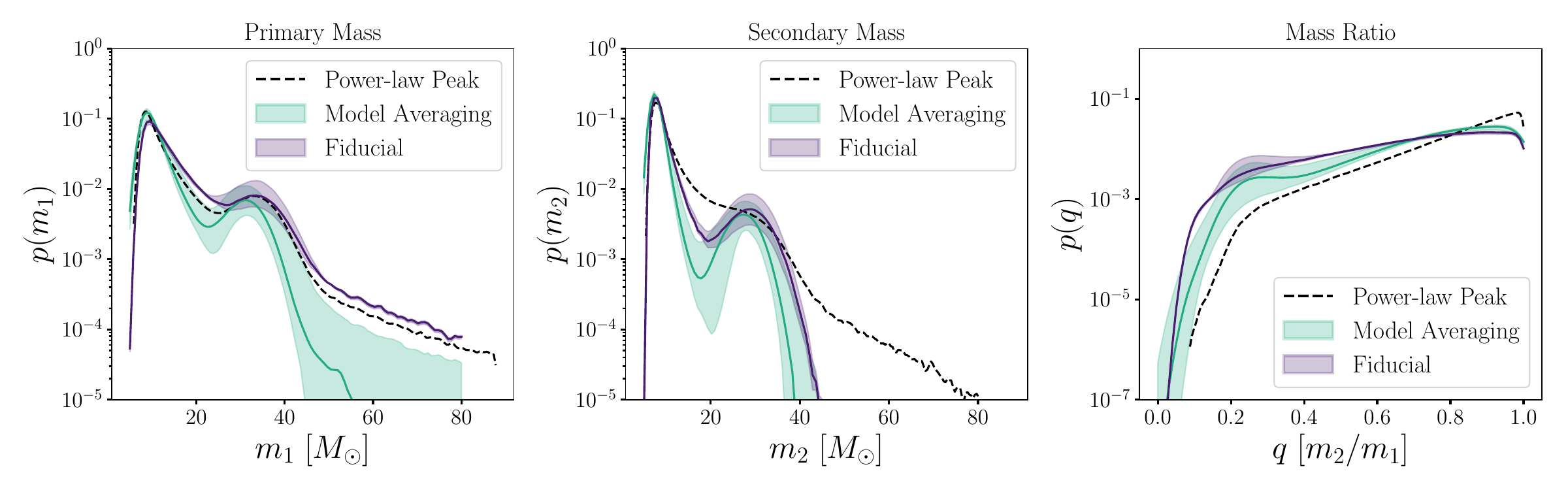}
  \caption{
  Predicted primary/secondary mass and mass ratio functions from the model averaging inference (in light green) and the fiducial inference (in purple).
  The solid lines represent the MAP and the shaded areas represent the $95\%$ confidence intervals.
  The light green lines represent predicted functions sampled from the posterior probability of both $\boldsymbol{\psi}$ and $\lambdavec$.
  The underlying black dashed lines represent the fiducial model of {\powerpeakmodel} from GWTC-3, with the corresponding fiducial parameters detailed in Table~\ref{table:fiducial}. The $95\%$ confidence interval for the ``Fiducial'' inference reflects only the posterior uncertainty in $\boldsymbol{\psi}$ and does not include uncertainty regarding $\lambdavec$. As we fix the minimum mass ($\mmin$) and the maximum mass ($\mmax$), the shape uncertainty at the low and high mass ends is not incorporated.
  }
  \label{fig:prediced_model_average_inference}
\end{figure*}

Figure~\ref{fig:prediced_model_average_inference} presents the predicted primary/secondary mass functions and mass ratio based on our fiducial inference. For comparison, we also include the {\powerpeakmodel} model with its fiducial parameters obtained from Ref.~\cite{2023PhRvX..13a1048A} (also see Table~\ref{table:fiducial}). The primary mass functions show good agreement, which is expected given that in Section~\ref{subsec:average_spectrum} we have constructed our average mass spectrum to match the {\powerpeakmodel} model. The secondary mass functions exhibit some differences. 
The bump in $m_2$ is approximately at $\sim 30 M_\odot$ for both population models. However, the power-law component in the {\powerpeakmodel} is comparatively flatter. This discrepancy might arise from an inherent difference between these two models.
The {\powerpeakmodel} models the $m_2$ via a power-law mass ratio, while our model assumes an average mass spectrum for both $m_1$ and $m_2$. It could mean that the secondary mass spectrum would appear much sharper under our model's assumptions.
Nevertheless, due to the limited dataset size of GW events, inferring the massive end of the mass spectrum remains highly uncertain. Since the primary focus of this paper is on estimating the mixing fraction, we do not emphasize the differences at the tail of the mass spectrum nor trying to infer $\mmax$.

\subsection{\label{subsec:model_averaging_inference} Model averaging}

Figure~\ref{fig:MCMC_model_average} compares the posterior of the mixing fractions, $p(\boldsymbol{\psi} \mid \{\data_i\}, \{\trigger\}, \Nobs)$, (``Model Averaging'' in green) with the posterior from the fiducial shape parameters (``Fiducial''), $p(\boldsymbol{\psi} \mid \lambdavec_\mathrm{fid}, \{\data_i\}, \{\trigger\}, \Nobs)$.
It shows a shift in the posterior mode to approximately the $68-95\%$ confidence contour.
Additionally, the posterior width for the mixing fractions narrows, suggesting that the fiducial shape values do not provide the best fit to the data and has a lower model evidence.
Otherwise, model averaging would result in an increased width of the posterior.
This is expected as our population model differs from the {\powerpeakmodel} model, so the fiducial shape parameters do not provide the best fit to the data.
The uncertainty in the predicted mass spectrum, as illustrated in Figure~\ref{fig:prediced_model_average_inference}, increases under the ``Model Averaging'' approach, particularly due to the varying spectral index of the power-law. This increase in uncertainty suggests that, with a flexible power-law model (with a varying spectral index), our mixture model gets a better fit to the data. This better fit is attributed to the fact that both PP and PG models can better explain the observed data, leading to narrower posteriors of the mixing fractions.

The posterior for the mixing fractions, $(\lambdapp$, $\lambdapg$, $\lambdagg) = (91.9^{+3.2}_{-6.8}\%, 3.1^{+5.0}_{-3.1}\%, 5.0^{+3.2}_{-1.7}\%)$, indicates that at a $95\%$ confidence level,
approximately $3.1\%$ of the binaries in the catalog can be attributed to the mixing of the populations. In Figure~\ref{fig:MCMC_model_average}, compared with the ``Completely separate'' scenario (red error bars) and the ``Co-located'' scenario (black error bars), we observe that, even with varying shape parameters, the model averaging result still shows a preference for the ``Completely separate'' scenario between the Gaussian and power-law populations. Nonetheless, there is a notable shift in the mode of PG mixing to a slightly higher value, $\lambdapg = 3.1^{+5.0}_{-3.1}\%$, which suggests that the PG mixing posterior now locates itself between the ``Completely separate'' and ``Co-located'' scenarios.

\begin{figure}
  \includegraphics[width=\columnwidth]{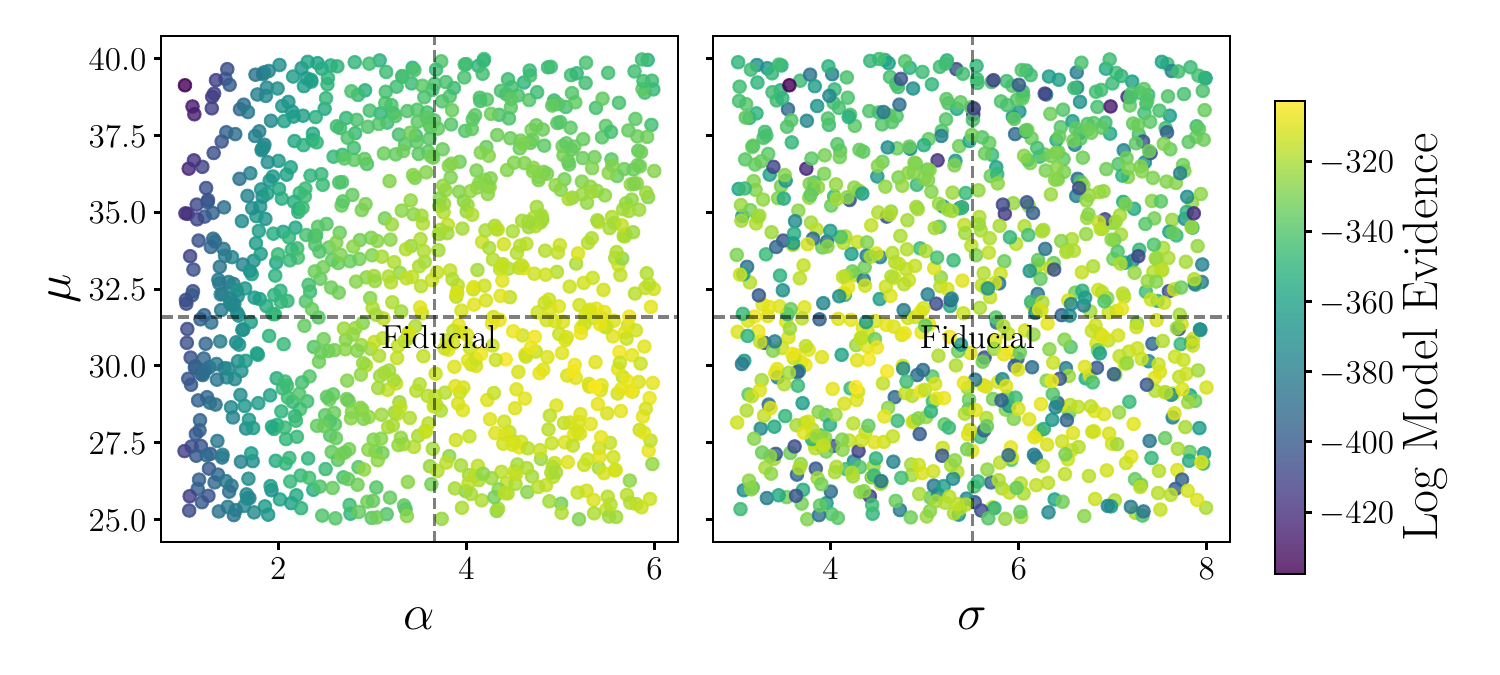}  
  \caption{
  The model evidences from 1,000 population inferences utilized in the ``Model Averaging'', namely, $p(\lambdavec \mid \{\data_i\}, \{\trigger\}, \Nobs)$, where we treat each set of $\lambdavec$ as a model.
  The colors indicate the model evidence for each of the shape parameters, $(-\alpha, \mu, \sigma)$.
  There is no obvious correlation between $\sigma$ and $\mu$, but there is some weak correlation between $\alpha$ and $\mu$ with correlation $\simeq - 0.3$.
  }
  \label{fig:model_average_2d_kde}
\end{figure}

Figure~\ref{fig:model_average_2d_kde} shows the model evidences for each population model used in the model averaging approach. The peaks of the model evidences, compared to the fiducial values from the average mass spectrum (which is at the center of the prior volume), are shifted towards slightly lower $\mu$ and a steeper spectral index (higher $-\alpha$).
The shape parameters at the maximum model evidence are $(-\alpha, \mu, \sigma) = (-5.3675, 29.3125, 4.2675)$, which represents a slight shift in the shape parameters compared to the fiducial values.
There is a $\simeq -0.3$ correlation between the $\alpha$ and the location of the Gaussian bump, $\mu$. Namely, with a steeper slope of the power-law, the Gaussian bump has to move to a lower mass to compensate.

\begin{figure*}
  \includegraphics[width=2\columnwidth]{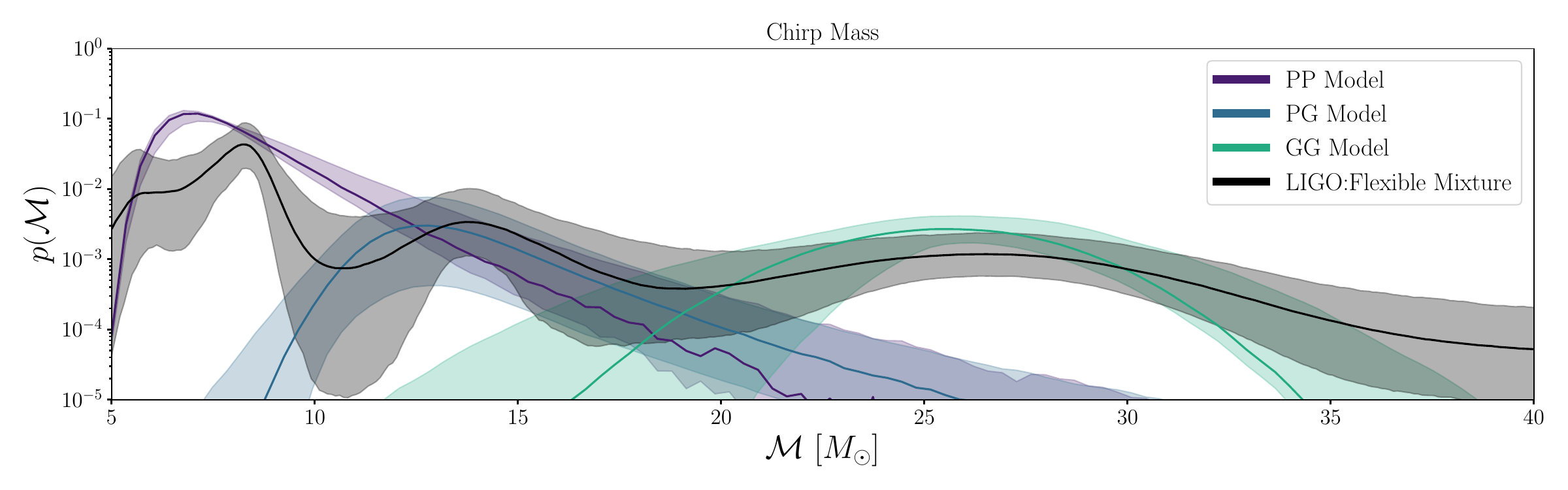}
  \caption{The chirp mass spectra predicted by the subpopulations in our population model: PP (purple), PG (blue), and GG (green).
  The shaded regions represent the $95\%$ confidence intervals, and the subpopulations are normalized to have $\lambdapp + \lambdapg + \lambdagg = 1$.
  The underlying mass spectrum (black) is from the Flexible Mixture model \cite{Tiwari:2021}, and utilizes fitting results from Ref.~\cite{2023PhRvX..13a1048A}.
  }
  \label{fig:predicted_chirp_mass_model_averaging}
\end{figure*}

Figure~\ref{fig:predicted_chirp_mass_model_averaging} presents the predicted chirp mass spectrum from each subpopulation model and compares these predictions with the Flexible Mixture model from Ref.~\cite{2023PhRvX..13a1048A}. The BBH chirp mass spectrum's three-peak structure is captured by the PP ($\mathcal{M} \sim 8 M_\odot$), PG ($\mathcal{M} \sim 14 M_\odot$), and GG ($\mathcal{M} \sim 28 M_\odot$) subpopulations. We do not vary $\mmin$, resulting in an overconfidence in the low-mass of the PP model compared with the predictions of the Flexible Mixture model. However, the relative abundances of both the PG and GG subpopulations align well with the second and third peaks of the chirp mass spectrum.

Our inference results suggest a high fraction of GG mixing and a low fraction of PG mixing ($\lambdagg > \lambdapg$),  indicating that the $35 M_\odot$ Gaussian bump BHs are likely separate from the rest of the population and forming BGBHs.
Existing theoretical models for the Gaussian bump thus need to account for the separation of the Gaussian bump black holes from the rest of the black hole population.

\section{\label{sec:discussion}Discussion}
\subsection{\label{sec:GGBH} High Proportion of Gaussian-Gaussian BBHs.}
Figure~\ref{fig:MCMC_model_average} shows a substantial fraction of Gaussian bump black holes merging with other objects from the Gaussian bump (BGBHs), implying that the Gaussian bump population is almost distinct from the power law population.
This high fraction of BGBHs presents challenges to existing black hole formation theories.
There are at least three explanations which could potentially explain the result:
(1). A cluster of stars forming simultaneously in cosmic time, undergoing supernova explosions, and consequently producing dense environments of $\sim 35 M_\odot$ stellar remnants.
(2). Mass segregation \cite{Binney:2008} in a stellar cluster, causing $\sim 35 M_\odot$ black holes to gravitate towards the cluster's core and merge predominantly with similar-mass black holes.
(3). A distinct population of $\sim 35 M_\odot$ black holes with a distinct spatial distribution, or a different set of host halos, from the black holes in the power-law subpopulation.

The prevailing explanation for the $\sim 35 M_\odot$ Gaussian bump is PPSNe, or the mass accumulation before the pair-instability cutoff at $\gtrsim 40 M_\odot$.
The population of BGBHs challenges our understanding of how PPSNe binaries form, particularly how PPSNe can generate BGBHs without becoming bound to lighter black holes.
One possible explanation is the formation of clusters composed exclusively of PPSNe black holes or PPSNe remnants. However, the specific process behind such cluster formation has not been thoroughly explored or discussed in the literature.
Another explanation is that these massive BGBHs could come from binary star systems (e.g., see Ref.~\cite{Belczynski:2014}), where both stars are already very massive and in the range to produce PPSNe.
However, the kicks from supernova explosion of massive stars will likely destroy the binary system and make them unbound.

Next, mass segregation (see Ref.~\cite{Hopman:2006}) within stellar clusters might lead to the formation of BGBHs. Heavier black holes sinking to the cluster's center would merge with similar-mass counterparts.
This requires an explanation for the production of a high abundance of $\sim 35 M_\odot$ black holes, potentially through PPSNe or PBHs acting as gravitational centers around which globular clusters form (e.g., Ref.~\cite{Lacroix:2018}).

Another possibility is Pop III stars \citep{Kinugawa:2014,Inayoshi:2016}, forming massive stars at high redshifts that evolve into $\sim 35 M_\odot$ black hole at the same time (e.g., Ref.~\citep{Inayoshi:2016}).
To form BGBHs, this scenario requires these stars to form in clusters and evolve simultaneously into supernovae, thus forming BBHs within the same population. This hypothesis could be further tested through its contributions to cosmic reionization around $z \simeq 6$.
However, it's unclear why Pop III stars would preferentially form black holes around $\sim 35 M_\odot$, but not beyond $40 M_\odot$, given the absence of pair-instability limitations at such low-metallicity.

Another explanation for a high fraction of BGBHs is PBHs, which are distributed more like the dark matter halo and thus distinct from luminous matter. If such PBHs exist with masses around $\sim 35 M_\odot$, they would predominantly merge within their group.
Gravitational microlensing constrains the fraction of dark matter in the form of PBHs to be less than $10\%$ of the halo \cite{Alcock:2001,Tisserand:2007,Wyrzykowski:2011,Blaineau:2022,Bird:2023}. However, the merger rate of PBHs remains highly uncertain (e.g., see Ref.\cite{Bird:2016,Sasaki:2018}), making it difficult to predict if $\lambdagg$ is consistent with PBHs.

\subsection{\label{subsec:limitations} Limitation on the Interpretation of the Mixing Fraction Posterior.}
A limitation in comparing two hypothetical scenarios using $\lambda_\mathrm{peak}$ to estimate mixing fractions is that the {\powerpeakmodel} model only measures the Gaussian bump's fraction in the primary mass spectrum, not across \textit{all black holes} in the Universe. Thus, using $\lambda_\mathrm{peak}$ as a proxy for the bump's overall abundance may not be directly applicable. However, this approach likely provides a conservative estimate for $\lambdagg$ of ``Completely separate,'' since the primary mass in a BBH is heavier, suggesting $\lambdagg$ could be overestimated using $\lambda_\mathrm{peak}$. Our interpretation of Figure~\ref{fig:MCMC_model_average} remains unchanged, the mixing fraction posterior aligning better with the ``Completely separate'' scenario.

Even if we assume there is no robust estimate of $\lambda_\mathrm{peak}$, by definition, the ``Completely separate'' scenario would yield $\lambdapg = 0\%$, which is more consistent with our inference results than the ``Co-located'' scenario, which requires $\lambdapg > \lambdagg$, given the power-law abundance is much higher than the Gaussian bump. To make our $\lambdagg = 5.0^{+3.2}_{-1.7}\%$ posterior consistent with the ``Co-located'' case, the relative abundance of the Gaussian bump would need to be approximately $18 - 28\%$ of all black holes which form BBHs, a significant difference from the $\lambda_\mathrm{peak}$ measurement which would be obvious in the inferred primary mass spectrum from GWTC-3. We therefore argue that our inference still suggests that a separate population causes the Gaussian bump in the GWTC-3 catalog.

We assume fixed $(\mmin, \mmax)$, which restricts the explanatory power of the PP and PG models. Secondly, we categorize the black hole population into either a Gaussian bump or a power-law, but the true BBH population might be more complex, containing more than two subpopulations.
Another limitation of our mixing approach is that it does not consider second generation mergers~\citep[e.g.,][]{Gerosa:2021mno}, which could be important for the massive end of the mass spectrum.
We also did not model the common envelope or isolated channel BBHs.
However, we can interpret PP, PG, and GG as isolated channels going through different IMF and metallicity environments, e.g., Ref.~\cite{Olejak:2020} can produce PG BBHs (30-10 $M_\odot$) with low-metallicity progenitors with initial $q < 0.5$.

\section{\label{sec:conclusion}Conclusion}

In this paper, we explore the substructure within the black hole mass spectrum, specifically focusing on the $m_1 \sim 35 M_\odot$ Gaussian bump in the primary mass spectrum and the $\mathcal{M} \sim 14 M_\odot$ peak in the BBH chirp mass spectrum. We investigate these substructures through a two-population mixing scenario, examining a power-law and Gaussian population of black holes in the Universe. 
We define three mixing scenarios: PP binaries, where the power-law population mixes with itself; PG binaries, involving a mix between the power-law population and the Gaussian bump black hole population; and GG binaries, where Gaussian bump black holes merge with themselves. A mixture model was developed to measure the relative abundance of each scenario. The fiducial inference results, aligning with the primary mass spectrum of the {\powerpeakmodel} model without varying the shape parameters of the power-law and Gaussian bump, suggest $(\lambdapp = 92.9^{+2.2}_{-11.1}\%, \lambdapg < 8.7\%, \lambdagg = 7.1^{+4.8}_{-3.0}\%)$. As we vary the shape parameters, including the spectral index of the power-law and the location and width of the bump, our model averaging results indicate $(\lambdapp = 91.9^{+3.2}_{-6.8}\%, \lambdapg = 3.1^{+5.0}_{-3.1}\%, \lambdagg = 5.0^{+3.2}_{-1.7}\%)$. Both sets of results highlight a relatively low PG mixing fraction and a high GG binary mixing fraction, indicating a preference in the GWTC-3 catalog data for a ``Completely separate'' scenario. This suggests that $35 M_\odot$ Gaussian bump black holes are likely separate from the rest of the population.

Our population model's predicted chirp mass spectrum and the relative abundance of each mixing scenario align well with the Flexible Mixture model results presented in Ref.~\cite{2023PhRvX..13a1048A}. The second chirp mass peak at $\mathcal{M} \sim 14 M_\odot$ closely matches the relative abundance of PG binaries, suggesting partial mixing between the power-law and Gaussian bump populations. Although these populations are likely separated, a fraction mixes, giving rise to the second chirp mass peak.

Most past formation channels explaining the $35 M_\odot$ Gaussian bump focus on the primary mass spectrum rather than the 2D BBH mass space. For instance, PPSNe are a popular mechanism for the Gaussian bump, facing challenges in explaining BGBH formation without pairing with lighter black holes. One possibility is that such PPSNe Gaussian bump black holes are typically found in star clusters, where mass segregation might facilitate their merger with similar black holes. However, the likelihood of mass segregation and the fraction of Gaussian bump black holes within star clusters remain uncertain.
Other formation channels, such as black holes originating from low-metallicity Pop III stars or primordial black holes, could also account for the high fraction of BGBHs, given their separation from other high-metallicity stellar-origin black holes. These channels might explain the separate population of BGBHs, but the precise mechanisms for producing $\sim 35 M_\odot$ black holes remain unknown and challenging to pinpoint.

We also acknowledge limitations in our population inference, such as the inflexibility of the power-law population model. However, we anticipate that enhancing model flexibility will likely not significantly alter our current interpretations, due to large error bars and the GWTC-3 catalog's limited size.
We suggest that future work on the formation of Gaussian bump black holes should consider the separation of this population, and the potential channels for forming BGBHs.

\begin{acknowledgments}
We are grateful to the LVC collaborations for publicly releasing their event posterior samples.
The authors would like to thank Natasha Abrams, Macy Huston and George Chapline for fruitful and enlightening conversations which helped during the writing of this article. 
MFH acknowledges the support of a National Aeronautics and Space Administration FINESST grant under No. ASTRO20-0022.
SB acknowledges funding from NASA ATP 80NSSC22K1897.
Computations were performed using the computer clusters and data storage resources of the HPCC, which were funded by grants from NSF (MRI-2215705, MRI-1429826) and NIH (1S10OD016290-01A1).
This work was performed under the auspices of the U.S. Department of Energy by Lawrence Livermore National Laboratory under Contract DE-AC52-07NA27344. The document number is \IMRELEASENO{}. This work was supported by the LLNL-LDRD Program under Project 22-ERD-037. 

\end{acknowledgments}
  
\textit{Facilities:} LIGO, Virgo\\
\textit{Software:} \texttt{PyMC5} \citep{pymc:2023}, \texttt{scipy} \citep{Virtanen:2020}, \texttt{bilby} \citep{Ashton:2019,Romero-Shaw:2020}, \texttt{pycbc} \citep{Biwer:2019}, \texttt{arviz} \citep{Kumar:2019}, \texttt{corner.py} \citep{corner:2016}, \texttt{matplotlib} \citep{Hunter:2007}.

\appendix
\section{\label{sec:appendix_average_mass_spectrum}Likelihood Function of Average Mass Spectrum Fitting}

In Section~\ref{subsec:average_spectrum}, we discuss how we obtain the fiducial parameters for our mixture model through fitting the average mass spectrum of the {\powerpeakmodel} model.
In this appendix, we describe the detailed procedures of this fitting.

We first forward sample the $(m_1, m_2)$ pairs using the {\powerpeakmodel} model (with the fiducial parameters in Table~\ref{table:fiducial}), consisting of a $m_1$ function in Eq.~\ref{eq:power_law_peak} and a power-law $q$ function in Eq.~\ref{eq:power_law_peak_mass_ratio}.
Then we concatenate a series of $(m_1, m_2)$ pairs to a 1-D array of black hole masses, assuming primary and secondary masses are arbitrary labels.

With a 1-D array of the forward sampled black hole masses, we apply a KDE to obtain the probability density function of this 1-D array, $p_\mathrm{KDE}(m)$.
We want to know how much the shape parameters, the spectral index and the location and standard deviation of the Gaussian bump change after we concatenate the $(m_1, m_2)$ into a 1-D array.
Then, we assume the average mass spectrum ($p_\mathrm{ave}(m)$) follows the power-law + peak structure and fit it to $p_\mathrm{KDE}(m)$:
\begin{equation}
  \begin{split}
    p_\mathrm{ave}(m &\mid -\alpha, \delta_m, \mmax, \mmin, \mu, \sigma, \lambda_p)
    =\\
    &(1 - \lambda_p) \mathcal{B}(m \mid -\alpha, \mmax, \mmin, \delta_m)
    \\
    &+\lambda_p \mathcal{G}(m \mid \mu, \sigma, \mmin, \delta_m).
  \end{split}
\end{equation}
Here, $m$ refers to the black hole mass regardless of the labels of primary/secondary $m_1$ and $m_2$.
The likelihood function for finding the best-fit shape parameters is:
\begin{equation}
  \begin{split}    
    \log \mathcal{L}(\thetavec) &=\\
    -\frac{1}{2}\sum
    &\left[
    \log (2\pi \sigma^2 )
    + \frac{(p_\mathrm{KDE}(m) - p_\mathrm{ave}(m \mid \thetavec))^2}{\sigma^2}
    \right].
  \end{split}
  \label{eq:average_mass_likelihood}
\end{equation}
Here, we fit the parameters of $\thetavec = (-\alpha, \delta_m, \mu, \sigma, \lambda_p)$ but fix $(\mmax, \mmin)$ to the input values to the forward sampling of the {\powerpeakmodel} model.
The prior for each shape parameter is listed in Table~\ref{table:average_mass_prior}.
The high mass end of the $p_\mathrm{KDE}(m)$ has numerical noise due to a lack of Monte Carlo samples to reconstruct the correct {\powerpeakmodel} through a KDE.
To avoid this artifact affecting the fit,
we let $\sigma$ scale as the Poisson error
\begin{equation}
  \sigma = \sigma_0 \sqrt{\frac{p_\mathrm{KDE}(m)}{N}},
\end{equation}
where $N$ is the total number of the Monte Carlo samples used to build the KDE probability density function.
Ideally, with a larger number of samples, the numerical uncertainty is smaller.
We assume a broad prior for the scaling constant $\sigma_0$ for this numerical uncertainty,
\begin{equation}
  \sigma_0 \sim \mathrm{LogNormal}(\mu=0, \sigma=1).
\end{equation}
The fitting gives $\sigma_0 = 15^{+2}_{-2}$ with 2-$\sigma$ error.

\begin{table*}
	\centering
	\caption{The prior for shape parameters in the average mass spectrum fitting.
  }
	\label{table:average_mass_prior}
	\begin{tabular}{lcccc}
		\hline
		Parameter & Description & Prior\\
		\hline
		$\delta_m$   & The $\delta m$ for the low-mass mass spectrum smoothing &  $\mathrm{U}(0, 10)$  \\
		$\mmax$   & Maximum mass bound for the power-law model  &  - \\
    $\mmin$   & Minimum mass bound for both power-law and Gaussian models   &  - \\
		$-\alpha$ & Spectral index of the power-law &  $\mathrm{U}(1, 6)$ \\
    $\mu$ & Mean of the Gaussian model & $\mathrm{U}(20, 50)$ \\
    $\sigma$ & Standard deviation of the Gaussian model &  $\mathrm{U}(1, 6)$ \\
    $\lambda_p$ & Mixing fraction of the Gaussian model &  $\mathrm{U}(0, 0.5)$ \\
		\hline
	\end{tabular}
\end{table*}





\section{\label{subsec:appendix_psd} Fiducial Inference with a Different Power Spectral Density}

In this work, we utilize the analytical Power Spectral Density (PSD), \texttt{AdVMidHighSensitivityP1200087}~\citep{lalsuite:2018} from \texttt{PyCBC}~\citep{Biwer:2019} and the IMRPhenomD \cite{Khan:2016,Husa:2016,Purrer:2023} waveform model, to calculate the detection efficiency of our population model across different subpopulations. The detection probability, $p_\text{det}(\theta)$, is computed following the approach from Ref.~\cite{Perkins:2021}. We employ a pre-marginalized version that excludes detector-dependent variables, focusing instead on primary mass, secondary mass, and luminosity distance, where $\theta = (m_1, m_2, L)$.
We set $\rho_\mathrm{threshold} = 8$, meaning that we consider a trigger to be a detection if the SNR is above 8.

Population models aim to establish physically motivated priors for black hole properties. Our work specifically targets the modeling of the black hole mass spectrum. For each subpopulation model, we sample black hole masses $(m_a, m_b)$ and convert these values to $(m_1, m_2)$. These Monte Carlo samples are utilized to determine detection efficiency. Given that our model does not account for luminosity distance, we introduce a prior on the luminosity distance, $p(L) \propto L^2$, within a range of $(5, 5000)$ Mpc, ensuring it is uniform across the survey volume.

\begin{figure}
  \includegraphics[width=\columnwidth]{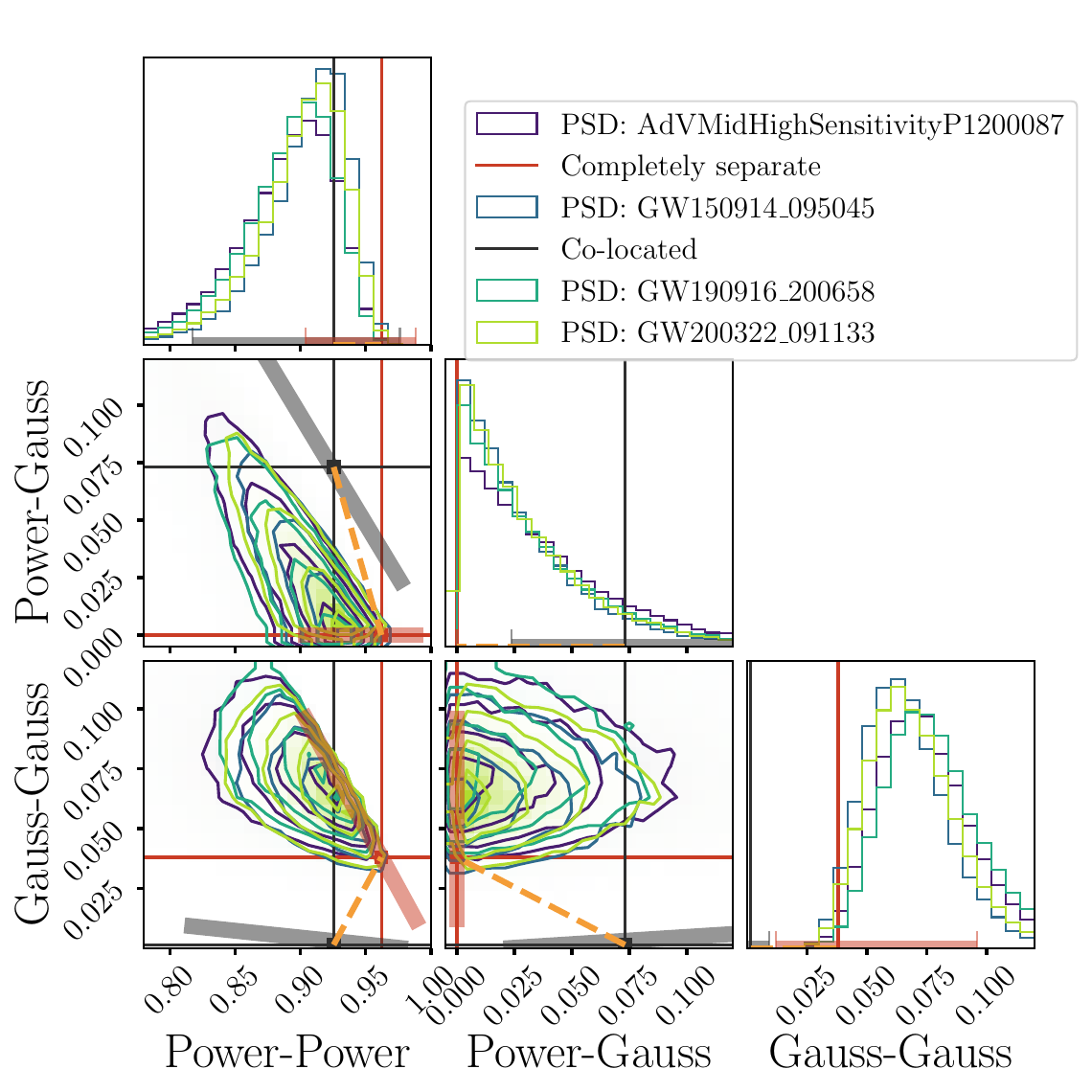}
  \caption{Fiducial MCMC chains using various Power Spectral Densities (PSDs) in the computation of the detection probability, $p_\mathrm{det}$. Four different PSDs are utilized: the analytical PSD (\texttt{AdVMidHighSensitivityP1200087}), the PSD from the GWTC-1 event (\texttt{GW150914\_095045}), the PSD from the GWTC-2 event (\texttt{GW190916\_200658}), and the PSD associated with the GWTC-3 event (\texttt{GW200322\_091133}). There is no evident shift in the posterior with different PSDs.
  }
  \label{fig:appendix_corner}
\end{figure}

In principle, the most robust approach for calculating detection efficiency requires marginalizing over the PSDs from various survey operational periods. Throughout this study, we have utilized an analytical PSD, \texttt{AdVMidHighSensitivityP1200087}. Therefore, our goal here is to assess the sensitivity of our inference results to different PSDs, ensuring that our conclusions are not significantly impacted by the choice of PSD and that our PSD approximation is sufficient.

Figure~\ref{fig:appendix_corner} presents the ``Fiducial'' inference results regarding the mixing fractions using various PSDs, including those from GWTC-1 (\texttt{GW150914\_095045}), GWTC-2 (\texttt{GW190916\_200658}), and GWTC-3 (\texttt{GW200322\_091133}). We observe minimal shifts in the mode of the posterior distribution (less than 1-sigma), with the exception that the width of the posterior for the \texttt{AdVMidHighSensitivityP1200087} PSD is slightly broader than that of the others. This suggests that the impact of using different PSDs on the main results (Figure~\ref{fig:MCMC_model_average}) presented in this paper is negligible.

\bibliography{ligo}

\providecommand{\noopsort}[1]{}\providecommand{\singleletter}[1]{#1}%
\begin{thebibliography}{117}%
\makeatletter
\providecommand \@ifxundefined [1]{%
 \@ifx{#1\undefined}
}%
\providecommand \@ifnum [1]{%
 \ifnum #1\expandafter \@firstoftwo
 \else \expandafter \@secondoftwo
 \fi
}%
\providecommand \@ifx [1]{%
 \ifx #1\expandafter \@firstoftwo
 \else \expandafter \@secondoftwo
 \fi
}%
\providecommand \natexlab [1]{#1}%
\providecommand \enquote  [1]{``#1''}%
\providecommand \bibnamefont  [1]{#1}%
\providecommand \bibfnamefont [1]{#1}%
\providecommand \citenamefont [1]{#1}%
\providecommand \href@noop [0]{\@secondoftwo}%
\providecommand \href [0]{\begingroup \@sanitize@url \@href}%
\providecommand \@href[1]{\@@startlink{#1}\@@href}%
\providecommand \@@href[1]{\endgroup#1\@@endlink}%
\providecommand \@sanitize@url [0]{\catcode `\\12\catcode `\$12\catcode
  `\&12\catcode `\#12\catcode `\^12\catcode `\_12\catcode `\%12\relax}%
\providecommand \@@startlink[1]{}%
\providecommand \@@endlink[0]{}%
\providecommand \url  [0]{\begingroup\@sanitize@url \@url }%
\providecommand \@url [1]{\endgroup\@href {#1}{\urlprefix }}%
\providecommand \urlprefix  [0]{URL }%
\providecommand \Eprint [0]{\href }%
\providecommand \doibase [0]{https://doi.org/}%
\providecommand \selectlanguage [0]{\@gobble}%
\providecommand \bibinfo  [0]{\@secondoftwo}%
\providecommand \bibfield  [0]{\@secondoftwo}%
\providecommand \translation [1]{[#1]}%
\providecommand \BibitemOpen [0]{}%
\providecommand \bibitemStop [0]{}%
\providecommand \bibitemNoStop [0]{.\EOS\space}%
\providecommand \EOS [0]{\spacefactor3000\relax}%
\providecommand \BibitemShut  [1]{\csname bibitem#1\endcsname}%
\let\auto@bib@innerbib\@empty
\bibitem [{\citenamefont {{Abbott}}\ \emph {et~al.}(2023)\citenamefont
  {{Abbott}}, \citenamefont {{Abbott}}, \citenamefont {{Acernese}} \emph
  {et~al.}}]{2023PhRvX..13a1048A}%
  \BibitemOpen
  \bibfield  {author} {\bibinfo {author} {\bibfnamefont {R.}~\bibnamefont
  {{Abbott}}}, \bibinfo {author} {\bibfnamefont {T.~D.}\ \bibnamefont
  {{Abbott}}}, \bibinfo {author} {\bibfnamefont {F.}~\bibnamefont
  {{Acernese}}}, \emph {et~al.},\ }\bibfield  {title} {\bibinfo {title}
  {{Population of Merging Compact Binaries Inferred Using Gravitational Waves
  through GWTC-3}},\ }\href {https://doi.org/10.1103/PhysRevX.13.011048}
  {\bibfield  {journal} {\bibinfo  {journal} {Physical Review X}\ }\textbf
  {\bibinfo {volume} {13}},\ \bibinfo {eid} {011048} (\bibinfo {year}
  {2023})},\ \Eprint {https://arxiv.org/abs/2111.03634} {arXiv:2111.03634
  [astro-ph.HE]} \BibitemShut {NoStop}%
\bibitem [{\citenamefont {{Nitz}}\ \emph {et~al.}(2023)\citenamefont {{Nitz}},
  \citenamefont {{Kumar}}, \citenamefont {{Wang}}, \citenamefont {{Kastha}},
  \citenamefont {{Wu}}, \citenamefont {{Sch{\"a}fer}}, \citenamefont
  {{Dhurkunde}},\ and\ \citenamefont {{Capano}}}]{Nitz:2023}%
  \BibitemOpen
  \bibfield  {author} {\bibinfo {author} {\bibfnamefont {A.~H.}\ \bibnamefont
  {{Nitz}}}, \bibinfo {author} {\bibfnamefont {S.}~\bibnamefont {{Kumar}}},
  \bibinfo {author} {\bibfnamefont {Y.-F.}\ \bibnamefont {{Wang}}}, \bibinfo
  {author} {\bibfnamefont {S.}~\bibnamefont {{Kastha}}}, \bibinfo {author}
  {\bibfnamefont {S.}~\bibnamefont {{Wu}}}, \bibinfo {author} {\bibfnamefont
  {M.}~\bibnamefont {{Sch{\"a}fer}}}, \bibinfo {author} {\bibfnamefont
  {R.}~\bibnamefont {{Dhurkunde}}},\ and\ \bibinfo {author} {\bibfnamefont
  {C.~D.}\ \bibnamefont {{Capano}}},\ }\bibfield  {title} {\bibinfo {title}
  {{4-OGC: Catalog of Gravitational Waves from Compact Binary Mergers}},\
  }\href {https://doi.org/10.3847/1538-4357/aca591} {\bibfield  {journal}
  {\bibinfo  {journal} {\apj}\ }\textbf {\bibinfo {volume} {946}},\ \bibinfo
  {eid} {59} (\bibinfo {year} {2023})},\ \Eprint
  {https://arxiv.org/abs/2112.06878} {arXiv:2112.06878 [astro-ph.HE]}
  \BibitemShut {NoStop}%
\bibitem [{\citenamefont {{The LIGO Scientific Collaboration}}\ \emph
  {et~al.}(2021{\natexlab{a}})\citenamefont {{The LIGO Scientific
  Collaboration}}, \citenamefont {{the Virgo Collaboration}}, \citenamefont
  {{the KAGRA Collaboration}}, \citenamefont {{Abbott}}, \citenamefont {{Abe}}
  \emph {et~al.}}]{2021arXiv211206861T}%
  \BibitemOpen
  \bibfield  {author} {\bibinfo {author} {\bibnamefont {{The LIGO Scientific
  Collaboration}}}, \bibinfo {author} {\bibnamefont {{the Virgo
  Collaboration}}}, \bibinfo {author} {\bibnamefont {{the KAGRA
  Collaboration}}}, \bibinfo {author} {\bibfnamefont {R.}~\bibnamefont
  {{Abbott}}}, \bibinfo {author} {\bibfnamefont {H.}~\bibnamefont {{Abe}}},
  \emph {et~al.},\ }\bibfield  {title} {\bibinfo {title} {{Tests of General
  Relativity with GWTC-3}},\ }\href {https://doi.org/10.48550/arXiv.2112.06861}
  {\bibfield  {journal} {\bibinfo  {journal} {arXiv e-prints}\ ,\ \bibinfo
  {eid} {arXiv:2112.06861}} (\bibinfo {year} {2021}{\natexlab{a}})},\ \Eprint
  {https://arxiv.org/abs/2112.06861} {arXiv:2112.06861 [gr-qc]} \BibitemShut
  {NoStop}%
\bibitem [{\citenamefont {{The LIGO Scientific Collaboration}}\ \emph
  {et~al.}(2021{\natexlab{b}})\citenamefont {{The LIGO Scientific
  Collaboration}}, \citenamefont {{the Virgo Collaboration}}, \citenamefont
  {{the KAGRA Collaboration}}, \citenamefont {{Abbott}}, \citenamefont {{Abe}}
  \emph {et~al.}}]{2021arXiv211103604T}%
  \BibitemOpen
  \bibfield  {author} {\bibinfo {author} {\bibnamefont {{The LIGO Scientific
  Collaboration}}}, \bibinfo {author} {\bibnamefont {{the Virgo
  Collaboration}}}, \bibinfo {author} {\bibnamefont {{the KAGRA
  Collaboration}}}, \bibinfo {author} {\bibfnamefont {R.}~\bibnamefont
  {{Abbott}}}, \bibinfo {author} {\bibfnamefont {H.}~\bibnamefont {{Abe}}},
  \emph {et~al.},\ }\bibfield  {title} {\bibinfo {title} {{Constraints on the
  cosmic expansion history from GWTC-3}},\ }\href
  {https://doi.org/10.48550/arXiv.2111.03604} {\bibfield  {journal} {\bibinfo
  {journal} {arXiv e-prints}\ ,\ \bibinfo {eid} {arXiv:2111.03604}} (\bibinfo
  {year} {2021}{\natexlab{b}})},\ \Eprint {https://arxiv.org/abs/2111.03604}
  {arXiv:2111.03604 [astro-ph.CO]} \BibitemShut {NoStop}%
\bibitem [{\citenamefont {{The LIGO Scientific Collaboration}}\ \emph
  {et~al.}(2021{\natexlab{c}})\citenamefont {{The LIGO Scientific
  Collaboration}}, \citenamefont {{the Virgo Collaboration}}, \citenamefont
  {{the KAGRA Collaboration}}, \citenamefont {{Abbott}}, \citenamefont
  {{Abbott}} \emph {et~al.}}]{gwtc3}%
  \BibitemOpen
  \bibfield  {author} {\bibinfo {author} {\bibnamefont {{The LIGO Scientific
  Collaboration}}}, \bibinfo {author} {\bibnamefont {{the Virgo
  Collaboration}}}, \bibinfo {author} {\bibnamefont {{the KAGRA
  Collaboration}}}, \bibinfo {author} {\bibfnamefont {R.}~\bibnamefont
  {{Abbott}}}, \bibinfo {author} {\bibfnamefont {T.~D.}\ \bibnamefont
  {{Abbott}}}, \emph {et~al.},\ }\bibfield  {title} {\bibinfo {title} {{GWTC-3:
  Compact Binary Coalescences Observed by LIGO and Virgo During the Second Part
  of the Third Observing Run}},\ }\href
  {https://doi.org/10.48550/arXiv.2111.03606} {\bibfield  {journal} {\bibinfo
  {journal} {arXiv e-prints}\ ,\ \bibinfo {eid} {arXiv:2111.03606}} (\bibinfo
  {year} {2021}{\natexlab{c}})},\ \Eprint {https://arxiv.org/abs/2111.03606}
  {arXiv:2111.03606 [gr-qc]} \BibitemShut {NoStop}%
\bibitem [{\citenamefont {{Acernese}}\ \emph {et~al.}(2015)\citenamefont
  {{Acernese}}, \citenamefont {{Agathos}}, \citenamefont {{Agatsuma}} \emph
  {et~al.}}]{2015CQGra..32b4001A}%
  \BibitemOpen
  \bibfield  {author} {\bibinfo {author} {\bibfnamefont {F.}~\bibnamefont
  {{Acernese}}}, \bibinfo {author} {\bibfnamefont {M.}~\bibnamefont
  {{Agathos}}}, \bibinfo {author} {\bibfnamefont {K.}~\bibnamefont
  {{Agatsuma}}}, \emph {et~al.},\ }\bibfield  {title} {\bibinfo {title}
  {{Advanced Virgo: a second-generation interferometric gravitational wave
  detector}},\ }\href {https://doi.org/10.1088/0264-9381/32/2/024001}
  {\bibfield  {journal} {\bibinfo  {journal} {Classical and Quantum Gravity}\
  }\textbf {\bibinfo {volume} {32}},\ \bibinfo {eid} {024001} (\bibinfo {year}
  {2015})},\ \Eprint {https://arxiv.org/abs/1408.3978} {arXiv:1408.3978
  [gr-qc]} \BibitemShut {NoStop}%
\bibitem [{\citenamefont {{Akutsu}}\ \emph {et~al.}(2021)\citenamefont
  {{Akutsu}}, \citenamefont {{Ando}}, \citenamefont {{Arai}} \emph
  {et~al.}}]{kagra}%
  \BibitemOpen
  \bibfield  {author} {\bibinfo {author} {\bibfnamefont {T.}~\bibnamefont
  {{Akutsu}}}, \bibinfo {author} {\bibfnamefont {M.}~\bibnamefont {{Ando}}},
  \bibinfo {author} {\bibfnamefont {K.}~\bibnamefont {{Arai}}}, \emph
  {et~al.},\ }\bibfield  {title} {\bibinfo {title} {{Overview of KAGRA: KAGRA
  science}},\ }\href {https://doi.org/10.1093/ptep/ptaa120} {\bibfield
  {journal} {\bibinfo  {journal} {Progress of Theoretical and Experimental
  Physics}\ }\textbf {\bibinfo {volume} {2021}},\ \bibinfo {eid} {05A103}
  (\bibinfo {year} {2021})},\ \Eprint {https://arxiv.org/abs/2008.02921}
  {arXiv:2008.02921 [gr-qc]} \BibitemShut {NoStop}%
\bibitem [{\citenamefont {{Abbott}}\ \emph {et~al.}(2019)\citenamefont
  {{Abbott}}, \citenamefont {{Abbott}}, \citenamefont {{Abbott}}, \citenamefont
  {{LIGO Scientific Collaboration}},\ and\ \citenamefont {{Virgo
  Collaboration}}}]{2019ApJ...882L..24A}%
  \BibitemOpen
  \bibfield  {author} {\bibinfo {author} {\bibfnamefont {B.~P.}\ \bibnamefont
  {{Abbott}}}, \bibinfo {author} {\bibfnamefont {R.}~\bibnamefont {{Abbott}}},
  \bibinfo {author} {\bibfnamefont {T.~D.}\ \bibnamefont {{Abbott}}}, \bibinfo
  {author} {\bibnamefont {{LIGO Scientific Collaboration}}},\ and\ \bibinfo
  {author} {\bibnamefont {{Virgo Collaboration}}},\ }\bibfield  {title}
  {\bibinfo {title} {{Binary Black Hole Population Properties Inferred from the
  First and Second Observing Runs of Advanced LIGO and Advanced Virgo}},\
  }\href {https://doi.org/10.3847/2041-8213/ab3800} {\bibfield  {journal}
  {\bibinfo  {journal} {\apjl}\ }\textbf {\bibinfo {volume} {882}},\ \bibinfo
  {eid} {L24} (\bibinfo {year} {2019})},\ \Eprint
  {https://arxiv.org/abs/1811.12940} {arXiv:1811.12940 [astro-ph.HE]}
  \BibitemShut {NoStop}%
\bibitem [{\citenamefont {{Venumadhav}}\ \emph {et~al.}(2019)\citenamefont
  {{Venumadhav}}, \citenamefont {{Zackay}}, \citenamefont {{Roulet}},
  \citenamefont {{Dai}},\ and\ \citenamefont
  {{Zaldarriaga}}}]{Venumadhav:2019}%
  \BibitemOpen
  \bibfield  {author} {\bibinfo {author} {\bibfnamefont {T.}~\bibnamefont
  {{Venumadhav}}}, \bibinfo {author} {\bibfnamefont {B.}~\bibnamefont
  {{Zackay}}}, \bibinfo {author} {\bibfnamefont {J.}~\bibnamefont {{Roulet}}},
  \bibinfo {author} {\bibfnamefont {L.}~\bibnamefont {{Dai}}},\ and\ \bibinfo
  {author} {\bibfnamefont {M.}~\bibnamefont {{Zaldarriaga}}},\ }\bibfield
  {title} {\bibinfo {title} {{New search pipeline for compact binary mergers:
  Results for binary black holes in the first observing run of Advanced
  LIGO}},\ }\href {https://doi.org/10.1103/PhysRevD.100.023011} {\bibfield
  {journal} {\bibinfo  {journal} {\prd}\ }\textbf {\bibinfo {volume} {100}},\
  \bibinfo {eid} {023011} (\bibinfo {year} {2019})},\ \Eprint
  {https://arxiv.org/abs/1902.10341} {arXiv:1902.10341 [astro-ph.IM]}
  \BibitemShut {NoStop}%
\bibitem [{\citenamefont {{Abbott}}\ \emph {et~al.}(2021)\citenamefont
  {{Abbott}}, \citenamefont {{Abbott}}, \citenamefont {{Abraham}},
  \citenamefont {{LIGO Scientific Collaboration}},\ and\ \citenamefont {{Virgo
  Collaboration}}}]{2021ApJ...913L...7A}%
  \BibitemOpen
  \bibfield  {author} {\bibinfo {author} {\bibfnamefont {R.}~\bibnamefont
  {{Abbott}}}, \bibinfo {author} {\bibfnamefont {T.~D.}\ \bibnamefont
  {{Abbott}}}, \bibinfo {author} {\bibfnamefont {S.}~\bibnamefont {{Abraham}}},
  \bibinfo {author} {\bibnamefont {{LIGO Scientific Collaboration}}},\ and\
  \bibinfo {author} {\bibnamefont {{Virgo Collaboration}}},\ }\bibfield
  {title} {\bibinfo {title} {{Population Properties of Compact Objects from the
  Second LIGO-Virgo Gravitational-Wave Transient Catalog}},\ }\href
  {https://doi.org/10.3847/2041-8213/abe949} {\bibfield  {journal} {\bibinfo
  {journal} {\apjl}\ }\textbf {\bibinfo {volume} {913}},\ \bibinfo {eid} {L7}
  (\bibinfo {year} {2021})},\ \Eprint {https://arxiv.org/abs/2010.14533}
  {arXiv:2010.14533 [astro-ph.HE]} \BibitemShut {NoStop}%
\bibitem [{\citenamefont {{Wadekar}}\ \emph {et~al.}(2023)\citenamefont
  {{Wadekar}}, \citenamefont {{Roulet}}, \citenamefont {{Venumadhav}},
  \citenamefont {{Mehta}}, \citenamefont {{Zackay}}, \citenamefont {{Mushkin}},
  \citenamefont {{Olsen}},\ and\ \citenamefont {{Zaldarriaga}}}]{Wadekar:2023}%
  \BibitemOpen
  \bibfield  {author} {\bibinfo {author} {\bibfnamefont {D.}~\bibnamefont
  {{Wadekar}}}, \bibinfo {author} {\bibfnamefont {J.}~\bibnamefont {{Roulet}}},
  \bibinfo {author} {\bibfnamefont {T.}~\bibnamefont {{Venumadhav}}}, \bibinfo
  {author} {\bibfnamefont {A.~K.}\ \bibnamefont {{Mehta}}}, \bibinfo {author}
  {\bibfnamefont {B.}~\bibnamefont {{Zackay}}}, \bibinfo {author}
  {\bibfnamefont {J.}~\bibnamefont {{Mushkin}}}, \bibinfo {author}
  {\bibfnamefont {S.}~\bibnamefont {{Olsen}}},\ and\ \bibinfo {author}
  {\bibfnamefont {M.}~\bibnamefont {{Zaldarriaga}}},\ }\bibfield  {title}
  {\bibinfo {title} {{New black hole mergers in the LIGO-Virgo O3 data from a
  gravitational wave search including higher-order harmonics}},\ }\href
  {https://doi.org/10.48550/arXiv.2312.06631} {\bibfield  {journal} {\bibinfo
  {journal} {arXiv e-prints}\ ,\ \bibinfo {eid} {arXiv:2312.06631}} (\bibinfo
  {year} {2023})},\ \Eprint {https://arxiv.org/abs/2312.06631}
  {arXiv:2312.06631 [gr-qc]} \BibitemShut {NoStop}%
\bibitem [{\citenamefont {{Zevin}}\ \emph {et~al.}(2017)\citenamefont
  {{Zevin}}, \citenamefont {{Pankow}}, \citenamefont {{Rodriguez}},
  \citenamefont {{Sampson}}, \citenamefont {{Chase}}, \citenamefont
  {{Kalogera}},\ and\ \citenamefont {{Rasio}}}]{Zevin:2017}%
  \BibitemOpen
  \bibfield  {author} {\bibinfo {author} {\bibfnamefont {M.}~\bibnamefont
  {{Zevin}}}, \bibinfo {author} {\bibfnamefont {C.}~\bibnamefont {{Pankow}}},
  \bibinfo {author} {\bibfnamefont {C.~L.}\ \bibnamefont {{Rodriguez}}},
  \bibinfo {author} {\bibfnamefont {L.}~\bibnamefont {{Sampson}}}, \bibinfo
  {author} {\bibfnamefont {E.}~\bibnamefont {{Chase}}}, \bibinfo {author}
  {\bibfnamefont {V.}~\bibnamefont {{Kalogera}}},\ and\ \bibinfo {author}
  {\bibfnamefont {F.~A.}\ \bibnamefont {{Rasio}}},\ }\bibfield  {title}
  {\bibinfo {title} {{Constraining Formation Models of Binary Black Holes with
  Gravitational-wave Observations}},\ }\href
  {https://doi.org/10.3847/1538-4357/aa8408} {\bibfield  {journal} {\bibinfo
  {journal} {\apj}\ }\textbf {\bibinfo {volume} {846}},\ \bibinfo {eid} {82}
  (\bibinfo {year} {2017})},\ \Eprint {https://arxiv.org/abs/1704.07379}
  {arXiv:1704.07379 [astro-ph.HE]} \BibitemShut {NoStop}%
\bibitem [{\citenamefont {Zevin}\ \emph {et~al.}(2021)\citenamefont {Zevin},
  \citenamefont {Bavera}, \citenamefont {Berry}, \citenamefont {Kalogera},
  \citenamefont {Fragos}, \citenamefont {Marchant}, \citenamefont {Rodriguez},
  \citenamefont {Antonini}, \citenamefont {Holz},\ and\ \citenamefont
  {Pankow}}]{Zevin:2020gbd}%
  \BibitemOpen
  \bibfield  {author} {\bibinfo {author} {\bibfnamefont {M.}~\bibnamefont
  {Zevin}}, \bibinfo {author} {\bibfnamefont {S.~S.}\ \bibnamefont {Bavera}},
  \bibinfo {author} {\bibfnamefont {C.~P.~L.}\ \bibnamefont {Berry}}, \bibinfo
  {author} {\bibfnamefont {V.}~\bibnamefont {Kalogera}}, \bibinfo {author}
  {\bibfnamefont {T.}~\bibnamefont {Fragos}}, \bibinfo {author} {\bibfnamefont
  {P.}~\bibnamefont {Marchant}}, \bibinfo {author} {\bibfnamefont {C.~L.}\
  \bibnamefont {Rodriguez}}, \bibinfo {author} {\bibfnamefont {F.}~\bibnamefont
  {Antonini}}, \bibinfo {author} {\bibfnamefont {D.~E.}\ \bibnamefont {Holz}},\
  and\ \bibinfo {author} {\bibfnamefont {C.}~\bibnamefont {Pankow}},\
  }\bibfield  {title} {\bibinfo {title} {{One Channel to Rule Them All?
  Constraining the Origins of Binary Black Holes Using Multiple Formation
  Pathways}},\ }\href {https://doi.org/10.3847/1538-4357/abe40e} {\bibfield
  {journal} {\bibinfo  {journal} {Astrophys. J.}\ }\textbf {\bibinfo {volume}
  {910}},\ \bibinfo {pages} {152} (\bibinfo {year} {2021})},\ \Eprint
  {https://arxiv.org/abs/2011.10057} {arXiv:2011.10057 [astro-ph.HE]}
  \BibitemShut {NoStop}%
\bibitem [{\citenamefont {{Edelman}}\ \emph {et~al.}(2022)\citenamefont
  {{Edelman}}, \citenamefont {{Doctor}}, \citenamefont {{Godfrey}},\ and\
  \citenamefont {{Farr}}}]{Edelman:2022}%
  \BibitemOpen
  \bibfield  {author} {\bibinfo {author} {\bibfnamefont {B.}~\bibnamefont
  {{Edelman}}}, \bibinfo {author} {\bibfnamefont {Z.}~\bibnamefont {{Doctor}}},
  \bibinfo {author} {\bibfnamefont {J.}~\bibnamefont {{Godfrey}}},\ and\
  \bibinfo {author} {\bibfnamefont {B.}~\bibnamefont {{Farr}}},\ }\bibfield
  {title} {\bibinfo {title} {{Ain't No Mountain High Enough: Semiparametric
  Modeling of LIGO-Virgo's Binary Black Hole Mass Distribution}},\ }\href
  {https://doi.org/10.3847/1538-4357/ac3667} {\bibfield  {journal} {\bibinfo
  {journal} {\apj}\ }\textbf {\bibinfo {volume} {924}},\ \bibinfo {eid} {101}
  (\bibinfo {year} {2022})},\ \Eprint {https://arxiv.org/abs/2109.06137}
  {arXiv:2109.06137 [astro-ph.HE]} \BibitemShut {NoStop}%
\bibitem [{\citenamefont {{Stevenson}}\ \emph {et~al.}(2015)\citenamefont
  {{Stevenson}}, \citenamefont {{Ohme}},\ and\ \citenamefont
  {{Fairhurst}}}]{Stevenson:2015}%
  \BibitemOpen
  \bibfield  {author} {\bibinfo {author} {\bibfnamefont {S.}~\bibnamefont
  {{Stevenson}}}, \bibinfo {author} {\bibfnamefont {F.}~\bibnamefont
  {{Ohme}}},\ and\ \bibinfo {author} {\bibfnamefont {S.}~\bibnamefont
  {{Fairhurst}}},\ }\bibfield  {title} {\bibinfo {title} {{Distinguishing
  Compact Binary Population Synthesis Models Using Gravitational Wave
  Observations of Coalescing Binary Black Holes}},\ }\href
  {https://doi.org/10.1088/0004-637X/810/1/58} {\bibfield  {journal} {\bibinfo
  {journal} {\apj}\ }\textbf {\bibinfo {volume} {810}},\ \bibinfo {eid} {58}
  (\bibinfo {year} {2015})},\ \Eprint {https://arxiv.org/abs/1504.07802}
  {arXiv:1504.07802 [astro-ph.HE]} \BibitemShut {NoStop}%
\bibitem [{\citenamefont {{{\"O}zel}}\ \emph {et~al.}(2010)\citenamefont
  {{{\"O}zel}}, \citenamefont {{Psaltis}}, \citenamefont {{Narayan}},\ and\
  \citenamefont {{McClintock}}}]{Ozel:2010}%
  \BibitemOpen
  \bibfield  {author} {\bibinfo {author} {\bibfnamefont {F.}~\bibnamefont
  {{{\"O}zel}}}, \bibinfo {author} {\bibfnamefont {D.}~\bibnamefont
  {{Psaltis}}}, \bibinfo {author} {\bibfnamefont {R.}~\bibnamefont
  {{Narayan}}},\ and\ \bibinfo {author} {\bibfnamefont {J.~E.}\ \bibnamefont
  {{McClintock}}},\ }\bibfield  {title} {\bibinfo {title} {{The Black Hole Mass
  Distribution in the Galaxy}},\ }\href
  {https://doi.org/10.1088/0004-637X/725/2/1918} {\bibfield  {journal}
  {\bibinfo  {journal} {\apj}\ }\textbf {\bibinfo {volume} {725}},\ \bibinfo
  {pages} {1918} (\bibinfo {year} {2010})},\ \Eprint
  {https://arxiv.org/abs/1006.2834} {arXiv:1006.2834 [astro-ph.GA]}
  \BibitemShut {NoStop}%
\bibitem [{\citenamefont {{Farr}}\ \emph {et~al.}(2011)\citenamefont {{Farr}},
  \citenamefont {{Sravan}}, \citenamefont {{Cantrell}}, \citenamefont
  {{Kreidberg}}, \citenamefont {{Bailyn}}, \citenamefont {{Mandel}},\ and\
  \citenamefont {{Kalogera}}}]{Farr:2011}%
  \BibitemOpen
  \bibfield  {author} {\bibinfo {author} {\bibfnamefont {W.~M.}\ \bibnamefont
  {{Farr}}}, \bibinfo {author} {\bibfnamefont {N.}~\bibnamefont {{Sravan}}},
  \bibinfo {author} {\bibfnamefont {A.}~\bibnamefont {{Cantrell}}}, \bibinfo
  {author} {\bibfnamefont {L.}~\bibnamefont {{Kreidberg}}}, \bibinfo {author}
  {\bibfnamefont {C.~D.}\ \bibnamefont {{Bailyn}}}, \bibinfo {author}
  {\bibfnamefont {I.}~\bibnamefont {{Mandel}}},\ and\ \bibinfo {author}
  {\bibfnamefont {V.}~\bibnamefont {{Kalogera}}},\ }\bibfield  {title}
  {\bibinfo {title} {{The Mass Distribution of Stellar-mass Black Holes}},\
  }\href {https://doi.org/10.1088/0004-637X/741/2/103} {\bibfield  {journal}
  {\bibinfo  {journal} {\apj}\ }\textbf {\bibinfo {volume} {741}},\ \bibinfo
  {eid} {103} (\bibinfo {year} {2011})},\ \Eprint
  {https://arxiv.org/abs/1011.1459} {arXiv:1011.1459 [astro-ph.GA]}
  \BibitemShut {NoStop}%
\bibitem [{\citenamefont {{Fishbach}}\ \emph {et~al.}(2020)\citenamefont
  {{Fishbach}}, \citenamefont {{Essick}},\ and\ \citenamefont
  {{Holz}}}]{Fishbach:2020}%
  \BibitemOpen
  \bibfield  {author} {\bibinfo {author} {\bibfnamefont {M.}~\bibnamefont
  {{Fishbach}}}, \bibinfo {author} {\bibfnamefont {R.}~\bibnamefont
  {{Essick}}},\ and\ \bibinfo {author} {\bibfnamefont {D.~E.}\ \bibnamefont
  {{Holz}}},\ }\bibfield  {title} {\bibinfo {title} {{Does Matter Matter? Using
  the Mass Distribution to Distinguish Neutron Stars and Black Holes}},\ }\href
  {https://doi.org/10.3847/2041-8213/aba7b6} {\bibfield  {journal} {\bibinfo
  {journal} {\apjl}\ }\textbf {\bibinfo {volume} {899}},\ \bibinfo {eid} {L8}
  (\bibinfo {year} {2020})},\ \Eprint {https://arxiv.org/abs/2006.13178}
  {arXiv:2006.13178 [astro-ph.HE]} \BibitemShut {NoStop}%
\bibitem [{\citenamefont {{Farah}}\ \emph {et~al.}(2022)\citenamefont
  {{Farah}}, \citenamefont {{Fishbach}}, \citenamefont {{Essick}},
  \citenamefont {{Holz}},\ and\ \citenamefont {{Galaudage}}}]{Farah:2021}%
  \BibitemOpen
  \bibfield  {author} {\bibinfo {author} {\bibfnamefont {A.}~\bibnamefont
  {{Farah}}}, \bibinfo {author} {\bibfnamefont {M.}~\bibnamefont {{Fishbach}}},
  \bibinfo {author} {\bibfnamefont {R.}~\bibnamefont {{Essick}}}, \bibinfo
  {author} {\bibfnamefont {D.~E.}\ \bibnamefont {{Holz}}},\ and\ \bibinfo
  {author} {\bibfnamefont {S.}~\bibnamefont {{Galaudage}}},\ }\bibfield
  {title} {\bibinfo {title} {{Bridging the Gap: Categorizing Gravitational-wave
  Events at the Transition between Neutron Stars and Black Holes}},\ }\href
  {https://doi.org/10.3847/1538-4357/ac5f03} {\bibfield  {journal} {\bibinfo
  {journal} {\apj}\ }\textbf {\bibinfo {volume} {931}},\ \bibinfo {eid} {108}
  (\bibinfo {year} {2022})},\ \Eprint {https://arxiv.org/abs/2111.03498}
  {arXiv:2111.03498 [astro-ph.HE]} \BibitemShut {NoStop}%
\bibitem [{\citenamefont {{Li}}\ \emph {et~al.}(2021)\citenamefont {{Li}},
  \citenamefont {{Miao}}, \citenamefont {{Han}},\ and\ \citenamefont
  {{Zhang}}}]{Li:2021}%
  \BibitemOpen
  \bibfield  {author} {\bibinfo {author} {\bibfnamefont {A.}~\bibnamefont
  {{Li}}}, \bibinfo {author} {\bibfnamefont {Z.}~\bibnamefont {{Miao}}},
  \bibinfo {author} {\bibfnamefont {S.}~\bibnamefont {{Han}}},\ and\ \bibinfo
  {author} {\bibfnamefont {B.}~\bibnamefont {{Zhang}}},\ }\bibfield  {title}
  {\bibinfo {title} {{Constraints on the Maximum Mass of Neutron Stars with a
  Quark Core from GW170817 and NICER PSR J0030+0451 Data}},\ }\href
  {https://doi.org/10.3847/1538-4357/abf355} {\bibfield  {journal} {\bibinfo
  {journal} {\apj}\ }\textbf {\bibinfo {volume} {913}},\ \bibinfo {eid} {27}
  (\bibinfo {year} {2021})},\ \Eprint {https://arxiv.org/abs/2103.15119}
  {arXiv:2103.15119 [astro-ph.HE]} \BibitemShut {NoStop}%
\bibitem [{\citenamefont {{Patton}}\ \emph {et~al.}(2022)\citenamefont
  {{Patton}}, \citenamefont {{Sukhbold}},\ and\ \citenamefont
  {{Eldridge}}}]{Patton:2022}%
  \BibitemOpen
  \bibfield  {author} {\bibinfo {author} {\bibfnamefont {R.~A.}\ \bibnamefont
  {{Patton}}}, \bibinfo {author} {\bibfnamefont {T.}~\bibnamefont
  {{Sukhbold}}},\ and\ \bibinfo {author} {\bibfnamefont {J.~J.}\ \bibnamefont
  {{Eldridge}}},\ }\bibfield  {title} {\bibinfo {title} {{Comparing compact
  object distributions from mass- and presupernova core structure-based
  prescriptions}},\ }\href {https://doi.org/10.1093/mnras/stab3797} {\bibfield
  {journal} {\bibinfo  {journal} {\mnras}\ }\textbf {\bibinfo {volume} {511}},\
  \bibinfo {pages} {903} (\bibinfo {year} {2022})},\ \Eprint
  {https://arxiv.org/abs/2106.05978} {arXiv:2106.05978 [astro-ph.HE]}
  \BibitemShut {NoStop}%
\bibitem [{\citenamefont {{Siegel}}\ \emph {et~al.}(2023)\citenamefont
  {{Siegel}}, \citenamefont {{Kiato}}, \citenamefont {{Kalogera}},
  \citenamefont {{Berry}}, \citenamefont {{Maccarone}}, \citenamefont
  {{Breivik}}, \citenamefont {{Andrews}}, \citenamefont {{Bavera}},
  \citenamefont {{Dotter}}, \citenamefont {{Fragos}}, \citenamefont
  {{Kovlakas}}, \citenamefont {{Misra}}, \citenamefont {{Rocha}}, \citenamefont
  {{Srivastava}}, \citenamefont {{Sun}}, \citenamefont {{Xing}},\ and\
  \citenamefont {{Zapartas}}}]{Siegel:2023}%
  \BibitemOpen
  \bibfield  {author} {\bibinfo {author} {\bibfnamefont {J.~C.}\ \bibnamefont
  {{Siegel}}}, \bibinfo {author} {\bibfnamefont {I.}~\bibnamefont {{Kiato}}},
  \bibinfo {author} {\bibfnamefont {V.}~\bibnamefont {{Kalogera}}}, \bibinfo
  {author} {\bibfnamefont {C.~P.~L.}\ \bibnamefont {{Berry}}}, \bibinfo
  {author} {\bibfnamefont {T.~J.}\ \bibnamefont {{Maccarone}}}, \bibinfo
  {author} {\bibfnamefont {K.}~\bibnamefont {{Breivik}}}, \bibinfo {author}
  {\bibfnamefont {J.~J.}\ \bibnamefont {{Andrews}}}, \bibinfo {author}
  {\bibfnamefont {S.~S.}\ \bibnamefont {{Bavera}}}, \bibinfo {author}
  {\bibfnamefont {A.}~\bibnamefont {{Dotter}}}, \bibinfo {author}
  {\bibfnamefont {T.}~\bibnamefont {{Fragos}}}, \bibinfo {author}
  {\bibfnamefont {K.}~\bibnamefont {{Kovlakas}}}, \bibinfo {author}
  {\bibfnamefont {D.}~\bibnamefont {{Misra}}}, \bibinfo {author} {\bibfnamefont
  {K.~A.}\ \bibnamefont {{Rocha}}}, \bibinfo {author} {\bibfnamefont {P.~M.}\
  \bibnamefont {{Srivastava}}}, \bibinfo {author} {\bibfnamefont
  {M.}~\bibnamefont {{Sun}}}, \bibinfo {author} {\bibfnamefont
  {Z.}~\bibnamefont {{Xing}}},\ and\ \bibinfo {author} {\bibfnamefont
  {E.}~\bibnamefont {{Zapartas}}},\ }\bibfield  {title} {\bibinfo {title}
  {{Investigating the Lower Mass Gap with Low-mass X-Ray Binary Population
  Synthesis}},\ }\href {https://doi.org/10.3847/1538-4357/ace9d9} {\bibfield
  {journal} {\bibinfo  {journal} {\apj}\ }\textbf {\bibinfo {volume} {954}},\
  \bibinfo {eid} {212} (\bibinfo {year} {2023})},\ \Eprint
  {https://arxiv.org/abs/2209.06844} {arXiv:2209.06844 [astro-ph.HE]}
  \BibitemShut {NoStop}%
\bibitem [{\citenamefont {{Fryer}}\ \emph {et~al.}(2012)\citenamefont
  {{Fryer}}, \citenamefont {{Belczynski}}, \citenamefont {{Wiktorowicz}},
  \citenamefont {{Dominik}}, \citenamefont {{Kalogera}},\ and\ \citenamefont
  {{Holz}}}]{Fryer:2012}%
  \BibitemOpen
  \bibfield  {author} {\bibinfo {author} {\bibfnamefont {C.~L.}\ \bibnamefont
  {{Fryer}}}, \bibinfo {author} {\bibfnamefont {K.}~\bibnamefont
  {{Belczynski}}}, \bibinfo {author} {\bibfnamefont {G.}~\bibnamefont
  {{Wiktorowicz}}}, \bibinfo {author} {\bibfnamefont {M.}~\bibnamefont
  {{Dominik}}}, \bibinfo {author} {\bibfnamefont {V.}~\bibnamefont
  {{Kalogera}}},\ and\ \bibinfo {author} {\bibfnamefont {D.~E.}\ \bibnamefont
  {{Holz}}},\ }\bibfield  {title} {\bibinfo {title} {{Compact Remnant Mass
  Function: Dependence on the Explosion Mechanism and Metallicity}},\ }\href
  {https://doi.org/10.1088/0004-637X/749/1/91} {\bibfield  {journal} {\bibinfo
  {journal} {\apj}\ }\textbf {\bibinfo {volume} {749}},\ \bibinfo {eid} {91}
  (\bibinfo {year} {2012})},\ \Eprint {https://arxiv.org/abs/1110.1726}
  {arXiv:1110.1726 [astro-ph.SR]} \BibitemShut {NoStop}%
\bibitem [{\citenamefont {{Zevin}}\ \emph {et~al.}(2020)\citenamefont
  {{Zevin}}, \citenamefont {{Spera}}, \citenamefont {{Berry}},\ and\
  \citenamefont {{Kalogera}}}]{Zevin:2020}%
  \BibitemOpen
  \bibfield  {author} {\bibinfo {author} {\bibfnamefont {M.}~\bibnamefont
  {{Zevin}}}, \bibinfo {author} {\bibfnamefont {M.}~\bibnamefont {{Spera}}},
  \bibinfo {author} {\bibfnamefont {C.~P.~L.}\ \bibnamefont {{Berry}}},\ and\
  \bibinfo {author} {\bibfnamefont {V.}~\bibnamefont {{Kalogera}}},\ }\bibfield
   {title} {\bibinfo {title} {{Exploring the Lower Mass Gap and Unequal Mass
  Regime in Compact Binary Evolution}},\ }\href
  {https://doi.org/10.3847/2041-8213/aba74e} {\bibfield  {journal} {\bibinfo
  {journal} {\apjl}\ }\textbf {\bibinfo {volume} {899}},\ \bibinfo {eid} {L1}
  (\bibinfo {year} {2020})},\ \Eprint {https://arxiv.org/abs/2006.14573}
  {arXiv:2006.14573 [astro-ph.HE]} \BibitemShut {NoStop}%
\bibitem [{\citenamefont {{Mandel}}\ and\ \citenamefont
  {{M{\"u}ller}}(2020)}]{Mandel:2020}%
  \BibitemOpen
  \bibfield  {author} {\bibinfo {author} {\bibfnamefont {I.}~\bibnamefont
  {{Mandel}}}\ and\ \bibinfo {author} {\bibfnamefont {B.}~\bibnamefont
  {{M{\"u}ller}}},\ }\bibfield  {title} {\bibinfo {title} {{Simple recipes for
  compact remnant masses and natal kicks}},\ }\href
  {https://doi.org/10.1093/mnras/staa3043} {\bibfield  {journal} {\bibinfo
  {journal} {\mnras}\ }\textbf {\bibinfo {volume} {499}},\ \bibinfo {pages}
  {3214} (\bibinfo {year} {2020})},\ \Eprint {https://arxiv.org/abs/2006.08360}
  {arXiv:2006.08360 [astro-ph.HE]} \BibitemShut {NoStop}%
\bibitem [{\citenamefont {{van Son}}\ \emph {et~al.}(2022)\citenamefont {{van
  Son}}, \citenamefont {{de Mink}}, \citenamefont {{Renzo}}, \citenamefont
  {{Justham}}, \citenamefont {{Zapartas}}, \citenamefont {{Breivik}},
  \citenamefont {{Callister}}, \citenamefont {{Farr}},\ and\ \citenamefont
  {{Conroy}}}]{vanSon:2022}%
  \BibitemOpen
  \bibfield  {author} {\bibinfo {author} {\bibfnamefont {L.~A.~C.}\
  \bibnamefont {{van Son}}}, \bibinfo {author} {\bibfnamefont {S.~E.}\
  \bibnamefont {{de Mink}}}, \bibinfo {author} {\bibfnamefont {M.}~\bibnamefont
  {{Renzo}}}, \bibinfo {author} {\bibfnamefont {S.}~\bibnamefont {{Justham}}},
  \bibinfo {author} {\bibfnamefont {E.}~\bibnamefont {{Zapartas}}}, \bibinfo
  {author} {\bibfnamefont {K.}~\bibnamefont {{Breivik}}}, \bibinfo {author}
  {\bibfnamefont {T.}~\bibnamefont {{Callister}}}, \bibinfo {author}
  {\bibfnamefont {W.~M.}\ \bibnamefont {{Farr}}},\ and\ \bibinfo {author}
  {\bibfnamefont {C.}~\bibnamefont {{Conroy}}},\ }\bibfield  {title} {\bibinfo
  {title} {{No Peaks without Valleys: The Stable Mass Transfer Channel for
  Gravitational-wave Sources in Light of the Neutron Star-Black Hole Mass
  Gap}},\ }\href {https://doi.org/10.3847/1538-4357/ac9b0a} {\bibfield
  {journal} {\bibinfo  {journal} {\apj}\ }\textbf {\bibinfo {volume} {940}},\
  \bibinfo {eid} {184} (\bibinfo {year} {2022})},\ \Eprint
  {https://arxiv.org/abs/2209.13609} {arXiv:2209.13609 [astro-ph.HE]}
  \BibitemShut {NoStop}%
\bibitem [{\citenamefont {{Schneider}}\ \emph {et~al.}(2023)\citenamefont
  {{Schneider}}, \citenamefont {{Podsiadlowski}},\ and\ \citenamefont
  {{Laplace}}}]{Schneider:2023}%
  \BibitemOpen
  \bibfield  {author} {\bibinfo {author} {\bibfnamefont {F.~R.~N.}\
  \bibnamefont {{Schneider}}}, \bibinfo {author} {\bibfnamefont
  {P.}~\bibnamefont {{Podsiadlowski}}},\ and\ \bibinfo {author} {\bibfnamefont
  {E.}~\bibnamefont {{Laplace}}},\ }\bibfield  {title} {\bibinfo {title}
  {{Bimodal Black Hole Mass Distribution and Chirp Masses of Binary Black Hole
  Mergers}},\ }\href {https://doi.org/10.3847/2041-8213/acd77a} {\bibfield
  {journal} {\bibinfo  {journal} {\apjl}\ }\textbf {\bibinfo {volume} {950}},\
  \bibinfo {eid} {L9} (\bibinfo {year} {2023})},\ \Eprint
  {https://arxiv.org/abs/2305.02380} {arXiv:2305.02380 [astro-ph.HE]}
  \BibitemShut {NoStop}%
\bibitem [{\citenamefont {{Fowler}}\ and\ \citenamefont
  {{Hoyle}}(1964)}]{Fowler:1964}%
  \BibitemOpen
  \bibfield  {author} {\bibinfo {author} {\bibfnamefont {W.~A.}\ \bibnamefont
  {{Fowler}}}\ and\ \bibinfo {author} {\bibfnamefont {F.}~\bibnamefont
  {{Hoyle}}},\ }\bibfield  {title} {\bibinfo {title} {{Neutrino Processes and
  Pair Formation in Massive Stars and Supernovae.}},\ }\href
  {https://doi.org/10.1086/190103} {\bibfield  {journal} {\bibinfo  {journal}
  {\apjs}\ }\textbf {\bibinfo {volume} {9}},\ \bibinfo {pages} {201} (\bibinfo
  {year} {1964})}\BibitemShut {NoStop}%
\bibitem [{\citenamefont {{Barkat}}\ \emph {et~al.}(1967)\citenamefont
  {{Barkat}}, \citenamefont {{Rakavy}},\ and\ \citenamefont
  {{Sack}}}]{Barkat:1967}%
  \BibitemOpen
  \bibfield  {author} {\bibinfo {author} {\bibfnamefont {Z.}~\bibnamefont
  {{Barkat}}}, \bibinfo {author} {\bibfnamefont {G.}~\bibnamefont {{Rakavy}}},\
  and\ \bibinfo {author} {\bibfnamefont {N.}~\bibnamefont {{Sack}}},\
  }\bibfield  {title} {\bibinfo {title} {{Dynamics of Supernova Explosion
  Resulting from Pair Formation}},\ }\href
  {https://doi.org/10.1103/PhysRevLett.18.379} {\bibfield  {journal} {\bibinfo
  {journal} {\prl}\ }\textbf {\bibinfo {volume} {18}},\ \bibinfo {pages} {379}
  (\bibinfo {year} {1967})}\BibitemShut {NoStop}%
\bibitem [{\citenamefont {{Heger}}\ and\ \citenamefont
  {{Woosley}}(2002)}]{Heger:2002}%
  \BibitemOpen
  \bibfield  {author} {\bibinfo {author} {\bibfnamefont {A.}~\bibnamefont
  {{Heger}}}\ and\ \bibinfo {author} {\bibfnamefont {S.~E.}\ \bibnamefont
  {{Woosley}}},\ }\bibfield  {title} {\bibinfo {title} {{The Nucleosynthetic
  Signature of Population III}},\ }\href {https://doi.org/10.1086/338487}
  {\bibfield  {journal} {\bibinfo  {journal} {\apj}\ }\textbf {\bibinfo
  {volume} {567}},\ \bibinfo {pages} {532} (\bibinfo {year} {2002})},\ \Eprint
  {https://arxiv.org/abs/astro-ph/0107037} {arXiv:astro-ph/0107037 [astro-ph]}
  \BibitemShut {NoStop}%
\bibitem [{\citenamefont {{Heger}}\ \emph {et~al.}(2003)\citenamefont
  {{Heger}}, \citenamefont {{Fryer}}, \citenamefont {{Woosley}}, \citenamefont
  {{Langer}},\ and\ \citenamefont {{Hartmann}}}]{Heger:2003}%
  \BibitemOpen
  \bibfield  {author} {\bibinfo {author} {\bibfnamefont {A.}~\bibnamefont
  {{Heger}}}, \bibinfo {author} {\bibfnamefont {C.~L.}\ \bibnamefont
  {{Fryer}}}, \bibinfo {author} {\bibfnamefont {S.~E.}\ \bibnamefont
  {{Woosley}}}, \bibinfo {author} {\bibfnamefont {N.}~\bibnamefont
  {{Langer}}},\ and\ \bibinfo {author} {\bibfnamefont {D.~H.}\ \bibnamefont
  {{Hartmann}}},\ }\bibfield  {title} {\bibinfo {title} {{How Massive Single
  Stars End Their Life}},\ }\href {https://doi.org/10.1086/375341} {\bibfield
  {journal} {\bibinfo  {journal} {\apj}\ }\textbf {\bibinfo {volume} {591}},\
  \bibinfo {pages} {288} (\bibinfo {year} {2003})},\ \Eprint
  {https://arxiv.org/abs/astro-ph/0212469} {arXiv:astro-ph/0212469 [astro-ph]}
  \BibitemShut {NoStop}%
\bibitem [{\citenamefont {{Woosley}}\ and\ \citenamefont
  {{Heger}}(2015)}]{Woosley:2015}%
  \BibitemOpen
  \bibfield  {author} {\bibinfo {author} {\bibfnamefont {S.~E.}\ \bibnamefont
  {{Woosley}}}\ and\ \bibinfo {author} {\bibfnamefont {A.}~\bibnamefont
  {{Heger}}},\ }\bibfield  {title} {\bibinfo {title} {{The Deaths of Very
  Massive Stars}},\ }in\ \href {https://doi.org/10.1007/978-3-319-09596-7_7}
  {\emph {\bibinfo {booktitle} {Very Massive Stars in the Local Universe}}},\
  \bibinfo {series} {Astrophysics and Space Science Library}, Vol.\ \bibinfo
  {volume} {412},\ \bibinfo {editor} {edited by\ \bibinfo {editor}
  {\bibfnamefont {J.~S.}\ \bibnamefont {{Vink}}}}\ (\bibinfo {year} {2015})\
  p.\ \bibinfo {pages} {199},\ \Eprint {https://arxiv.org/abs/1406.5657}
  {arXiv:1406.5657 [astro-ph.SR]} \BibitemShut {NoStop}%
\bibitem [{\citenamefont {{Belczynski}}\ \emph {et~al.}(2016)\citenamefont
  {{Belczynski}}, \citenamefont {{Heger}}, \citenamefont {{Gladysz}},
  \citenamefont {{Ruiter}}, \citenamefont {{Woosley}}, \citenamefont
  {{Wiktorowicz}}, \citenamefont {{Chen}}, \citenamefont {{Bulik}},
  \citenamefont {{O'Shaughnessy}}, \citenamefont {{Holz}}, \citenamefont
  {{Fryer}},\ and\ \citenamefont {{Berti}}}]{Belczynski:2016}%
  \BibitemOpen
  \bibfield  {author} {\bibinfo {author} {\bibfnamefont {K.}~\bibnamefont
  {{Belczynski}}}, \bibinfo {author} {\bibfnamefont {A.}~\bibnamefont
  {{Heger}}}, \bibinfo {author} {\bibfnamefont {W.}~\bibnamefont {{Gladysz}}},
  \bibinfo {author} {\bibfnamefont {A.~J.}\ \bibnamefont {{Ruiter}}}, \bibinfo
  {author} {\bibfnamefont {S.}~\bibnamefont {{Woosley}}}, \bibinfo {author}
  {\bibfnamefont {G.}~\bibnamefont {{Wiktorowicz}}}, \bibinfo {author}
  {\bibfnamefont {H.~Y.}\ \bibnamefont {{Chen}}}, \bibinfo {author}
  {\bibfnamefont {T.}~\bibnamefont {{Bulik}}}, \bibinfo {author} {\bibfnamefont
  {R.}~\bibnamefont {{O'Shaughnessy}}}, \bibinfo {author} {\bibfnamefont
  {D.~E.}\ \bibnamefont {{Holz}}}, \bibinfo {author} {\bibfnamefont {C.~L.}\
  \bibnamefont {{Fryer}}},\ and\ \bibinfo {author} {\bibfnamefont
  {E.}~\bibnamefont {{Berti}}},\ }\bibfield  {title} {\bibinfo {title} {{The
  effect of pair-instability mass loss on black-hole mergers}},\ }\href
  {https://doi.org/10.1051/0004-6361/201628980} {\bibfield  {journal} {\bibinfo
   {journal} {\aap}\ }\textbf {\bibinfo {volume} {594}},\ \bibinfo {eid} {A97}
  (\bibinfo {year} {2016})},\ \Eprint {https://arxiv.org/abs/1607.03116}
  {arXiv:1607.03116 [astro-ph.HE]} \BibitemShut {NoStop}%
\bibitem [{\citenamefont {{Talbot}}\ and\ \citenamefont
  {{Thrane}}(2018)}]{Talbot:2018}%
  \BibitemOpen
  \bibfield  {author} {\bibinfo {author} {\bibfnamefont {C.}~\bibnamefont
  {{Talbot}}}\ and\ \bibinfo {author} {\bibfnamefont {E.}~\bibnamefont
  {{Thrane}}},\ }\bibfield  {title} {\bibinfo {title} {{Measuring the Binary
  Black Hole Mass Spectrum with an Astrophysically Motivated
  Parameterization}},\ }\href {https://doi.org/10.3847/1538-4357/aab34c}
  {\bibfield  {journal} {\bibinfo  {journal} {\apj}\ }\textbf {\bibinfo
  {volume} {856}},\ \bibinfo {eid} {173} (\bibinfo {year} {2018})},\ \Eprint
  {https://arxiv.org/abs/1801.02699} {arXiv:1801.02699 [astro-ph.HE]}
  \BibitemShut {NoStop}%
\bibitem [{\citenamefont {{Marchant}}\ \emph {et~al.}(2019)\citenamefont
  {{Marchant}}, \citenamefont {{Renzo}}, \citenamefont {{Farmer}},
  \citenamefont {{Pappas}}, \citenamefont {{Taam}}, \citenamefont {{de Mink}},\
  and\ \citenamefont {{Kalogera}}}]{Marchant:2019}%
  \BibitemOpen
  \bibfield  {author} {\bibinfo {author} {\bibfnamefont {P.}~\bibnamefont
  {{Marchant}}}, \bibinfo {author} {\bibfnamefont {M.}~\bibnamefont {{Renzo}}},
  \bibinfo {author} {\bibfnamefont {R.}~\bibnamefont {{Farmer}}}, \bibinfo
  {author} {\bibfnamefont {K.~M.~W.}\ \bibnamefont {{Pappas}}}, \bibinfo
  {author} {\bibfnamefont {R.~E.}\ \bibnamefont {{Taam}}}, \bibinfo {author}
  {\bibfnamefont {S.~E.}\ \bibnamefont {{de Mink}}},\ and\ \bibinfo {author}
  {\bibfnamefont {V.}~\bibnamefont {{Kalogera}}},\ }\bibfield  {title}
  {\bibinfo {title} {{Pulsational Pair-instability Supernovae in Very Close
  Binaries}},\ }\href {https://doi.org/10.3847/1538-4357/ab3426} {\bibfield
  {journal} {\bibinfo  {journal} {\apj}\ }\textbf {\bibinfo {volume} {882}},\
  \bibinfo {eid} {36} (\bibinfo {year} {2019})},\ \Eprint
  {https://arxiv.org/abs/1810.13412} {arXiv:1810.13412 [astro-ph.HE]}
  \BibitemShut {NoStop}%
\bibitem [{\citenamefont {{Woosley}}(2019)}]{Woosley:2019}%
  \BibitemOpen
  \bibfield  {author} {\bibinfo {author} {\bibfnamefont {S.~E.}\ \bibnamefont
  {{Woosley}}},\ }\bibfield  {title} {\bibinfo {title} {{The Evolution of
  Massive Helium Stars, Including Mass Loss}},\ }\href
  {https://doi.org/10.3847/1538-4357/ab1b41} {\bibfield  {journal} {\bibinfo
  {journal} {\apj}\ }\textbf {\bibinfo {volume} {878}},\ \bibinfo {eid} {49}
  (\bibinfo {year} {2019})},\ \Eprint {https://arxiv.org/abs/1901.00215}
  {arXiv:1901.00215 [astro-ph.SR]} \BibitemShut {NoStop}%
\bibitem [{\citenamefont {{Renzo}}\ \emph {et~al.}(2020)\citenamefont
  {{Renzo}}, \citenamefont {{Farmer}}, \citenamefont {{Justham}}, \citenamefont
  {{G{\"o}tberg}}, \citenamefont {{de Mink}}, \citenamefont {{Zapartas}},
  \citenamefont {{Marchant}},\ and\ \citenamefont {{Smith}}}]{Renzo:2020}%
  \BibitemOpen
  \bibfield  {author} {\bibinfo {author} {\bibfnamefont {M.}~\bibnamefont
  {{Renzo}}}, \bibinfo {author} {\bibfnamefont {R.}~\bibnamefont {{Farmer}}},
  \bibinfo {author} {\bibfnamefont {S.}~\bibnamefont {{Justham}}}, \bibinfo
  {author} {\bibfnamefont {Y.}~\bibnamefont {{G{\"o}tberg}}}, \bibinfo {author}
  {\bibfnamefont {S.~E.}\ \bibnamefont {{de Mink}}}, \bibinfo {author}
  {\bibfnamefont {E.}~\bibnamefont {{Zapartas}}}, \bibinfo {author}
  {\bibfnamefont {P.}~\bibnamefont {{Marchant}}},\ and\ \bibinfo {author}
  {\bibfnamefont {N.}~\bibnamefont {{Smith}}},\ }\bibfield  {title} {\bibinfo
  {title} {{Predictions for the hydrogen-free ejecta of pulsational
  pair-instability supernovae}},\ }\href
  {https://doi.org/10.1051/0004-6361/202037710} {\bibfield  {journal} {\bibinfo
   {journal} {\aap}\ }\textbf {\bibinfo {volume} {640}},\ \bibinfo {eid} {A56}
  (\bibinfo {year} {2020})},\ \Eprint {https://arxiv.org/abs/2002.05077}
  {arXiv:2002.05077 [astro-ph.SR]} \BibitemShut {NoStop}%
\bibitem [{\citenamefont {{Farmer}}\ \emph {et~al.}(2019)\citenamefont
  {{Farmer}}, \citenamefont {{Renzo}}, \citenamefont {{de Mink}}, \citenamefont
  {{Marchant}},\ and\ \citenamefont {{Justham}}}]{Farmer:2019}%
  \BibitemOpen
  \bibfield  {author} {\bibinfo {author} {\bibfnamefont {R.}~\bibnamefont
  {{Farmer}}}, \bibinfo {author} {\bibfnamefont {M.}~\bibnamefont {{Renzo}}},
  \bibinfo {author} {\bibfnamefont {S.~E.}\ \bibnamefont {{de Mink}}}, \bibinfo
  {author} {\bibfnamefont {P.}~\bibnamefont {{Marchant}}},\ and\ \bibinfo
  {author} {\bibfnamefont {S.}~\bibnamefont {{Justham}}},\ }\bibfield  {title}
  {\bibinfo {title} {{Mind the Gap: The Location of the Lower Edge of the
  Pair-instability Supernova Black Hole Mass Gap}},\ }\href
  {https://doi.org/10.3847/1538-4357/ab518b} {\bibfield  {journal} {\bibinfo
  {journal} {\apj}\ }\textbf {\bibinfo {volume} {887}},\ \bibinfo {eid} {53}
  (\bibinfo {year} {2019})},\ \Eprint {https://arxiv.org/abs/1910.12874}
  {arXiv:1910.12874 [astro-ph.SR]} \BibitemShut {NoStop}%
\bibitem [{\citenamefont {{Hendriks}}\ \emph {et~al.}(2023)\citenamefont
  {{Hendriks}}, \citenamefont {{van Son}}, \citenamefont {{Renzo}},
  \citenamefont {{Izzard}},\ and\ \citenamefont {{Farmer}}}]{Hendriks:2023}%
  \BibitemOpen
  \bibfield  {author} {\bibinfo {author} {\bibfnamefont {D.~D.}\ \bibnamefont
  {{Hendriks}}}, \bibinfo {author} {\bibfnamefont {L.~A.~C.}\ \bibnamefont
  {{van Son}}}, \bibinfo {author} {\bibfnamefont {M.}~\bibnamefont {{Renzo}}},
  \bibinfo {author} {\bibfnamefont {R.~G.}\ \bibnamefont {{Izzard}}},\ and\
  \bibinfo {author} {\bibfnamefont {R.}~\bibnamefont {{Farmer}}},\ }\bibfield
  {title} {\bibinfo {title} {{Pulsational pair-instability supernovae in
  gravitational-wave and electromagnetic transients}},\ }\href
  {https://doi.org/10.1093/mnras/stad2857} {\bibfield  {journal} {\bibinfo
  {journal} {\mnras}\ }\textbf {\bibinfo {volume} {526}},\ \bibinfo {pages}
  {4130} (\bibinfo {year} {2023})},\ \Eprint {https://arxiv.org/abs/2309.09339}
  {arXiv:2309.09339 [astro-ph.HE]} \BibitemShut {NoStop}%
\bibitem [{\citenamefont {{Briel}}\ \emph {et~al.}(2023)\citenamefont
  {{Briel}}, \citenamefont {{Stevance}},\ and\ \citenamefont
  {{Eldridge}}}]{Briel:2023}%
  \BibitemOpen
  \bibfield  {author} {\bibinfo {author} {\bibfnamefont {M.~M.}\ \bibnamefont
  {{Briel}}}, \bibinfo {author} {\bibfnamefont {H.~F.}\ \bibnamefont
  {{Stevance}}},\ and\ \bibinfo {author} {\bibfnamefont {J.~J.}\ \bibnamefont
  {{Eldridge}}},\ }\bibfield  {title} {\bibinfo {title} {{Understanding the
  high-mass binary black hole population from stable mass transfer and
  super-Eddington accretion in BPASS}},\ }\href
  {https://doi.org/10.1093/mnras/stad399} {\bibfield  {journal} {\bibinfo
  {journal} {\mnras}\ }\textbf {\bibinfo {volume} {520}},\ \bibinfo {pages}
  {5724} (\bibinfo {year} {2023})},\ \Eprint {https://arxiv.org/abs/2206.13842}
  {arXiv:2206.13842 [astro-ph.HE]} \BibitemShut {NoStop}%
\bibitem [{\citenamefont {{Chapline}}(1975)}]{George:1975}%
  \BibitemOpen
  \bibfield  {author} {\bibinfo {author} {\bibfnamefont {G.~F.}\ \bibnamefont
  {{Chapline}}},\ }\bibfield  {title} {\bibinfo {title} {{Cosmological effects
  of primordial black holes}},\ }\href {https://doi.org/10.1038/253251a0}
  {\bibfield  {journal} {\bibinfo  {journal} {\nat}\ }\textbf {\bibinfo
  {volume} {253}},\ \bibinfo {pages} {251} (\bibinfo {year}
  {1975})}\BibitemShut {NoStop}%
\bibitem [{\citenamefont {{Carr}}(1975)}]{Carr:1975}%
  \BibitemOpen
  \bibfield  {author} {\bibinfo {author} {\bibfnamefont {B.~J.}\ \bibnamefont
  {{Carr}}},\ }\bibfield  {title} {\bibinfo {title} {{The primordial black hole
  mass spectrum.}},\ }\href {https://doi.org/10.1086/153853} {\bibfield
  {journal} {\bibinfo  {journal} {\apj}\ }\textbf {\bibinfo {volume} {201}},\
  \bibinfo {pages} {1} (\bibinfo {year} {1975})}\BibitemShut {NoStop}%
\bibitem [{\citenamefont {{Bird}}\ \emph {et~al.}(2016)\citenamefont {{Bird}},
  \citenamefont {{Cholis}}, \citenamefont {{Mu{\~n}oz}}, \citenamefont
  {{Ali-Ha{\"\i}moud}}, \citenamefont {{Kamionkowski}}, \citenamefont
  {{Kovetz}}, \citenamefont {{Raccanelli}},\ and\ \citenamefont
  {{Riess}}}]{Bird:2016}%
  \BibitemOpen
  \bibfield  {author} {\bibinfo {author} {\bibfnamefont {S.}~\bibnamefont
  {{Bird}}}, \bibinfo {author} {\bibfnamefont {I.}~\bibnamefont {{Cholis}}},
  \bibinfo {author} {\bibfnamefont {J.~B.}\ \bibnamefont {{Mu{\~n}oz}}},
  \bibinfo {author} {\bibfnamefont {Y.}~\bibnamefont {{Ali-Ha{\"\i}moud}}},
  \bibinfo {author} {\bibfnamefont {M.}~\bibnamefont {{Kamionkowski}}},
  \bibinfo {author} {\bibfnamefont {E.~D.}\ \bibnamefont {{Kovetz}}}, \bibinfo
  {author} {\bibfnamefont {A.}~\bibnamefont {{Raccanelli}}},\ and\ \bibinfo
  {author} {\bibfnamefont {A.~G.}\ \bibnamefont {{Riess}}},\ }\bibfield
  {title} {\bibinfo {title} {{Did LIGO Detect Dark Matter?}},\ }\href
  {https://doi.org/10.1103/PhysRevLett.116.201301} {\bibfield  {journal}
  {\bibinfo  {journal} {\prl}\ }\textbf {\bibinfo {volume} {116}},\ \bibinfo
  {eid} {201301} (\bibinfo {year} {2016})},\ \Eprint
  {https://arxiv.org/abs/1603.00464} {arXiv:1603.00464 [astro-ph.CO]}
  \BibitemShut {NoStop}%
\bibitem [{\citenamefont {{Sasaki}}\ \emph {et~al.}(2018)\citenamefont
  {{Sasaki}}, \citenamefont {{Suyama}}, \citenamefont {{Tanaka}},\ and\
  \citenamefont {{Yokoyama}}}]{Sasaki:2018}%
  \BibitemOpen
  \bibfield  {author} {\bibinfo {author} {\bibfnamefont {M.}~\bibnamefont
  {{Sasaki}}}, \bibinfo {author} {\bibfnamefont {T.}~\bibnamefont {{Suyama}}},
  \bibinfo {author} {\bibfnamefont {T.}~\bibnamefont {{Tanaka}}},\ and\
  \bibinfo {author} {\bibfnamefont {S.}~\bibnamefont {{Yokoyama}}},\ }\bibfield
   {title} {\bibinfo {title} {{Primordial black holes{\textemdash}perspectives
  in gravitational wave astronomy}},\ }\href
  {https://doi.org/10.1088/1361-6382/aaa7b4} {\bibfield  {journal} {\bibinfo
  {journal} {Classical and Quantum Gravity}\ }\textbf {\bibinfo {volume}
  {35}},\ \bibinfo {eid} {063001} (\bibinfo {year} {2018})},\ \Eprint
  {https://arxiv.org/abs/1801.05235} {arXiv:1801.05235 [astro-ph.CO]}
  \BibitemShut {NoStop}%
\bibitem [{\citenamefont {{Antonini}}\ \emph {et~al.}(2017)\citenamefont
  {{Antonini}}, \citenamefont {{Toonen}},\ and\ \citenamefont
  {{Hamers}}}]{Antonini:2017}%
  \BibitemOpen
  \bibfield  {author} {\bibinfo {author} {\bibfnamefont {F.}~\bibnamefont
  {{Antonini}}}, \bibinfo {author} {\bibfnamefont {S.}~\bibnamefont
  {{Toonen}}},\ and\ \bibinfo {author} {\bibfnamefont {A.~S.}\ \bibnamefont
  {{Hamers}}},\ }\bibfield  {title} {\bibinfo {title} {{Binary Black Hole
  Mergers from Field Triples: Properties, Rates, and the Impact of Stellar
  Evolution}},\ }\href {https://doi.org/10.3847/1538-4357/aa6f5e} {\bibfield
  {journal} {\bibinfo  {journal} {\apj}\ }\textbf {\bibinfo {volume} {841}},\
  \bibinfo {eid} {77} (\bibinfo {year} {2017})},\ \Eprint
  {https://arxiv.org/abs/1703.06614} {arXiv:1703.06614 [astro-ph.GA]}
  \BibitemShut {NoStop}%
\bibitem [{\citenamefont {{Silsbee}}\ and\ \citenamefont
  {{Tremaine}}(2017)}]{Silsbee:2017}%
  \BibitemOpen
  \bibfield  {author} {\bibinfo {author} {\bibfnamefont {K.}~\bibnamefont
  {{Silsbee}}}\ and\ \bibinfo {author} {\bibfnamefont {S.}~\bibnamefont
  {{Tremaine}}},\ }\bibfield  {title} {\bibinfo {title} {{Lidov-Kozai Cycles
  with Gravitational Radiation: Merging Black Holes in Isolated Triple
  Systems}},\ }\href {https://doi.org/10.3847/1538-4357/aa5729} {\bibfield
  {journal} {\bibinfo  {journal} {\apj}\ }\textbf {\bibinfo {volume} {836}},\
  \bibinfo {eid} {39} (\bibinfo {year} {2017})},\ \Eprint
  {https://arxiv.org/abs/1608.07642} {arXiv:1608.07642 [astro-ph.HE]}
  \BibitemShut {NoStop}%
\bibitem [{\citenamefont {{Inayoshi}}\ \emph {et~al.}(2016)\citenamefont
  {{Inayoshi}}, \citenamefont {{Kashiyama}}, \citenamefont {{Visbal}},\ and\
  \citenamefont {{Haiman}}}]{Inayoshi:2016}%
  \BibitemOpen
  \bibfield  {author} {\bibinfo {author} {\bibfnamefont {K.}~\bibnamefont
  {{Inayoshi}}}, \bibinfo {author} {\bibfnamefont {K.}~\bibnamefont
  {{Kashiyama}}}, \bibinfo {author} {\bibfnamefont {E.}~\bibnamefont
  {{Visbal}}},\ and\ \bibinfo {author} {\bibfnamefont {Z.}~\bibnamefont
  {{Haiman}}},\ }\bibfield  {title} {\bibinfo {title} {{Gravitational wave
  background from Population III binary black holes consistent with cosmic
  reionization}},\ }\href {https://doi.org/10.1093/mnras/stw1431} {\bibfield
  {journal} {\bibinfo  {journal} {\mnras}\ }\textbf {\bibinfo {volume} {461}},\
  \bibinfo {pages} {2722} (\bibinfo {year} {2016})},\ \Eprint
  {https://arxiv.org/abs/1603.06921} {arXiv:1603.06921 [astro-ph.GA]}
  \BibitemShut {NoStop}%
\bibitem [{\citenamefont {{Tiwari}}\ and\ \citenamefont
  {{Fairhurst}}(2021)}]{Tiwari:2021b}%
  \BibitemOpen
  \bibfield  {author} {\bibinfo {author} {\bibfnamefont {V.}~\bibnamefont
  {{Tiwari}}}\ and\ \bibinfo {author} {\bibfnamefont {S.}~\bibnamefont
  {{Fairhurst}}},\ }\bibfield  {title} {\bibinfo {title} {{The Emergence of
  Structure in the Binary Black Hole Mass Distribution}},\ }\href
  {https://doi.org/10.3847/2041-8213/abfbe7} {\bibfield  {journal} {\bibinfo
  {journal} {\apjl}\ }\textbf {\bibinfo {volume} {913}},\ \bibinfo {eid} {L19}
  (\bibinfo {year} {2021})},\ \Eprint {https://arxiv.org/abs/2011.04502}
  {arXiv:2011.04502 [astro-ph.HE]} \BibitemShut {NoStop}%
\bibitem [{\citenamefont {{Rodriguez}}\ \emph {et~al.}(2016)\citenamefont
  {{Rodriguez}}, \citenamefont {{Chatterjee}},\ and\ \citenamefont
  {{Rasio}}}]{Rodriguez:2016PhRvD..93h4029R}%
  \BibitemOpen
  \bibfield  {author} {\bibinfo {author} {\bibfnamefont {C.~L.}\ \bibnamefont
  {{Rodriguez}}}, \bibinfo {author} {\bibfnamefont {S.}~\bibnamefont
  {{Chatterjee}}},\ and\ \bibinfo {author} {\bibfnamefont {F.~A.}\ \bibnamefont
  {{Rasio}}},\ }\bibfield  {title} {\bibinfo {title} {{Binary black hole
  mergers from globular clusters: Masses, merger rates, and the impact of
  stellar evolution}},\ }\href {https://doi.org/10.1103/PhysRevD.93.084029}
  {\bibfield  {journal} {\bibinfo  {journal} {\prd}\ }\textbf {\bibinfo
  {volume} {93}},\ \bibinfo {eid} {084029} (\bibinfo {year} {2016})},\ \Eprint
  {https://arxiv.org/abs/1602.02444} {arXiv:1602.02444 [astro-ph.HE]}
  \BibitemShut {NoStop}%
\bibitem [{\citenamefont {{Amaro-Seoane}}\ and\ \citenamefont
  {{Chen}}(2016)}]{Amaro-Seoane:2016}%
  \BibitemOpen
  \bibfield  {author} {\bibinfo {author} {\bibfnamefont {P.}~\bibnamefont
  {{Amaro-Seoane}}}\ and\ \bibinfo {author} {\bibfnamefont {X.}~\bibnamefont
  {{Chen}}},\ }\bibfield  {title} {\bibinfo {title} {{Relativistic mergers of
  black hole binaries have large, similar masses, low spins and are
  circular}},\ }\href {https://doi.org/10.1093/mnras/stw503} {\bibfield
  {journal} {\bibinfo  {journal} {\mnras}\ }\textbf {\bibinfo {volume} {458}},\
  \bibinfo {pages} {3075} (\bibinfo {year} {2016})},\ \Eprint
  {https://arxiv.org/abs/1512.04897} {arXiv:1512.04897 [astro-ph.CO]}
  \BibitemShut {NoStop}%
\bibitem [{\citenamefont {{Farah}}\ \emph
  {et~al.}(2023{\natexlab{a}})\citenamefont {{Farah}}, \citenamefont
  {{Fishbach}},\ and\ \citenamefont {{Holz}}}]{Farah:2023b}%
  \BibitemOpen
  \bibfield  {author} {\bibinfo {author} {\bibfnamefont {A.~M.}\ \bibnamefont
  {{Farah}}}, \bibinfo {author} {\bibfnamefont {M.}~\bibnamefont
  {{Fishbach}}},\ and\ \bibinfo {author} {\bibfnamefont {D.~E.}\ \bibnamefont
  {{Holz}}},\ }\bibfield  {title} {\bibinfo {title} {{Two of a Kind: Comparing
  big and small black holes in binaries with gravitational waves}},\ }\href
  {https://doi.org/10.48550/arXiv.2308.05102} {\bibfield  {journal} {\bibinfo
  {journal} {arXiv e-prints}\ ,\ \bibinfo {eid} {arXiv:2308.05102}} (\bibinfo
  {year} {2023}{\natexlab{a}})},\ \Eprint {https://arxiv.org/abs/2308.05102}
  {arXiv:2308.05102 [astro-ph.HE]} \BibitemShut {NoStop}%
\bibitem [{\citenamefont {{Michaely}}\ and\ \citenamefont
  {{Perets}}(2019)}]{Michaely:2019}%
  \BibitemOpen
  \bibfield  {author} {\bibinfo {author} {\bibfnamefont {E.}~\bibnamefont
  {{Michaely}}}\ and\ \bibinfo {author} {\bibfnamefont {H.~B.}\ \bibnamefont
  {{Perets}}},\ }\bibfield  {title} {\bibinfo {title} {{Gravitational-wave
  Sources from Mergers of Binary Black Holes Catalyzed by Flyby Interactions in
  the Field}},\ }\href {https://doi.org/10.3847/2041-8213/ab5b9b} {\bibfield
  {journal} {\bibinfo  {journal} {\apjl}\ }\textbf {\bibinfo {volume} {887}},\
  \bibinfo {eid} {L36} (\bibinfo {year} {2019})},\ \Eprint
  {https://arxiv.org/abs/1902.01864} {arXiv:1902.01864 [astro-ph.SR]}
  \BibitemShut {NoStop}%
\bibitem [{\citenamefont {{Dominik}}\ \emph {et~al.}(2015)\citenamefont
  {{Dominik}}, \citenamefont {{Berti}}, \citenamefont {{O'Shaughnessy}},
  \citenamefont {{Mandel}}, \citenamefont {{Belczynski}}, \citenamefont
  {{Fryer}}, \citenamefont {{Holz}}, \citenamefont {{Bulik}},\ and\
  \citenamefont {{Pannarale}}}]{Dominik:2015}%
  \BibitemOpen
  \bibfield  {author} {\bibinfo {author} {\bibfnamefont {M.}~\bibnamefont
  {{Dominik}}}, \bibinfo {author} {\bibfnamefont {E.}~\bibnamefont {{Berti}}},
  \bibinfo {author} {\bibfnamefont {R.}~\bibnamefont {{O'Shaughnessy}}},
  \bibinfo {author} {\bibfnamefont {I.}~\bibnamefont {{Mandel}}}, \bibinfo
  {author} {\bibfnamefont {K.}~\bibnamefont {{Belczynski}}}, \bibinfo {author}
  {\bibfnamefont {C.}~\bibnamefont {{Fryer}}}, \bibinfo {author} {\bibfnamefont
  {D.~E.}\ \bibnamefont {{Holz}}}, \bibinfo {author} {\bibfnamefont
  {T.}~\bibnamefont {{Bulik}}},\ and\ \bibinfo {author} {\bibfnamefont
  {F.}~\bibnamefont {{Pannarale}}},\ }\bibfield  {title} {\bibinfo {title}
  {{Double Compact Objects III: Gravitational-wave Detection Rates}},\ }\href
  {https://doi.org/10.1088/0004-637X/806/2/263} {\bibfield  {journal} {\bibinfo
   {journal} {\apj}\ }\textbf {\bibinfo {volume} {806}},\ \bibinfo {eid} {263}
  (\bibinfo {year} {2015})},\ \Eprint {https://arxiv.org/abs/1405.7016}
  {arXiv:1405.7016 [astro-ph.HE]} \BibitemShut {NoStop}%
\bibitem [{\citenamefont {{Giacobbo}}\ \emph {et~al.}(2018)\citenamefont
  {{Giacobbo}}, \citenamefont {{Mapelli}},\ and\ \citenamefont
  {{Spera}}}]{Giacobbo:2018}%
  \BibitemOpen
  \bibfield  {author} {\bibinfo {author} {\bibfnamefont {N.}~\bibnamefont
  {{Giacobbo}}}, \bibinfo {author} {\bibfnamefont {M.}~\bibnamefont
  {{Mapelli}}},\ and\ \bibinfo {author} {\bibfnamefont {M.}~\bibnamefont
  {{Spera}}},\ }\bibfield  {title} {\bibinfo {title} {{Merging black hole
  binaries: the effects of progenitor's metallicity, mass-loss rate and
  Eddington factor}},\ }\href {https://doi.org/10.1093/mnras/stx2933}
  {\bibfield  {journal} {\bibinfo  {journal} {\mnras}\ }\textbf {\bibinfo
  {volume} {474}},\ \bibinfo {pages} {2959} (\bibinfo {year} {2018})},\ \Eprint
  {https://arxiv.org/abs/1711.03556} {arXiv:1711.03556 [astro-ph.SR]}
  \BibitemShut {NoStop}%
\bibitem [{\citenamefont {{Spera}}\ \emph {et~al.}(2019)\citenamefont
  {{Spera}}, \citenamefont {{Mapelli}}, \citenamefont {{Giacobbo}},
  \citenamefont {{Trani}}, \citenamefont {{Bressan}},\ and\ \citenamefont
  {{Costa}}}]{Spera:2019}%
  \BibitemOpen
  \bibfield  {author} {\bibinfo {author} {\bibfnamefont {M.}~\bibnamefont
  {{Spera}}}, \bibinfo {author} {\bibfnamefont {M.}~\bibnamefont {{Mapelli}}},
  \bibinfo {author} {\bibfnamefont {N.}~\bibnamefont {{Giacobbo}}}, \bibinfo
  {author} {\bibfnamefont {A.~A.}\ \bibnamefont {{Trani}}}, \bibinfo {author}
  {\bibfnamefont {A.}~\bibnamefont {{Bressan}}},\ and\ \bibinfo {author}
  {\bibfnamefont {G.}~\bibnamefont {{Costa}}},\ }\bibfield  {title} {\bibinfo
  {title} {{Merging black hole binaries with the SEVN code}},\ }\href
  {https://doi.org/10.1093/mnras/stz359} {\bibfield  {journal} {\bibinfo
  {journal} {\mnras}\ }\textbf {\bibinfo {volume} {485}},\ \bibinfo {pages}
  {889} (\bibinfo {year} {2019})},\ \Eprint {https://arxiv.org/abs/1809.04605}
  {arXiv:1809.04605 [astro-ph.HE]} \BibitemShut {NoStop}%
\bibitem [{\citenamefont {{Grudi{\'c}}}\ \emph {et~al.}(2023)\citenamefont
  {{Grudi{\'c}}}, \citenamefont {{Offner}}, \citenamefont {{Guszejnov}},
  \citenamefont {{Faucher-Gigu{\`e}re}},\ and\ \citenamefont
  {{Hopkins}}}]{Grudic:2023}%
  \BibitemOpen
  \bibfield  {author} {\bibinfo {author} {\bibfnamefont {M.~Y.}\ \bibnamefont
  {{Grudi{\'c}}}}, \bibinfo {author} {\bibfnamefont {S.~S.~R.}\ \bibnamefont
  {{Offner}}}, \bibinfo {author} {\bibfnamefont {D.}~\bibnamefont
  {{Guszejnov}}}, \bibinfo {author} {\bibfnamefont {C.-A.}\ \bibnamefont
  {{Faucher-Gigu{\`e}re}}},\ and\ \bibinfo {author} {\bibfnamefont {P.~F.}\
  \bibnamefont {{Hopkins}}},\ }\bibfield  {title} {\bibinfo {title} {{Does God
  play dice with star clusters?}},\ }\href
  {https://doi.org/10.48550/arXiv.2307.00052} {\bibfield  {journal} {\bibinfo
  {journal} {arXiv e-prints}\ ,\ \bibinfo {eid} {arXiv:2307.00052}} (\bibinfo
  {year} {2023})},\ \Eprint {https://arxiv.org/abs/2307.00052}
  {arXiv:2307.00052 [astro-ph.GA]} \BibitemShut {NoStop}%
\bibitem [{\citenamefont {{Dominik}}\ \emph {et~al.}(2012)\citenamefont
  {{Dominik}}, \citenamefont {{Belczynski}}, \citenamefont {{Fryer}},
  \citenamefont {{Holz}}, \citenamefont {{Berti}}, \citenamefont {{Bulik}},
  \citenamefont {{Mandel}},\ and\ \citenamefont
  {{O'Shaughnessy}}}]{Dominik:2012}%
  \BibitemOpen
  \bibfield  {author} {\bibinfo {author} {\bibfnamefont {M.}~\bibnamefont
  {{Dominik}}}, \bibinfo {author} {\bibfnamefont {K.}~\bibnamefont
  {{Belczynski}}}, \bibinfo {author} {\bibfnamefont {C.}~\bibnamefont
  {{Fryer}}}, \bibinfo {author} {\bibfnamefont {D.~E.}\ \bibnamefont {{Holz}}},
  \bibinfo {author} {\bibfnamefont {E.}~\bibnamefont {{Berti}}}, \bibinfo
  {author} {\bibfnamefont {T.}~\bibnamefont {{Bulik}}}, \bibinfo {author}
  {\bibfnamefont {I.}~\bibnamefont {{Mandel}}},\ and\ \bibinfo {author}
  {\bibfnamefont {R.}~\bibnamefont {{O'Shaughnessy}}},\ }\bibfield  {title}
  {\bibinfo {title} {{Double Compact Objects. I. The Significance of the Common
  Envelope on Merger Rates}},\ }\href
  {https://doi.org/10.1088/0004-637X/759/1/52} {\bibfield  {journal} {\bibinfo
  {journal} {\apj}\ }\textbf {\bibinfo {volume} {759}},\ \bibinfo {eid} {52}
  (\bibinfo {year} {2012})},\ \Eprint {https://arxiv.org/abs/1202.4901}
  {arXiv:1202.4901 [astro-ph.HE]} \BibitemShut {NoStop}%
\bibitem [{\citenamefont {{Stevenson}}\ \emph {et~al.}(2017)\citenamefont
  {{Stevenson}}, \citenamefont {{Vigna-G{\'o}mez}}, \citenamefont {{Mandel}},
  \citenamefont {{Barrett}}, \citenamefont {{Neijssel}}, \citenamefont
  {{Perkins}},\ and\ \citenamefont {{de Mink}}}]{Stevenson:2017}%
  \BibitemOpen
  \bibfield  {author} {\bibinfo {author} {\bibfnamefont {S.}~\bibnamefont
  {{Stevenson}}}, \bibinfo {author} {\bibfnamefont {A.}~\bibnamefont
  {{Vigna-G{\'o}mez}}}, \bibinfo {author} {\bibfnamefont {I.}~\bibnamefont
  {{Mandel}}}, \bibinfo {author} {\bibfnamefont {J.~W.}\ \bibnamefont
  {{Barrett}}}, \bibinfo {author} {\bibfnamefont {C.~J.}\ \bibnamefont
  {{Neijssel}}}, \bibinfo {author} {\bibfnamefont {D.}~\bibnamefont
  {{Perkins}}},\ and\ \bibinfo {author} {\bibfnamefont {S.~E.}\ \bibnamefont
  {{de Mink}}},\ }\bibfield  {title} {\bibinfo {title} {{Formation of the first
  three gravitational-wave observations through isolated binary evolution}},\
  }\href {https://doi.org/10.1038/ncomms14906} {\bibfield  {journal} {\bibinfo
  {journal} {Nature Communications}\ }\textbf {\bibinfo {volume} {8}},\
  \bibinfo {eid} {14906} (\bibinfo {year} {2017})},\ \Eprint
  {https://arxiv.org/abs/1704.01352} {arXiv:1704.01352 [astro-ph.HE]}
  \BibitemShut {NoStop}%
\bibitem [{\citenamefont {{Laplace}}\ \emph {et~al.}(2021)\citenamefont
  {{Laplace}}, \citenamefont {{Justham}}, \citenamefont {{Renzo}},
  \citenamefont {{G{\"o}tberg}}, \citenamefont {{Farmer}}, \citenamefont
  {{Vartanyan}},\ and\ \citenamefont {{de Mink}}}]{Laplace:2021}%
  \BibitemOpen
  \bibfield  {author} {\bibinfo {author} {\bibfnamefont {E.}~\bibnamefont
  {{Laplace}}}, \bibinfo {author} {\bibfnamefont {S.}~\bibnamefont
  {{Justham}}}, \bibinfo {author} {\bibfnamefont {M.}~\bibnamefont {{Renzo}}},
  \bibinfo {author} {\bibfnamefont {Y.}~\bibnamefont {{G{\"o}tberg}}}, \bibinfo
  {author} {\bibfnamefont {R.}~\bibnamefont {{Farmer}}}, \bibinfo {author}
  {\bibfnamefont {D.}~\bibnamefont {{Vartanyan}}},\ and\ \bibinfo {author}
  {\bibfnamefont {S.~E.}\ \bibnamefont {{de Mink}}},\ }\bibfield  {title}
  {\bibinfo {title} {{Different to the core: The pre-supernova structures of
  massive single and binary-stripped stars}},\ }\href
  {https://doi.org/10.1051/0004-6361/202140506} {\bibfield  {journal} {\bibinfo
   {journal} {\aap}\ }\textbf {\bibinfo {volume} {656}},\ \bibinfo {eid} {A58}
  (\bibinfo {year} {2021})},\ \Eprint {https://arxiv.org/abs/2102.05036}
  {arXiv:2102.05036 [astro-ph.SR]} \BibitemShut {NoStop}%
\bibitem [{\citenamefont {{Olejak}}\ and\ \citenamefont
  {{Belczynski}}(2021)}]{Olejak:2021}%
  \BibitemOpen
  \bibfield  {author} {\bibinfo {author} {\bibfnamefont {A.}~\bibnamefont
  {{Olejak}}}\ and\ \bibinfo {author} {\bibfnamefont {K.}~\bibnamefont
  {{Belczynski}}},\ }\bibfield  {title} {\bibinfo {title} {{The Implications of
  High Black Hole Spins for the Origin of Binary Black Hole Mergers}},\ }\href
  {https://doi.org/10.3847/2041-8213/ac2f48} {\bibfield  {journal} {\bibinfo
  {journal} {\apjl}\ }\textbf {\bibinfo {volume} {921}},\ \bibinfo {eid} {L2}
  (\bibinfo {year} {2021})},\ \Eprint {https://arxiv.org/abs/2109.06872}
  {arXiv:2109.06872 [astro-ph.HE]} \BibitemShut {NoStop}%
\bibitem [{\citenamefont {{Broekgaarden}}\ \emph {et~al.}(2022)\citenamefont
  {{Broekgaarden}}, \citenamefont {{Stevenson}},\ and\ \citenamefont
  {{Thrane}}}]{Broekgaarden:2022}%
  \BibitemOpen
  \bibfield  {author} {\bibinfo {author} {\bibfnamefont {F.~S.}\ \bibnamefont
  {{Broekgaarden}}}, \bibinfo {author} {\bibfnamefont {S.}~\bibnamefont
  {{Stevenson}}},\ and\ \bibinfo {author} {\bibfnamefont {E.}~\bibnamefont
  {{Thrane}}},\ }\bibfield  {title} {\bibinfo {title} {{Signatures of Mass
  Ratio Reversal in Gravitational Waves from Merging Binary Black Holes}},\
  }\href {https://doi.org/10.3847/1538-4357/ac8879} {\bibfield  {journal}
  {\bibinfo  {journal} {\apj}\ }\textbf {\bibinfo {volume} {938}},\ \bibinfo
  {eid} {45} (\bibinfo {year} {2022})},\ \Eprint
  {https://arxiv.org/abs/2205.01693} {arXiv:2205.01693 [astro-ph.HE]}
  \BibitemShut {NoStop}%
\bibitem [{\citenamefont {{Zevin}}\ and\ \citenamefont
  {{Bavera}}(2022)}]{Zevin:2022}%
  \BibitemOpen
  \bibfield  {author} {\bibinfo {author} {\bibfnamefont {M.}~\bibnamefont
  {{Zevin}}}\ and\ \bibinfo {author} {\bibfnamefont {S.~S.}\ \bibnamefont
  {{Bavera}}},\ }\bibfield  {title} {\bibinfo {title} {{Suspicious Siblings:
  The Distribution of Mass and Spin across Component Black Holes in Isolated
  Binary Evolution}},\ }\href {https://doi.org/10.3847/1538-4357/ac6f5d}
  {\bibfield  {journal} {\bibinfo  {journal} {\apj}\ }\textbf {\bibinfo
  {volume} {933}},\ \bibinfo {eid} {86} (\bibinfo {year} {2022})},\ \Eprint
  {https://arxiv.org/abs/2203.02515} {arXiv:2203.02515 [astro-ph.HE]}
  \BibitemShut {NoStop}%
\bibitem [{\citenamefont {{Kovetz}}\ \emph {et~al.}(2017)\citenamefont
  {{Kovetz}}, \citenamefont {{Cholis}}, \citenamefont {{Breysse}},\ and\
  \citenamefont {{Kamionkowski}}}]{Kovetz:2017}%
  \BibitemOpen
  \bibfield  {author} {\bibinfo {author} {\bibfnamefont {E.~D.}\ \bibnamefont
  {{Kovetz}}}, \bibinfo {author} {\bibfnamefont {I.}~\bibnamefont {{Cholis}}},
  \bibinfo {author} {\bibfnamefont {P.~C.}\ \bibnamefont {{Breysse}}},\ and\
  \bibinfo {author} {\bibfnamefont {M.}~\bibnamefont {{Kamionkowski}}},\
  }\bibfield  {title} {\bibinfo {title} {{Black hole mass function from
  gravitational wave measurements}},\ }\href
  {https://doi.org/10.1103/PhysRevD.95.103010} {\bibfield  {journal} {\bibinfo
  {journal} {\prd}\ }\textbf {\bibinfo {volume} {95}},\ \bibinfo {eid} {103010}
  (\bibinfo {year} {2017})},\ \Eprint {https://arxiv.org/abs/1611.01157}
  {arXiv:1611.01157 [astro-ph.CO]} \BibitemShut {NoStop}%
\bibitem [{\citenamefont {{Fishbach}}\ and\ \citenamefont
  {{Holz}}(2017)}]{Fishbach:2017}%
  \BibitemOpen
  \bibfield  {author} {\bibinfo {author} {\bibfnamefont {M.}~\bibnamefont
  {{Fishbach}}}\ and\ \bibinfo {author} {\bibfnamefont {D.~E.}\ \bibnamefont
  {{Holz}}},\ }\bibfield  {title} {\bibinfo {title} {{Where Are
  LIGO{\textquoteright}s Big Black Holes?}},\ }\href
  {https://doi.org/10.3847/2041-8213/aa9bf6} {\bibfield  {journal} {\bibinfo
  {journal} {\apjl}\ }\textbf {\bibinfo {volume} {851}},\ \bibinfo {eid} {L25}
  (\bibinfo {year} {2017})},\ \Eprint {https://arxiv.org/abs/1709.08584}
  {arXiv:1709.08584 [astro-ph.HE]} \BibitemShut {NoStop}%
\bibitem [{\citenamefont {{Tiwari}}(2021)}]{Tiwari:2021}%
  \BibitemOpen
  \bibfield  {author} {\bibinfo {author} {\bibfnamefont {V.}~\bibnamefont
  {{Tiwari}}},\ }\bibfield  {title} {\bibinfo {title} {{VAMANA: modeling binary
  black hole population with minimal assumptions}},\ }\href
  {https://doi.org/10.1088/1361-6382/ac0b54} {\bibfield  {journal} {\bibinfo
  {journal} {Classical and Quantum Gravity}\ }\textbf {\bibinfo {volume}
  {38}},\ \bibinfo {eid} {155007} (\bibinfo {year} {2021})},\ \Eprint
  {https://arxiv.org/abs/2006.15047} {arXiv:2006.15047 [astro-ph.HE]}
  \BibitemShut {NoStop}%
\bibitem [{\citenamefont {{Callister}}\ and\ \citenamefont
  {{Farr}}(2023)}]{Callister:2023}%
  \BibitemOpen
  \bibfield  {author} {\bibinfo {author} {\bibfnamefont {T.~A.}\ \bibnamefont
  {{Callister}}}\ and\ \bibinfo {author} {\bibfnamefont {W.~M.}\ \bibnamefont
  {{Farr}}},\ }\bibfield  {title} {\bibinfo {title} {{A Parameter-Free Tour of
  the Binary Black Hole Population}},\ }\href
  {https://doi.org/10.48550/arXiv.2302.07289} {\bibfield  {journal} {\bibinfo
  {journal} {arXiv e-prints}\ ,\ \bibinfo {eid} {arXiv:2302.07289}} (\bibinfo
  {year} {2023})},\ \Eprint {https://arxiv.org/abs/2302.07289}
  {arXiv:2302.07289 [astro-ph.HE]} \BibitemShut {NoStop}%
\bibitem [{\citenamefont {{Godfrey}}\ \emph {et~al.}(2023)\citenamefont
  {{Godfrey}}, \citenamefont {{Edelman}},\ and\ \citenamefont
  {{Farr}}}]{Godfrey:2023}%
  \BibitemOpen
  \bibfield  {author} {\bibinfo {author} {\bibfnamefont {J.}~\bibnamefont
  {{Godfrey}}}, \bibinfo {author} {\bibfnamefont {B.}~\bibnamefont
  {{Edelman}}},\ and\ \bibinfo {author} {\bibfnamefont {B.}~\bibnamefont
  {{Farr}}},\ }\bibfield  {title} {\bibinfo {title} {{Cosmic Cousins:
  Identification of a Subpopulation of Binary Black Holes Consistent with
  Isolated Binary Evolution}},\ }\href
  {https://doi.org/10.48550/arXiv.2304.01288} {\bibfield  {journal} {\bibinfo
  {journal} {arXiv e-prints}\ ,\ \bibinfo {eid} {arXiv:2304.01288}} (\bibinfo
  {year} {2023})},\ \Eprint {https://arxiv.org/abs/2304.01288}
  {arXiv:2304.01288 [astro-ph.HE]} \BibitemShut {NoStop}%
\bibitem [{\citenamefont {{Fishbach}}\ and\ \citenamefont
  {{Holz}}(2020)}]{Fishbach:2020b}%
  \BibitemOpen
  \bibfield  {author} {\bibinfo {author} {\bibfnamefont {M.}~\bibnamefont
  {{Fishbach}}}\ and\ \bibinfo {author} {\bibfnamefont {D.~E.}\ \bibnamefont
  {{Holz}}},\ }\bibfield  {title} {\bibinfo {title} {{Picky Partners: The
  Pairing of Component Masses in Binary Black Hole Mergers}},\ }\href
  {https://doi.org/10.3847/2041-8213/ab7247} {\bibfield  {journal} {\bibinfo
  {journal} {\apjl}\ }\textbf {\bibinfo {volume} {891}},\ \bibinfo {eid} {L27}
  (\bibinfo {year} {2020})},\ \Eprint {https://arxiv.org/abs/1905.12669}
  {arXiv:1905.12669 [astro-ph.HE]} \BibitemShut {NoStop}%
\bibitem [{\citenamefont {{Wong}}\ and\ \citenamefont
  {{Cranmer}}(2022)}]{Wong:2022}%
  \BibitemOpen
  \bibfield  {author} {\bibinfo {author} {\bibfnamefont {K.}~\bibnamefont
  {{Wong}}}\ and\ \bibinfo {author} {\bibfnamefont {M.}~\bibnamefont
  {{Cranmer}}},\ }\bibfield  {title} {\bibinfo {title} {{Automated discovery of
  interpretable gravitational-wave population models}},\ }in\ \href
  {https://doi.org/10.48550/arXiv.2207.12409} {\emph {\bibinfo {booktitle}
  {Machine Learning for Astrophysics}}}\ (\bibinfo {year} {2022})\ p.~\bibinfo
  {pages} {25},\ \Eprint {https://arxiv.org/abs/2207.12409} {arXiv:2207.12409
  [astro-ph.IM]} \BibitemShut {NoStop}%
\bibitem [{\citenamefont {{Farah}}\ \emph
  {et~al.}(2023{\natexlab{b}})\citenamefont {{Farah}}, \citenamefont
  {{Edelman}}, \citenamefont {{Zevin}}, \citenamefont {{Fishbach}},
  \citenamefont {{Mar{\'\i}a Ezquiaga}}, \citenamefont {{Farr}},\ and\
  \citenamefont {{Holz}}}]{Farah:2023a}%
  \BibitemOpen
  \bibfield  {author} {\bibinfo {author} {\bibfnamefont {A.~M.}\ \bibnamefont
  {{Farah}}}, \bibinfo {author} {\bibfnamefont {B.}~\bibnamefont {{Edelman}}},
  \bibinfo {author} {\bibfnamefont {M.}~\bibnamefont {{Zevin}}}, \bibinfo
  {author} {\bibfnamefont {M.}~\bibnamefont {{Fishbach}}}, \bibinfo {author}
  {\bibfnamefont {J.}~\bibnamefont {{Mar{\'\i}a Ezquiaga}}}, \bibinfo {author}
  {\bibfnamefont {B.}~\bibnamefont {{Farr}}},\ and\ \bibinfo {author}
  {\bibfnamefont {D.~E.}\ \bibnamefont {{Holz}}},\ }\bibfield  {title}
  {\bibinfo {title} {{Things That Might Go Bump in the Night: Assessing
  Structure in the Binary Black Hole Mass Spectrum}},\ }\href
  {https://doi.org/10.3847/1538-4357/aced02} {\bibfield  {journal} {\bibinfo
  {journal} {\apj}\ }\textbf {\bibinfo {volume} {955}},\ \bibinfo {eid} {107}
  (\bibinfo {year} {2023}{\natexlab{b}})},\ \Eprint
  {https://arxiv.org/abs/2301.00834} {arXiv:2301.00834 [astro-ph.HE]}
  \BibitemShut {NoStop}%
\bibitem [{\citenamefont {{Tiwari}}(2024)}]{Tiwari:2024}%
  \BibitemOpen
  \bibfield  {author} {\bibinfo {author} {\bibfnamefont {V.}~\bibnamefont
  {{Tiwari}}},\ }\bibfield  {title} {\bibinfo {title} {{What's in a binary
  black hole's mass parameter?}},\ }\href
  {https://doi.org/10.1093/mnras/stad3155} {\bibfield  {journal} {\bibinfo
  {journal} {\mnras}\ }\textbf {\bibinfo {volume} {527}},\ \bibinfo {pages}
  {298} (\bibinfo {year} {2024})},\ \Eprint {https://arxiv.org/abs/2304.03498}
  {arXiv:2304.03498 [astro-ph.HE]} \BibitemShut {NoStop}%
\bibitem [{Note1()}]{Note1}%
  \BibitemOpen
  \bibinfo {note} {\protect \url
  {https://zenodo.org/records/6513631}}\BibitemShut {NoStop}%
\bibitem [{\citenamefont {{The LIGO Scientific Collaboration}}\ \emph
  {et~al.}(2021{\natexlab{d}})\citenamefont {{The LIGO Scientific
  Collaboration}}, \citenamefont {{the Virgo Collaboration}}, \citenamefont
  {{Abbott}}, \citenamefont {{Abbott}} \emph {et~al.}}]{gwtc2_1:2021}%
  \BibitemOpen
  \bibfield  {author} {\bibinfo {author} {\bibnamefont {{The LIGO Scientific
  Collaboration}}}, \bibinfo {author} {\bibnamefont {{the Virgo
  Collaboration}}}, \bibinfo {author} {\bibfnamefont {R.}~\bibnamefont
  {{Abbott}}}, \bibinfo {author} {\bibfnamefont {T.~D.}\ \bibnamefont
  {{Abbott}}}, \emph {et~al.},\ }\bibfield  {title} {\bibinfo {title}
  {{GWTC-2.1: Deep Extended Catalog of Compact Binary Coalescences Observed by
  LIGO and Virgo During the First Half of the Third Observing Run}},\ }\href
  {https://doi.org/10.48550/arXiv.2108.01045} {\bibfield  {journal} {\bibinfo
  {journal} {arXiv e-prints}\ ,\ \bibinfo {eid} {arXiv:2108.01045}} (\bibinfo
  {year} {2021}{\natexlab{d}})},\ \Eprint {https://arxiv.org/abs/2108.01045}
  {arXiv:2108.01045 [gr-qc]} \BibitemShut {NoStop}%
\bibitem [{\citenamefont {{Pratten}}\ \emph {et~al.}(2021)\citenamefont
  {{Pratten}}, \citenamefont {{Garc{\'\i}a-Quir{\'o}s}}, \citenamefont
  {{Colleoni}}, \citenamefont {{Ramos-Buades}}, \citenamefont {{Estell{\'e}s}},
  \citenamefont {{Mateu-Lucena}}, \citenamefont {{Jaume}}, \citenamefont
  {{Haney}}, \citenamefont {{Keitel}}, \citenamefont {{Thompson}},\ and\
  \citenamefont {{Husa}}}]{Pratten:2021}%
  \BibitemOpen
  \bibfield  {author} {\bibinfo {author} {\bibfnamefont {G.}~\bibnamefont
  {{Pratten}}}, \bibinfo {author} {\bibfnamefont {C.}~\bibnamefont
  {{Garc{\'\i}a-Quir{\'o}s}}}, \bibinfo {author} {\bibfnamefont
  {M.}~\bibnamefont {{Colleoni}}}, \bibinfo {author} {\bibfnamefont
  {A.}~\bibnamefont {{Ramos-Buades}}}, \bibinfo {author} {\bibfnamefont
  {H.}~\bibnamefont {{Estell{\'e}s}}}, \bibinfo {author} {\bibfnamefont
  {M.}~\bibnamefont {{Mateu-Lucena}}}, \bibinfo {author} {\bibfnamefont
  {R.}~\bibnamefont {{Jaume}}}, \bibinfo {author} {\bibfnamefont
  {M.}~\bibnamefont {{Haney}}}, \bibinfo {author} {\bibfnamefont
  {D.}~\bibnamefont {{Keitel}}}, \bibinfo {author} {\bibfnamefont {J.~E.}\
  \bibnamefont {{Thompson}}},\ and\ \bibinfo {author} {\bibfnamefont
  {S.}~\bibnamefont {{Husa}}},\ }\bibfield  {title} {\bibinfo {title}
  {{Computationally efficient models for the dominant and subdominant harmonic
  modes of precessing binary black holes}},\ }\href
  {https://doi.org/10.1103/PhysRevD.103.104056} {\bibfield  {journal} {\bibinfo
   {journal} {\prd}\ }\textbf {\bibinfo {volume} {103}},\ \bibinfo {eid}
  {104056} (\bibinfo {year} {2021})},\ \Eprint
  {https://arxiv.org/abs/2004.06503} {arXiv:2004.06503 [gr-qc]} \BibitemShut
  {NoStop}%
\bibitem [{\citenamefont {{Ossokine}}\ \emph {et~al.}(2020)\citenamefont
  {{Ossokine}}, \citenamefont {{Buonanno}}, \citenamefont {{Marsat}},
  \citenamefont {{Cotesta}}, \citenamefont {{Babak}}, \citenamefont
  {{Dietrich}}, \citenamefont {{Haas}}, \citenamefont {{Hinder}}, \citenamefont
  {{Pfeiffer}}, \citenamefont {{P{\"u}rrer}}, \citenamefont {{Woodford}},
  \citenamefont {{Boyle}}, \citenamefont {{Kidder}}, \citenamefont {{Scheel}},\
  and\ \citenamefont {{Szil{\'a}gyi}}}]{Ossokine:2020}%
  \BibitemOpen
  \bibfield  {author} {\bibinfo {author} {\bibfnamefont {S.}~\bibnamefont
  {{Ossokine}}}, \bibinfo {author} {\bibfnamefont {A.}~\bibnamefont
  {{Buonanno}}}, \bibinfo {author} {\bibfnamefont {S.}~\bibnamefont
  {{Marsat}}}, \bibinfo {author} {\bibfnamefont {R.}~\bibnamefont {{Cotesta}}},
  \bibinfo {author} {\bibfnamefont {S.}~\bibnamefont {{Babak}}}, \bibinfo
  {author} {\bibfnamefont {T.}~\bibnamefont {{Dietrich}}}, \bibinfo {author}
  {\bibfnamefont {R.}~\bibnamefont {{Haas}}}, \bibinfo {author} {\bibfnamefont
  {I.}~\bibnamefont {{Hinder}}}, \bibinfo {author} {\bibfnamefont {H.~P.}\
  \bibnamefont {{Pfeiffer}}}, \bibinfo {author} {\bibfnamefont
  {M.}~\bibnamefont {{P{\"u}rrer}}}, \bibinfo {author} {\bibfnamefont {C.~J.}\
  \bibnamefont {{Woodford}}}, \bibinfo {author} {\bibfnamefont
  {M.}~\bibnamefont {{Boyle}}}, \bibinfo {author} {\bibfnamefont {L.~E.}\
  \bibnamefont {{Kidder}}}, \bibinfo {author} {\bibfnamefont {M.~A.}\
  \bibnamefont {{Scheel}}},\ and\ \bibinfo {author} {\bibfnamefont
  {B.}~\bibnamefont {{Szil{\'a}gyi}}},\ }\bibfield  {title} {\bibinfo {title}
  {{Multipolar effective-one-body waveforms for precessing binary black holes:
  Construction and validation}},\ }\href
  {https://doi.org/10.1103/PhysRevD.102.044055} {\bibfield  {journal} {\bibinfo
   {journal} {\prd}\ }\textbf {\bibinfo {volume} {102}},\ \bibinfo {eid}
  {044055} (\bibinfo {year} {2020})},\ \Eprint
  {https://arxiv.org/abs/2004.09442} {arXiv:2004.09442 [gr-qc]} \BibitemShut
  {NoStop}%
\bibitem [{\citenamefont {Loredo}(2004)}]{Loredo:2004}%
  \BibitemOpen
  \bibfield  {author} {\bibinfo {author} {\bibfnamefont {T.~J.}\ \bibnamefont
  {Loredo}},\ }\bibfield  {title} {\bibinfo {title} {Accounting for source
  uncertainties in analyses of astronomical survey data},\ }\href
  {https://doi.org/10.1063/1.1835214} {\bibfield  {journal} {\bibinfo
  {journal} {AIP Conference Proceedings}\ }\textbf {\bibinfo {volume} {735}},\
  \bibinfo {pages} {195} (\bibinfo {year} {2004})},\ \Eprint
  {https://arxiv.org/abs/https://aip.scitation.org/doi/pdf/10.1063/1.1835214}
  {https://aip.scitation.org/doi/pdf/10.1063/1.1835214} \BibitemShut {NoStop}%
\bibitem [{\citenamefont {Vitale}\ \emph {et~al.}(2020)\citenamefont {Vitale},
  \citenamefont {Gerosa}, \citenamefont {Farr},\ and\ \citenamefont
  {Taylor}}]{Vitale:2020aaz}%
  \BibitemOpen
  \bibfield  {author} {\bibinfo {author} {\bibfnamefont {S.}~\bibnamefont
  {Vitale}}, \bibinfo {author} {\bibfnamefont {D.}~\bibnamefont {Gerosa}},
  \bibinfo {author} {\bibfnamefont {W.~M.}\ \bibnamefont {Farr}},\ and\
  \bibinfo {author} {\bibfnamefont {S.~R.}\ \bibnamefont {Taylor}},\ }\href
  {https://doi.org/10.1007/978-981-15-4702-7_45-1} {\emph {\bibinfo {title}
  {{Inferring the properties of a population of compact binaries in presence of
  selection effects}}}}\ (\bibinfo {year} {2020})\ \Eprint
  {https://arxiv.org/abs/2007.05579} {arXiv:2007.05579 [astro-ph.IM]}
  \BibitemShut {NoStop}%
\bibitem [{\citenamefont {Taylor}\ and\ \citenamefont
  {Gerosa}(2018)}]{Taylor:2018iat}%
  \BibitemOpen
  \bibfield  {author} {\bibinfo {author} {\bibfnamefont {S.~R.}\ \bibnamefont
  {Taylor}}\ and\ \bibinfo {author} {\bibfnamefont {D.}~\bibnamefont
  {Gerosa}},\ }\bibfield  {title} {\bibinfo {title} {{Mining Gravitational-wave
  Catalogs To Understand Binary Stellar Evolution: A New Hierarchical Bayesian
  Framework}},\ }\href {https://doi.org/10.1103/PhysRevD.98.083017} {\bibfield
  {journal} {\bibinfo  {journal} {Phys. Rev. D}\ }\textbf {\bibinfo {volume}
  {98}},\ \bibinfo {pages} {083017} (\bibinfo {year} {2018})},\ \Eprint
  {https://arxiv.org/abs/1806.08365} {arXiv:1806.08365 [astro-ph.HE]}
  \BibitemShut {NoStop}%
\bibitem [{\citenamefont {Mandel}\ \emph {et~al.}(2019)\citenamefont {Mandel},
  \citenamefont {Farr},\ and\ \citenamefont {Gair}}]{Mandel:2018mve}%
  \BibitemOpen
  \bibfield  {author} {\bibinfo {author} {\bibfnamefont {I.}~\bibnamefont
  {Mandel}}, \bibinfo {author} {\bibfnamefont {W.~M.}\ \bibnamefont {Farr}},\
  and\ \bibinfo {author} {\bibfnamefont {J.~R.}\ \bibnamefont {Gair}},\
  }\bibfield  {title} {\bibinfo {title} {{Extracting distribution parameters
  from multiple uncertain observations with selection biases}},\ }\href
  {https://doi.org/10.1093/mnras/stz896} {\bibfield  {journal} {\bibinfo
  {journal} {Mon. Not. Roy. Astron. Soc.}\ }\textbf {\bibinfo {volume} {486}},\
  \bibinfo {pages} {1086} (\bibinfo {year} {2019})},\ \Eprint
  {https://arxiv.org/abs/1809.02063} {arXiv:1809.02063 [physics.data-an]}
  \BibitemShut {NoStop}%
\bibitem [{\citenamefont {Fishbach}\ \emph {et~al.}(2018)\citenamefont
  {Fishbach}, \citenamefont {Holz},\ and\ \citenamefont
  {Farr}}]{Fishbach:2018edt}%
  \BibitemOpen
  \bibfield  {author} {\bibinfo {author} {\bibfnamefont {M.}~\bibnamefont
  {Fishbach}}, \bibinfo {author} {\bibfnamefont {D.~E.}\ \bibnamefont {Holz}},\
  and\ \bibinfo {author} {\bibfnamefont {W.~M.}\ \bibnamefont {Farr}},\
  }\bibfield  {title} {\bibinfo {title} {{Does the Black Hole Merger Rate
  Evolve with Redshift?}},\ }\href {https://doi.org/10.3847/2041-8213/aad800}
  {\bibfield  {journal} {\bibinfo  {journal} {Astrophys. J. Lett.}\ }\textbf
  {\bibinfo {volume} {863}},\ \bibinfo {pages} {L41} (\bibinfo {year}
  {2018})},\ \Eprint {https://arxiv.org/abs/1805.10270} {arXiv:1805.10270
  [astro-ph.HE]} \BibitemShut {NoStop}%
\bibitem [{\citenamefont {{Ellis Perkins}}\ \emph {et~al.}(2023)\citenamefont
  {{Ellis Perkins}}, \citenamefont {{McGill}}, \citenamefont {{Dawson}},
  \citenamefont {{Abrams}}, \citenamefont {{Lam}}, \citenamefont {{Ho}},
  \citenamefont {{Lu}}, \citenamefont {{Bird}}, \citenamefont {{Pruett}},
  \citenamefont {{Golovich}},\ and\ \citenamefont {{Chapline}}}]{Perkins:2023}%
  \BibitemOpen
  \bibfield  {author} {\bibinfo {author} {\bibfnamefont {S.}~\bibnamefont
  {{Ellis Perkins}}}, \bibinfo {author} {\bibfnamefont {P.}~\bibnamefont
  {{McGill}}}, \bibinfo {author} {\bibfnamefont {W.}~\bibnamefont {{Dawson}}},
  \bibinfo {author} {\bibfnamefont {N.~S.}\ \bibnamefont {{Abrams}}}, \bibinfo
  {author} {\bibfnamefont {C.~Y.}\ \bibnamefont {{Lam}}}, \bibinfo {author}
  {\bibfnamefont {M.-F.}\ \bibnamefont {{Ho}}}, \bibinfo {author}
  {\bibfnamefont {J.~R.}\ \bibnamefont {{Lu}}}, \bibinfo {author}
  {\bibfnamefont {S.}~\bibnamefont {{Bird}}}, \bibinfo {author} {\bibfnamefont
  {K.}~\bibnamefont {{Pruett}}}, \bibinfo {author} {\bibfnamefont
  {N.}~\bibnamefont {{Golovich}}},\ and\ \bibinfo {author} {\bibfnamefont
  {G.}~\bibnamefont {{Chapline}}},\ }\bibfield  {title} {\bibinfo {title}
  {{Disentangling the Black Hole Mass Spectrum with Photometric Microlensing
  Surveys}},\ }\href {https://doi.org/10.48550/arXiv.2310.03943} {\bibfield
  {journal} {\bibinfo  {journal} {arXiv e-prints}\ ,\ \bibinfo {eid}
  {arXiv:2310.03943}} (\bibinfo {year} {2023})},\ \Eprint
  {https://arxiv.org/abs/2310.03943} {arXiv:2310.03943 [astro-ph.IM]}
  \BibitemShut {NoStop}%
\bibitem [{\citenamefont {{Perkins}}\ \emph {et~al.}(2021)\citenamefont
  {{Perkins}}, \citenamefont {{Yunes}},\ and\ \citenamefont
  {{Berti}}}]{Perkins:2021}%
  \BibitemOpen
  \bibfield  {author} {\bibinfo {author} {\bibfnamefont {S.~E.}\ \bibnamefont
  {{Perkins}}}, \bibinfo {author} {\bibfnamefont {N.}~\bibnamefont {{Yunes}}},\
  and\ \bibinfo {author} {\bibfnamefont {E.}~\bibnamefont {{Berti}}},\
  }\bibfield  {title} {\bibinfo {title} {{Probing fundamental physics with
  gravitational waves: The next generation}},\ }\href
  {https://doi.org/10.1103/PhysRevD.103.044024} {\bibfield  {journal} {\bibinfo
   {journal} {\prd}\ }\textbf {\bibinfo {volume} {103}},\ \bibinfo {eid}
  {044024} (\bibinfo {year} {2021})},\ \Eprint
  {https://arxiv.org/abs/2010.09010} {arXiv:2010.09010 [gr-qc]} \BibitemShut
  {NoStop}%
\bibitem [{\citenamefont {Finn}(1996)}]{Finn:1995ah}%
  \BibitemOpen
  \bibfield  {author} {\bibinfo {author} {\bibfnamefont {L.~S.}\ \bibnamefont
  {Finn}},\ }\bibfield  {title} {\bibinfo {title} {{Binary inspiral,
  gravitational radiation, and cosmology}},\ }\href
  {https://doi.org/10.1103/PhysRevD.53.2878} {\bibfield  {journal} {\bibinfo
  {journal} {Phys. Rev. D}\ }\textbf {\bibinfo {volume} {53}},\ \bibinfo
  {pages} {2878} (\bibinfo {year} {1996})},\ \Eprint
  {https://arxiv.org/abs/gr-qc/9601048} {arXiv:gr-qc/9601048} \BibitemShut
  {NoStop}%
\bibitem [{\citenamefont {Finn}\ and\ \citenamefont
  {Chernoff}(1993)}]{Finn:1992xs}%
  \BibitemOpen
  \bibfield  {author} {\bibinfo {author} {\bibfnamefont {L.~S.}\ \bibnamefont
  {Finn}}\ and\ \bibinfo {author} {\bibfnamefont {D.~F.}\ \bibnamefont
  {Chernoff}},\ }\bibfield  {title} {\bibinfo {title} {{Observing binary
  inspiral in gravitational radiation: One interferometer}},\ }\href
  {https://doi.org/10.1103/PhysRevD.47.2198} {\bibfield  {journal} {\bibinfo
  {journal} {Phys. Rev. D}\ }\textbf {\bibinfo {volume} {47}},\ \bibinfo
  {pages} {2198} (\bibinfo {year} {1993})},\ \Eprint
  {https://arxiv.org/abs/gr-qc/9301003} {arXiv:gr-qc/9301003} \BibitemShut
  {NoStop}%
\bibitem [{\citenamefont {Dominik}\ \emph {et~al.}(2015)\citenamefont
  {Dominik}, \citenamefont {Berti}, \citenamefont {O'Shaughnessy},
  \citenamefont {Mandel}, \citenamefont {Belczynski}, \citenamefont {Fryer},
  \citenamefont {Holz}, \citenamefont {Bulik},\ and\ \citenamefont
  {Pannarale}}]{Dominik:2014yma}%
  \BibitemOpen
  \bibfield  {author} {\bibinfo {author} {\bibfnamefont {M.}~\bibnamefont
  {Dominik}}, \bibinfo {author} {\bibfnamefont {E.}~\bibnamefont {Berti}},
  \bibinfo {author} {\bibfnamefont {R.}~\bibnamefont {O'Shaughnessy}}, \bibinfo
  {author} {\bibfnamefont {I.}~\bibnamefont {Mandel}}, \bibinfo {author}
  {\bibfnamefont {K.}~\bibnamefont {Belczynski}}, \bibinfo {author}
  {\bibfnamefont {C.}~\bibnamefont {Fryer}}, \bibinfo {author} {\bibfnamefont
  {D.~E.}\ \bibnamefont {Holz}}, \bibinfo {author} {\bibfnamefont
  {T.}~\bibnamefont {Bulik}},\ and\ \bibinfo {author} {\bibfnamefont
  {F.}~\bibnamefont {Pannarale}},\ }\bibfield  {title} {\bibinfo {title}
  {{Double Compact Objects III: Gravitational Wave Detection Rates}},\ }\href
  {https://doi.org/10.1088/0004-637X/806/2/263} {\bibfield  {journal} {\bibinfo
   {journal} {Astrophys. J.}\ }\textbf {\bibinfo {volume} {806}},\ \bibinfo
  {pages} {263} (\bibinfo {year} {2015})},\ \Eprint
  {https://arxiv.org/abs/1405.7016} {arXiv:1405.7016 [astro-ph.HE]}
  \BibitemShut {NoStop}%
\bibitem [{\citenamefont {{Tiwari}}\ \emph {et~al.}(2018)\citenamefont
  {{Tiwari}}, \citenamefont {{Fairhurst}},\ and\ \citenamefont
  {{Hannam}}}]{Tiwari:2018}%
  \BibitemOpen
  \bibfield  {author} {\bibinfo {author} {\bibfnamefont {V.}~\bibnamefont
  {{Tiwari}}}, \bibinfo {author} {\bibfnamefont {S.}~\bibnamefont
  {{Fairhurst}}},\ and\ \bibinfo {author} {\bibfnamefont {M.}~\bibnamefont
  {{Hannam}}},\ }\bibfield  {title} {\bibinfo {title} {{Constraining Black Hole
  Spins with Gravitational-wave Observations}},\ }\href
  {https://doi.org/10.3847/1538-4357/aae8df} {\bibfield  {journal} {\bibinfo
  {journal} {\apj}\ }\textbf {\bibinfo {volume} {868}},\ \bibinfo {eid} {140}
  (\bibinfo {year} {2018})},\ \Eprint {https://arxiv.org/abs/1809.01401}
  {arXiv:1809.01401 [gr-qc]} \BibitemShut {NoStop}%
\bibitem [{\citenamefont {Campanelli}\ \emph {et~al.}(2006)\citenamefont
  {Campanelli}, \citenamefont {Lousto},\ and\ \citenamefont
  {Zlochower}}]{PhysRevD.74.041501}%
  \BibitemOpen
  \bibfield  {author} {\bibinfo {author} {\bibfnamefont {M.}~\bibnamefont
  {Campanelli}}, \bibinfo {author} {\bibfnamefont {C.~O.}\ \bibnamefont
  {Lousto}},\ and\ \bibinfo {author} {\bibfnamefont {Y.}~\bibnamefont
  {Zlochower}},\ }\bibfield  {title} {\bibinfo {title} {Spinning-black-hole
  binaries: The orbital hang-up},\ }\href
  {https://doi.org/10.1103/PhysRevD.74.041501} {\bibfield  {journal} {\bibinfo
  {journal} {Phys. Rev. D}\ }\textbf {\bibinfo {volume} {74}},\ \bibinfo
  {pages} {041501} (\bibinfo {year} {2006})}\BibitemShut {NoStop}%
\bibitem [{\citenamefont {{LIGO Scientific Collaboration}}\ \emph
  {et~al.}(2018)\citenamefont {{LIGO Scientific Collaboration}}, \citenamefont
  {{Virgo Collaboration}},\ and\ \citenamefont {{KAGRA
  Collaboration}}}]{lalsuite:2018}%
  \BibitemOpen
  \bibfield  {author} {\bibinfo {author} {\bibnamefont {{LIGO Scientific
  Collaboration}}}, \bibinfo {author} {\bibnamefont {{Virgo Collaboration}}},\
  and\ \bibinfo {author} {\bibnamefont {{KAGRA Collaboration}}},\ }\href
  {https://doi.org/10.7935/GT1W-FZ16} {\bibinfo {title} {{LVK} {A}lgorithm
  {L}ibrary - {LALS}uite}},\ \bibinfo {howpublished} {Free software (GPL)}
  (\bibinfo {year} {2018})\BibitemShut {NoStop}%
\bibitem [{\citenamefont {{Biwer}}\ \emph {et~al.}(2019)\citenamefont
  {{Biwer}}, \citenamefont {{Capano}}, \citenamefont {{De}}, \citenamefont
  {{Cabero}}, \citenamefont {{Brown}}, \citenamefont {{Nitz}},\ and\
  \citenamefont {{Raymond}}}]{Biwer:2019}%
  \BibitemOpen
  \bibfield  {author} {\bibinfo {author} {\bibfnamefont {C.~M.}\ \bibnamefont
  {{Biwer}}}, \bibinfo {author} {\bibfnamefont {C.~D.}\ \bibnamefont
  {{Capano}}}, \bibinfo {author} {\bibfnamefont {S.}~\bibnamefont {{De}}},
  \bibinfo {author} {\bibfnamefont {M.}~\bibnamefont {{Cabero}}}, \bibinfo
  {author} {\bibfnamefont {D.~A.}\ \bibnamefont {{Brown}}}, \bibinfo {author}
  {\bibfnamefont {A.~H.}\ \bibnamefont {{Nitz}}},\ and\ \bibinfo {author}
  {\bibfnamefont {V.}~\bibnamefont {{Raymond}}},\ }\bibfield  {title} {\bibinfo
  {title} {{PyCBC Inference: A Python-based Parameter Estimation Toolkit for
  Compact Binary Coalescence Signal}},\ }\href
  {https://doi.org/10.1088/1538-3873/aaef0b} {\bibfield  {journal} {\bibinfo
  {journal} {\pasp}\ }\textbf {\bibinfo {volume} {131}},\ \bibinfo {pages}
  {024503} (\bibinfo {year} {2019})},\ \Eprint
  {https://arxiv.org/abs/1807.10312} {arXiv:1807.10312 [astro-ph.IM]}
  \BibitemShut {NoStop}%
\bibitem [{\citenamefont {{Finn}}\ and\ \citenamefont
  {{Chernoff}}(1993)}]{Finn:1993}%
  \BibitemOpen
  \bibfield  {author} {\bibinfo {author} {\bibfnamefont {L.~S.}\ \bibnamefont
  {{Finn}}}\ and\ \bibinfo {author} {\bibfnamefont {D.~F.}\ \bibnamefont
  {{Chernoff}}},\ }\bibfield  {title} {\bibinfo {title} {{Observing binary
  inspiral in gravitational radiation: One interferometer}},\ }\href
  {https://doi.org/10.1103/PhysRevD.47.2198} {\bibfield  {journal} {\bibinfo
  {journal} {\prd}\ }\textbf {\bibinfo {volume} {47}},\ \bibinfo {pages} {2198}
  (\bibinfo {year} {1993})},\ \Eprint {https://arxiv.org/abs/gr-qc/9301003}
  {arXiv:gr-qc/9301003 [gr-qc]} \BibitemShut {NoStop}%
\bibitem [{\citenamefont {{Finn}}(1996)}]{Finn:1996}%
  \BibitemOpen
  \bibfield  {author} {\bibinfo {author} {\bibfnamefont {L.~S.}\ \bibnamefont
  {{Finn}}},\ }\bibfield  {title} {\bibinfo {title} {{Binary inspiral,
  gravitational radiation, and cosmology}},\ }\href
  {https://doi.org/10.1103/PhysRevD.53.2878} {\bibfield  {journal} {\bibinfo
  {journal} {\prd}\ }\textbf {\bibinfo {volume} {53}},\ \bibinfo {pages} {2878}
  (\bibinfo {year} {1996})},\ \Eprint {https://arxiv.org/abs/gr-qc/9601048}
  {arXiv:gr-qc/9601048 [gr-qc]} \BibitemShut {NoStop}%
\bibitem [{\citenamefont {{Khan}}\ \emph {et~al.}(2016)\citenamefont {{Khan}},
  \citenamefont {{Husa}}, \citenamefont {{Hannam}}, \citenamefont {{Ohme}},
  \citenamefont {{P{\"u}rrer}}, \citenamefont {{Forteza}},\ and\ \citenamefont
  {{Boh{\'e}}}}]{Khan:2016}%
  \BibitemOpen
  \bibfield  {author} {\bibinfo {author} {\bibfnamefont {S.}~\bibnamefont
  {{Khan}}}, \bibinfo {author} {\bibfnamefont {S.}~\bibnamefont {{Husa}}},
  \bibinfo {author} {\bibfnamefont {M.}~\bibnamefont {{Hannam}}}, \bibinfo
  {author} {\bibfnamefont {F.}~\bibnamefont {{Ohme}}}, \bibinfo {author}
  {\bibfnamefont {M.}~\bibnamefont {{P{\"u}rrer}}}, \bibinfo {author}
  {\bibfnamefont {X.~J.}\ \bibnamefont {{Forteza}}},\ and\ \bibinfo {author}
  {\bibfnamefont {A.}~\bibnamefont {{Boh{\'e}}}},\ }\bibfield  {title}
  {\bibinfo {title} {{Frequency-domain gravitational waves from nonprecessing
  black-hole binaries. II. A phenomenological model for the advanced detector
  era}},\ }\href {https://doi.org/10.1103/PhysRevD.93.044007} {\bibfield
  {journal} {\bibinfo  {journal} {\prd}\ }\textbf {\bibinfo {volume} {93}},\
  \bibinfo {eid} {044007} (\bibinfo {year} {2016})},\ \Eprint
  {https://arxiv.org/abs/1508.07253} {arXiv:1508.07253 [gr-qc]} \BibitemShut
  {NoStop}%
\bibitem [{\citenamefont {{Husa}}\ \emph {et~al.}(2016)\citenamefont {{Husa}},
  \citenamefont {{Khan}}, \citenamefont {{Hannam}}, \citenamefont
  {{P{\"u}rrer}}, \citenamefont {{Ohme}}, \citenamefont {{Forteza}},\ and\
  \citenamefont {{Boh{\'e}}}}]{Husa:2016}%
  \BibitemOpen
  \bibfield  {author} {\bibinfo {author} {\bibfnamefont {S.}~\bibnamefont
  {{Husa}}}, \bibinfo {author} {\bibfnamefont {S.}~\bibnamefont {{Khan}}},
  \bibinfo {author} {\bibfnamefont {M.}~\bibnamefont {{Hannam}}}, \bibinfo
  {author} {\bibfnamefont {M.}~\bibnamefont {{P{\"u}rrer}}}, \bibinfo {author}
  {\bibfnamefont {F.}~\bibnamefont {{Ohme}}}, \bibinfo {author} {\bibfnamefont
  {X.~J.}\ \bibnamefont {{Forteza}}},\ and\ \bibinfo {author} {\bibfnamefont
  {A.}~\bibnamefont {{Boh{\'e}}}},\ }\bibfield  {title} {\bibinfo {title}
  {{Frequency-domain gravitational waves from nonprecessing black-hole
  binaries. I. New numerical waveforms and anatomy of the signal}},\ }\href
  {https://doi.org/10.1103/PhysRevD.93.044006} {\bibfield  {journal} {\bibinfo
  {journal} {\prd}\ }\textbf {\bibinfo {volume} {93}},\ \bibinfo {eid} {044006}
  (\bibinfo {year} {2016})},\ \Eprint {https://arxiv.org/abs/1508.07250}
  {arXiv:1508.07250 [gr-qc]} \BibitemShut {NoStop}%
\bibitem [{\citenamefont {{P{\"u}rrer}}\ \emph {et~al.}(2023)\citenamefont
  {{P{\"u}rrer}}, \citenamefont {{Khan}}, \citenamefont {{Ohme}}, \citenamefont
  {{Birnholtz}},\ and\ \citenamefont {{London}}}]{Purrer:2023}%
  \BibitemOpen
  \bibfield  {author} {\bibinfo {author} {\bibfnamefont {M.}~\bibnamefont
  {{P{\"u}rrer}}}, \bibinfo {author} {\bibfnamefont {S.}~\bibnamefont
  {{Khan}}}, \bibinfo {author} {\bibfnamefont {F.}~\bibnamefont {{Ohme}}},
  \bibinfo {author} {\bibfnamefont {O.}~\bibnamefont {{Birnholtz}}},\ and\
  \bibinfo {author} {\bibfnamefont {L.}~\bibnamefont {{London}}},\ }\href@noop
  {} {\bibinfo {title} {{IMRPhenomD: Phenomenological waveform model}}},\
  \bibinfo {howpublished} {Astrophysics Source Code Library, record
  ascl:2307.019} (\bibinfo {year} {2023})\BibitemShut {NoStop}%
\bibitem [{\citenamefont {Gerosa}\ \emph {et~al.}(2019)\citenamefont {Gerosa},
  \citenamefont {Ma}, \citenamefont {Wong}, \citenamefont {Berti},
  \citenamefont {O'Shaughnessy}, \citenamefont {Chen},\ and\ \citenamefont
  {Belczynski}}]{Gerosa:2019dbe}%
  \BibitemOpen
  \bibfield  {author} {\bibinfo {author} {\bibfnamefont {D.}~\bibnamefont
  {Gerosa}}, \bibinfo {author} {\bibfnamefont {S.}~\bibnamefont {Ma}}, \bibinfo
  {author} {\bibfnamefont {K.~W.~K.}\ \bibnamefont {Wong}}, \bibinfo {author}
  {\bibfnamefont {E.}~\bibnamefont {Berti}}, \bibinfo {author} {\bibfnamefont
  {R.}~\bibnamefont {O'Shaughnessy}}, \bibinfo {author} {\bibfnamefont
  {Y.}~\bibnamefont {Chen}},\ and\ \bibinfo {author} {\bibfnamefont
  {K.}~\bibnamefont {Belczynski}},\ }\bibfield  {title} {\bibinfo {title}
  {{Multiband gravitational-wave event rates and stellar physics}},\ }\href
  {https://doi.org/10.1103/PhysRevD.99.103004} {\bibfield  {journal} {\bibinfo
  {journal} {Phys. Rev. D}\ }\textbf {\bibinfo {volume} {99}},\ \bibinfo
  {pages} {103004} (\bibinfo {year} {2019})},\ \Eprint
  {https://arxiv.org/abs/1902.00021} {arXiv:1902.00021 [astro-ph.HE]}
  \BibitemShut {NoStop}%
\bibitem [{\citenamefont {Abbott}\ \emph
  {et~al.}(2016{\natexlab{a}})\citenamefont {Abbott} \emph
  {et~al.}}]{LIGOScientific:2016ebi}%
  \BibitemOpen
  \bibfield  {author} {\bibinfo {author} {\bibfnamefont {B.~P.}\ \bibnamefont
  {Abbott}} \emph {et~al.} (\bibinfo {collaboration} {LIGO Scientific,
  Virgo}),\ }\bibfield  {title} {\bibinfo {title} {{Supplement: The Rate of
  Binary Black Hole Mergers Inferred from Advanced LIGO Observations
  Surrounding GW150914}},\ }\href {https://doi.org/10.3847/0067-0049/227/2/14}
  {\bibfield  {journal} {\bibinfo  {journal} {Astrophys. J. Suppl.}\ }\textbf
  {\bibinfo {volume} {227}},\ \bibinfo {pages} {14} (\bibinfo {year}
  {2016}{\natexlab{a}})},\ \Eprint {https://arxiv.org/abs/1606.03939}
  {arXiv:1606.03939 [astro-ph.HE]} \BibitemShut {NoStop}%
\bibitem [{\citenamefont {Abbott}\ \emph
  {et~al.}(2016{\natexlab{b}})\citenamefont {Abbott} \emph
  {et~al.}}]{LIGOScientific:2016kwr}%
  \BibitemOpen
  \bibfield  {author} {\bibinfo {author} {\bibfnamefont {B.~P.}\ \bibnamefont
  {Abbott}} \emph {et~al.} (\bibinfo {collaboration} {LIGO Scientific,
  Virgo}),\ }\bibfield  {title} {\bibinfo {title} {{The Rate of Binary Black
  Hole Mergers Inferred from Advanced LIGO Observations Surrounding
  GW150914}},\ }\href {https://doi.org/10.3847/2041-8205/833/1/L1} {\bibfield
  {journal} {\bibinfo  {journal} {Astrophys. J. Lett.}\ }\textbf {\bibinfo
  {volume} {833}},\ \bibinfo {pages} {L1} (\bibinfo {year}
  {2016}{\natexlab{b}})},\ \Eprint {https://arxiv.org/abs/1602.03842}
  {arXiv:1602.03842 [astro-ph.HE]} \BibitemShut {NoStop}%
\bibitem [{\citenamefont {{Hogg}}\ \emph {et~al.}(2010)\citenamefont {{Hogg}},
  \citenamefont {{Myers}},\ and\ \citenamefont {{Bovy}}}]{Hogg:2010}%
  \BibitemOpen
  \bibfield  {author} {\bibinfo {author} {\bibfnamefont {D.~W.}\ \bibnamefont
  {{Hogg}}}, \bibinfo {author} {\bibfnamefont {A.~D.}\ \bibnamefont
  {{Myers}}},\ and\ \bibinfo {author} {\bibfnamefont {J.}~\bibnamefont
  {{Bovy}}},\ }\bibfield  {title} {\bibinfo {title} {{Inferring the
  Eccentricity Distribution}},\ }\href
  {https://doi.org/10.1088/0004-637X/725/2/2166} {\bibfield  {journal}
  {\bibinfo  {journal} {\apj}\ }\textbf {\bibinfo {volume} {725}},\ \bibinfo
  {pages} {2166} (\bibinfo {year} {2010})},\ \Eprint
  {https://arxiv.org/abs/1008.4146} {arXiv:1008.4146 [astro-ph.SR]}
  \BibitemShut {NoStop}%
\bibitem [{\citenamefont {{Binney}}\ and\ \citenamefont
  {{Tremaine}}(2008)}]{Binney:2008}%
  \BibitemOpen
  \bibfield  {author} {\bibinfo {author} {\bibfnamefont {J.}~\bibnamefont
  {{Binney}}}\ and\ \bibinfo {author} {\bibfnamefont {S.}~\bibnamefont
  {{Tremaine}}},\ }\href@noop {} {\emph {\bibinfo {title} {{Galactic Dynamics:
  Second Edition}}}}\ (\bibinfo {year} {2008})\BibitemShut {NoStop}%
\bibitem [{\citenamefont {{Belczynski}}\ \emph {et~al.}(2014)\citenamefont
  {{Belczynski}}, \citenamefont {{Buonanno}}, \citenamefont {{Cantiello}},
  \citenamefont {{Fryer}}, \citenamefont {{Holz}}, \citenamefont {{Mandel}},
  \citenamefont {{Miller}},\ and\ \citenamefont {{Walczak}}}]{Belczynski:2014}%
  \BibitemOpen
  \bibfield  {author} {\bibinfo {author} {\bibfnamefont {K.}~\bibnamefont
  {{Belczynski}}}, \bibinfo {author} {\bibfnamefont {A.}~\bibnamefont
  {{Buonanno}}}, \bibinfo {author} {\bibfnamefont {M.}~\bibnamefont
  {{Cantiello}}}, \bibinfo {author} {\bibfnamefont {C.~L.}\ \bibnamefont
  {{Fryer}}}, \bibinfo {author} {\bibfnamefont {D.~E.}\ \bibnamefont {{Holz}}},
  \bibinfo {author} {\bibfnamefont {I.}~\bibnamefont {{Mandel}}}, \bibinfo
  {author} {\bibfnamefont {M.~C.}\ \bibnamefont {{Miller}}},\ and\ \bibinfo
  {author} {\bibfnamefont {M.}~\bibnamefont {{Walczak}}},\ }\bibfield  {title}
  {\bibinfo {title} {{The Formation and Gravitational-wave Detection of Massive
  Stellar Black Hole Binaries}},\ }\href
  {https://doi.org/10.1088/0004-637X/789/2/120} {\bibfield  {journal} {\bibinfo
   {journal} {\apj}\ }\textbf {\bibinfo {volume} {789}},\ \bibinfo {eid} {120}
  (\bibinfo {year} {2014})},\ \Eprint {https://arxiv.org/abs/1403.0677}
  {arXiv:1403.0677 [astro-ph.HE]} \BibitemShut {NoStop}%
\bibitem [{\citenamefont {{Hopman}}\ and\ \citenamefont
  {{Alexander}}(2006)}]{Hopman:2006}%
  \BibitemOpen
  \bibfield  {author} {\bibinfo {author} {\bibfnamefont {C.}~\bibnamefont
  {{Hopman}}}\ and\ \bibinfo {author} {\bibfnamefont {T.}~\bibnamefont
  {{Alexander}}},\ }\bibfield  {title} {\bibinfo {title} {{The Effect of Mass
  Segregation on Gravitational Wave Sources near Massive Black Holes}},\ }\href
  {https://doi.org/10.1086/506273} {\bibfield  {journal} {\bibinfo  {journal}
  {\apjl}\ }\textbf {\bibinfo {volume} {645}},\ \bibinfo {pages} {L133}
  (\bibinfo {year} {2006})},\ \Eprint {https://arxiv.org/abs/astro-ph/0603324}
  {arXiv:astro-ph/0603324 [astro-ph]} \BibitemShut {NoStop}%
\bibitem [{\citenamefont {{Lacroix}}\ and\ \citenamefont
  {{Silk}}(2018)}]{Lacroix:2018}%
  \BibitemOpen
  \bibfield  {author} {\bibinfo {author} {\bibfnamefont {T.}~\bibnamefont
  {{Lacroix}}}\ and\ \bibinfo {author} {\bibfnamefont {J.}~\bibnamefont
  {{Silk}}},\ }\bibfield  {title} {\bibinfo {title} {{Intermediate-mass Black
  Holes and Dark Matter at the Galactic Center}},\ }\href
  {https://doi.org/10.3847/2041-8213/aaa775} {\bibfield  {journal} {\bibinfo
  {journal} {\apjl}\ }\textbf {\bibinfo {volume} {853}},\ \bibinfo {eid} {L16}
  (\bibinfo {year} {2018})},\ \Eprint {https://arxiv.org/abs/1712.00452}
  {arXiv:1712.00452 [astro-ph.GA]} \BibitemShut {NoStop}%
\bibitem [{\citenamefont {{Kinugawa}}\ \emph {et~al.}(2014)\citenamefont
  {{Kinugawa}}, \citenamefont {{Inayoshi}}, \citenamefont {{Hotokezaka}},
  \citenamefont {{Nakauchi}},\ and\ \citenamefont
  {{Nakamura}}}]{Kinugawa:2014}%
  \BibitemOpen
  \bibfield  {author} {\bibinfo {author} {\bibfnamefont {T.}~\bibnamefont
  {{Kinugawa}}}, \bibinfo {author} {\bibfnamefont {K.}~\bibnamefont
  {{Inayoshi}}}, \bibinfo {author} {\bibfnamefont {K.}~\bibnamefont
  {{Hotokezaka}}}, \bibinfo {author} {\bibfnamefont {D.}~\bibnamefont
  {{Nakauchi}}},\ and\ \bibinfo {author} {\bibfnamefont {T.}~\bibnamefont
  {{Nakamura}}},\ }\bibfield  {title} {\bibinfo {title} {{Possible indirect
  confirmation of the existence of Pop III massive stars by gravitational
  wave}},\ }\href {https://doi.org/10.1093/mnras/stu1022} {\bibfield  {journal}
  {\bibinfo  {journal} {\mnras}\ }\textbf {\bibinfo {volume} {442}},\ \bibinfo
  {pages} {2963} (\bibinfo {year} {2014})},\ \Eprint
  {https://arxiv.org/abs/1402.6672} {arXiv:1402.6672 [astro-ph.HE]}
  \BibitemShut {NoStop}%
\bibitem [{\citenamefont {{Alcock}}\ \emph {et~al.}(2001)\citenamefont
  {{Alcock}}, \citenamefont {{Allsman}}, \citenamefont {{Alves}}, \citenamefont
  {{Axelrod}}, \citenamefont {{Becker}}, \citenamefont {{Bennett}},
  \citenamefont {{Cook}}, \citenamefont {{Dalal}}, \citenamefont {{Drake}},
  \citenamefont {{Freeman}}, \citenamefont {{Geha}}, \citenamefont {{Griest}},
  \citenamefont {{Lehner}}, \citenamefont {{Marshall}}, \citenamefont
  {{Minniti}}, \citenamefont {{Nelson}}, \citenamefont {{Peterson}},
  \citenamefont {{Popowski}}, \citenamefont {{Pratt}}, \citenamefont {{Quinn}},
  \citenamefont {{Stubbs}}, \citenamefont {{Sutherland}}, \citenamefont
  {{Tomaney}}, \citenamefont {{Vandehei}},\ and\ \citenamefont
  {{Welch}}}]{Alcock:2001}%
  \BibitemOpen
  \bibfield  {author} {\bibinfo {author} {\bibfnamefont {C.}~\bibnamefont
  {{Alcock}}}, \bibinfo {author} {\bibfnamefont {R.~A.}\ \bibnamefont
  {{Allsman}}}, \bibinfo {author} {\bibfnamefont {D.~R.}\ \bibnamefont
  {{Alves}}}, \bibinfo {author} {\bibfnamefont {T.~S.}\ \bibnamefont
  {{Axelrod}}}, \bibinfo {author} {\bibfnamefont {A.~C.}\ \bibnamefont
  {{Becker}}}, \bibinfo {author} {\bibfnamefont {D.~P.}\ \bibnamefont
  {{Bennett}}}, \bibinfo {author} {\bibfnamefont {K.~H.}\ \bibnamefont
  {{Cook}}}, \bibinfo {author} {\bibfnamefont {N.}~\bibnamefont {{Dalal}}},
  \bibinfo {author} {\bibfnamefont {A.~J.}\ \bibnamefont {{Drake}}}, \bibinfo
  {author} {\bibfnamefont {K.~C.}\ \bibnamefont {{Freeman}}}, \bibinfo {author}
  {\bibfnamefont {M.}~\bibnamefont {{Geha}}}, \bibinfo {author} {\bibfnamefont
  {K.}~\bibnamefont {{Griest}}}, \bibinfo {author} {\bibfnamefont {M.~J.}\
  \bibnamefont {{Lehner}}}, \bibinfo {author} {\bibfnamefont {S.~L.}\
  \bibnamefont {{Marshall}}}, \bibinfo {author} {\bibfnamefont
  {D.}~\bibnamefont {{Minniti}}}, \bibinfo {author} {\bibfnamefont {C.~A.}\
  \bibnamefont {{Nelson}}}, \bibinfo {author} {\bibfnamefont {B.~A.}\
  \bibnamefont {{Peterson}}}, \bibinfo {author} {\bibfnamefont
  {P.}~\bibnamefont {{Popowski}}}, \bibinfo {author} {\bibfnamefont {M.~R.}\
  \bibnamefont {{Pratt}}}, \bibinfo {author} {\bibfnamefont {P.~J.}\
  \bibnamefont {{Quinn}}}, \bibinfo {author} {\bibfnamefont {C.~W.}\
  \bibnamefont {{Stubbs}}}, \bibinfo {author} {\bibfnamefont {W.}~\bibnamefont
  {{Sutherland}}}, \bibinfo {author} {\bibfnamefont {A.~B.}\ \bibnamefont
  {{Tomaney}}}, \bibinfo {author} {\bibfnamefont {T.}~\bibnamefont
  {{Vandehei}}},\ and\ \bibinfo {author} {\bibfnamefont {D.~L.}\ \bibnamefont
  {{Welch}}},\ }\bibfield  {title} {\bibinfo {title} {{MACHO Project Limits on
  Black Hole Dark Matter in the 1-30 M$_{solar}$ Range}},\ }\href
  {https://doi.org/10.1086/319636} {\bibfield  {journal} {\bibinfo  {journal}
  {\apjl}\ }\textbf {\bibinfo {volume} {550}},\ \bibinfo {pages} {L169}
  (\bibinfo {year} {2001})},\ \Eprint {https://arxiv.org/abs/astro-ph/0011506}
  {arXiv:astro-ph/0011506 [astro-ph]} \BibitemShut {NoStop}%
\bibitem [{\citenamefont {{Tisserand}}\ \emph {et~al.}(2007)\citenamefont
  {{Tisserand}}, \citenamefont {{Le Guillou}}, \citenamefont {{Afonso}},
  \citenamefont {{Albert}}, \citenamefont {{Andersen}}, \citenamefont
  {{Ansari}}, \citenamefont {{Aubourg}}, \citenamefont {{Bareyre}},
  \citenamefont {{Beaulieu}}, \citenamefont {{Charlot}}, \citenamefont
  {{Coutures}}, \citenamefont {{Ferlet}}, \citenamefont {{Fouqu{\'e}}},
  \citenamefont {{Glicenstein}}, \citenamefont {{Goldman}}, \citenamefont
  {{Gould}}, \citenamefont {{Graff}}, \citenamefont {{Gros}}, \citenamefont
  {{Haissinski}}, \citenamefont {{Hamadache}}, \citenamefont {{de Kat}},
  \citenamefont {{Lasserre}}, \citenamefont {{Lesquoy}}, \citenamefont
  {{Loup}}, \citenamefont {{Magneville}}, \citenamefont {{Marquette}},
  \citenamefont {{Maurice}}, \citenamefont {{Maury}}, \citenamefont
  {{Milsztajn}}, \citenamefont {{Moniez}}, \citenamefont
  {{Palanque-Delabrouille}}, \citenamefont {{Perdereau}}, \citenamefont
  {{Rahal}}, \citenamefont {{Rich}}, \citenamefont {{Spiro}}, \citenamefont
  {{Vidal-Madjar}}, \citenamefont {{Vigroux}}, \citenamefont {{Zylberajch}},\
  and\ \citenamefont {{EROS-2 Collaboration}}}]{Tisserand:2007}%
  \BibitemOpen
  \bibfield  {author} {\bibinfo {author} {\bibfnamefont {P.}~\bibnamefont
  {{Tisserand}}}, \bibinfo {author} {\bibfnamefont {L.}~\bibnamefont {{Le
  Guillou}}}, \bibinfo {author} {\bibfnamefont {C.}~\bibnamefont {{Afonso}}},
  \bibinfo {author} {\bibfnamefont {J.~N.}\ \bibnamefont {{Albert}}}, \bibinfo
  {author} {\bibfnamefont {J.}~\bibnamefont {{Andersen}}}, \bibinfo {author}
  {\bibfnamefont {R.}~\bibnamefont {{Ansari}}}, \bibinfo {author}
  {\bibfnamefont {{\'E}.}~\bibnamefont {{Aubourg}}}, \bibinfo {author}
  {\bibfnamefont {P.}~\bibnamefont {{Bareyre}}}, \bibinfo {author}
  {\bibfnamefont {J.~P.}\ \bibnamefont {{Beaulieu}}}, \bibinfo {author}
  {\bibfnamefont {X.}~\bibnamefont {{Charlot}}}, \bibinfo {author}
  {\bibfnamefont {C.}~\bibnamefont {{Coutures}}}, \bibinfo {author}
  {\bibfnamefont {R.}~\bibnamefont {{Ferlet}}}, \bibinfo {author}
  {\bibfnamefont {P.}~\bibnamefont {{Fouqu{\'e}}}}, \bibinfo {author}
  {\bibfnamefont {J.~F.}\ \bibnamefont {{Glicenstein}}}, \bibinfo {author}
  {\bibfnamefont {B.}~\bibnamefont {{Goldman}}}, \bibinfo {author}
  {\bibfnamefont {A.}~\bibnamefont {{Gould}}}, \bibinfo {author} {\bibfnamefont
  {D.}~\bibnamefont {{Graff}}}, \bibinfo {author} {\bibfnamefont
  {M.}~\bibnamefont {{Gros}}}, \bibinfo {author} {\bibfnamefont
  {J.}~\bibnamefont {{Haissinski}}}, \bibinfo {author} {\bibfnamefont
  {C.}~\bibnamefont {{Hamadache}}}, \bibinfo {author} {\bibfnamefont
  {J.}~\bibnamefont {{de Kat}}}, \bibinfo {author} {\bibfnamefont
  {T.}~\bibnamefont {{Lasserre}}}, \bibinfo {author} {\bibfnamefont
  {{\'E}.}~\bibnamefont {{Lesquoy}}}, \bibinfo {author} {\bibfnamefont
  {C.}~\bibnamefont {{Loup}}}, \bibinfo {author} {\bibfnamefont
  {C.}~\bibnamefont {{Magneville}}}, \bibinfo {author} {\bibfnamefont {J.~B.}\
  \bibnamefont {{Marquette}}}, \bibinfo {author} {\bibfnamefont
  {{\'E}.}~\bibnamefont {{Maurice}}}, \bibinfo {author} {\bibfnamefont
  {A.}~\bibnamefont {{Maury}}}, \bibinfo {author} {\bibfnamefont
  {A.}~\bibnamefont {{Milsztajn}}}, \bibinfo {author} {\bibfnamefont
  {M.}~\bibnamefont {{Moniez}}}, \bibinfo {author} {\bibfnamefont
  {N.}~\bibnamefont {{Palanque-Delabrouille}}}, \bibinfo {author}
  {\bibfnamefont {O.}~\bibnamefont {{Perdereau}}}, \bibinfo {author}
  {\bibfnamefont {Y.~R.}\ \bibnamefont {{Rahal}}}, \bibinfo {author}
  {\bibfnamefont {J.}~\bibnamefont {{Rich}}}, \bibinfo {author} {\bibfnamefont
  {M.}~\bibnamefont {{Spiro}}}, \bibinfo {author} {\bibfnamefont
  {A.}~\bibnamefont {{Vidal-Madjar}}}, \bibinfo {author} {\bibfnamefont
  {L.}~\bibnamefont {{Vigroux}}}, \bibinfo {author} {\bibfnamefont
  {S.}~\bibnamefont {{Zylberajch}}},\ and\ \bibinfo {author} {\bibnamefont
  {{EROS-2 Collaboration}}},\ }\bibfield  {title} {\bibinfo {title} {{Limits on
  the Macho content of the Galactic Halo from the EROS-2 Survey of the
  Magellanic Clouds}},\ }\href {https://doi.org/10.1051/0004-6361:20066017}
  {\bibfield  {journal} {\bibinfo  {journal} {\aap}\ }\textbf {\bibinfo
  {volume} {469}},\ \bibinfo {pages} {387} (\bibinfo {year} {2007})},\ \Eprint
  {https://arxiv.org/abs/astro-ph/0607207} {arXiv:astro-ph/0607207 [astro-ph]}
  \BibitemShut {NoStop}%
\bibitem [{\citenamefont {{Wyrzykowski}}\ \emph {et~al.}(2011)\citenamefont
  {{Wyrzykowski}}, \citenamefont {{Skowron}}, \citenamefont {{Koz{\l}owski}},
  \citenamefont {{Udalski}}, \citenamefont {{Szyma{\'n}ski}}, \citenamefont
  {{Kubiak}}, \citenamefont {{Pietrzy{\'n}ski}}, \citenamefont
  {{Soszy{\'n}ski}}, \citenamefont {{Szewczyk}}, \citenamefont {{Ulaczyk}},
  \citenamefont {{Poleski}},\ and\ \citenamefont
  {{Tisserand}}}]{Wyrzykowski:2011}%
  \BibitemOpen
  \bibfield  {author} {\bibinfo {author} {\bibfnamefont {L.}~\bibnamefont
  {{Wyrzykowski}}}, \bibinfo {author} {\bibfnamefont {J.}~\bibnamefont
  {{Skowron}}}, \bibinfo {author} {\bibfnamefont {S.}~\bibnamefont
  {{Koz{\l}owski}}}, \bibinfo {author} {\bibfnamefont {A.}~\bibnamefont
  {{Udalski}}}, \bibinfo {author} {\bibfnamefont {M.~K.}\ \bibnamefont
  {{Szyma{\'n}ski}}}, \bibinfo {author} {\bibfnamefont {M.}~\bibnamefont
  {{Kubiak}}}, \bibinfo {author} {\bibfnamefont {G.}~\bibnamefont
  {{Pietrzy{\'n}ski}}}, \bibinfo {author} {\bibfnamefont {I.}~\bibnamefont
  {{Soszy{\'n}ski}}}, \bibinfo {author} {\bibfnamefont {O.}~\bibnamefont
  {{Szewczyk}}}, \bibinfo {author} {\bibfnamefont {K.}~\bibnamefont
  {{Ulaczyk}}}, \bibinfo {author} {\bibfnamefont {R.}~\bibnamefont
  {{Poleski}}},\ and\ \bibinfo {author} {\bibfnamefont {P.}~\bibnamefont
  {{Tisserand}}},\ }\bibfield  {title} {\bibinfo {title} {{The OGLE view of
  microlensing towards the Magellanic Clouds - IV. OGLE-III SMC data and final
  conclusions on MACHOs}},\ }\href
  {https://doi.org/10.1111/j.1365-2966.2011.19243.x} {\bibfield  {journal}
  {\bibinfo  {journal} {\mnras}\ }\textbf {\bibinfo {volume} {416}},\ \bibinfo
  {pages} {2949} (\bibinfo {year} {2011})},\ \Eprint
  {https://arxiv.org/abs/1106.2925} {arXiv:1106.2925 [astro-ph.GA]}
  \BibitemShut {NoStop}%
\bibitem [{\citenamefont {{Blaineau}}\ \emph {et~al.}(2022)\citenamefont
  {{Blaineau}}, \citenamefont {{Moniez}}, \citenamefont {{Afonso}},
  \citenamefont {{Albert}}, \citenamefont {{Ansari}}, \citenamefont
  {{Aubourg}}, \citenamefont {{Coutures}}, \citenamefont {{Glicenstein}},
  \citenamefont {{Goldman}}, \citenamefont {{Hamadache}}, \citenamefont
  {{Lasserre}}, \citenamefont {{Le Guillou}}, \citenamefont {{Lesquoy}},
  \citenamefont {{Magneville}}, \citenamefont {{Marquette}}, \citenamefont
  {{Palanque-Delabrouille}}, \citenamefont {{Perdereau}}, \citenamefont
  {{Rich}}, \citenamefont {{Spiro}},\ and\ \citenamefont
  {{Tisserand}}}]{Blaineau:2022}%
  \BibitemOpen
  \bibfield  {author} {\bibinfo {author} {\bibfnamefont {T.}~\bibnamefont
  {{Blaineau}}}, \bibinfo {author} {\bibfnamefont {M.}~\bibnamefont
  {{Moniez}}}, \bibinfo {author} {\bibfnamefont {C.}~\bibnamefont {{Afonso}}},
  \bibinfo {author} {\bibfnamefont {J.~N.}\ \bibnamefont {{Albert}}}, \bibinfo
  {author} {\bibfnamefont {R.}~\bibnamefont {{Ansari}}}, \bibinfo {author}
  {\bibfnamefont {E.}~\bibnamefont {{Aubourg}}}, \bibinfo {author}
  {\bibfnamefont {C.}~\bibnamefont {{Coutures}}}, \bibinfo {author}
  {\bibfnamefont {J.~F.}\ \bibnamefont {{Glicenstein}}}, \bibinfo {author}
  {\bibfnamefont {B.}~\bibnamefont {{Goldman}}}, \bibinfo {author}
  {\bibfnamefont {C.}~\bibnamefont {{Hamadache}}}, \bibinfo {author}
  {\bibfnamefont {T.}~\bibnamefont {{Lasserre}}}, \bibinfo {author}
  {\bibfnamefont {L.}~\bibnamefont {{Le Guillou}}}, \bibinfo {author}
  {\bibfnamefont {E.}~\bibnamefont {{Lesquoy}}}, \bibinfo {author}
  {\bibfnamefont {C.}~\bibnamefont {{Magneville}}}, \bibinfo {author}
  {\bibfnamefont {J.~B.}\ \bibnamefont {{Marquette}}}, \bibinfo {author}
  {\bibfnamefont {N.}~\bibnamefont {{Palanque-Delabrouille}}}, \bibinfo
  {author} {\bibfnamefont {O.}~\bibnamefont {{Perdereau}}}, \bibinfo {author}
  {\bibfnamefont {J.}~\bibnamefont {{Rich}}}, \bibinfo {author} {\bibfnamefont
  {M.}~\bibnamefont {{Spiro}}},\ and\ \bibinfo {author} {\bibfnamefont
  {P.}~\bibnamefont {{Tisserand}}},\ }\bibfield  {title} {\bibinfo {title}
  {{New limits from microlensing on Galactic black holes in the mass range 10
  M$_{{\ensuremath{\odot}}}$ < M < 1000 M$_{{\ensuremath{\odot}}}$}},\ }\href
  {https://doi.org/10.1051/0004-6361/202243430} {\bibfield  {journal} {\bibinfo
   {journal} {\aap}\ }\textbf {\bibinfo {volume} {664}},\ \bibinfo {eid} {A106}
  (\bibinfo {year} {2022})},\ \Eprint {https://arxiv.org/abs/2202.13819}
  {arXiv:2202.13819 [astro-ph.GA]} \BibitemShut {NoStop}%
\bibitem [{\citenamefont {{Bird}}\ \emph {et~al.}(2023)\citenamefont {{Bird}},
  \citenamefont {{Albert}}, \citenamefont {{Dawson}}, \citenamefont
  {{Ali-Ha{\"\i}moud}}, \citenamefont {{Coogan}}, \citenamefont
  {{Drlica-Wagner}}, \citenamefont {{Feng}}, \citenamefont {{Inman}},
  \citenamefont {{Inomata}}, \citenamefont {{Kovetz}}, \citenamefont
  {{Kusenko}}, \citenamefont {{Lehmann}}, \citenamefont {{Mu{\~n}oz}},
  \citenamefont {{Singh}}, \citenamefont {{Takhistov}},\ and\ \citenamefont
  {{Tsai}}}]{Bird:2023}%
  \BibitemOpen
  \bibfield  {author} {\bibinfo {author} {\bibfnamefont {S.}~\bibnamefont
  {{Bird}}}, \bibinfo {author} {\bibfnamefont {A.}~\bibnamefont {{Albert}}},
  \bibinfo {author} {\bibfnamefont {W.}~\bibnamefont {{Dawson}}}, \bibinfo
  {author} {\bibfnamefont {Y.}~\bibnamefont {{Ali-Ha{\"\i}moud}}}, \bibinfo
  {author} {\bibfnamefont {A.}~\bibnamefont {{Coogan}}}, \bibinfo {author}
  {\bibfnamefont {A.}~\bibnamefont {{Drlica-Wagner}}}, \bibinfo {author}
  {\bibfnamefont {Q.}~\bibnamefont {{Feng}}}, \bibinfo {author} {\bibfnamefont
  {D.}~\bibnamefont {{Inman}}}, \bibinfo {author} {\bibfnamefont
  {K.}~\bibnamefont {{Inomata}}}, \bibinfo {author} {\bibfnamefont
  {E.}~\bibnamefont {{Kovetz}}}, \bibinfo {author} {\bibfnamefont
  {A.}~\bibnamefont {{Kusenko}}}, \bibinfo {author} {\bibfnamefont {B.~V.}\
  \bibnamefont {{Lehmann}}}, \bibinfo {author} {\bibfnamefont {J.~B.}\
  \bibnamefont {{Mu{\~n}oz}}}, \bibinfo {author} {\bibfnamefont
  {R.}~\bibnamefont {{Singh}}}, \bibinfo {author} {\bibfnamefont
  {V.}~\bibnamefont {{Takhistov}}},\ and\ \bibinfo {author} {\bibfnamefont
  {Y.-D.}\ \bibnamefont {{Tsai}}},\ }\bibfield  {title} {\bibinfo {title}
  {{Snowmass2021 Cosmic Frontier White Paper: Primordial black hole dark
  matter}},\ }\href {https://doi.org/10.1016/j.dark.2023.101231} {\bibfield
  {journal} {\bibinfo  {journal} {Physics of the Dark Universe}\ }\textbf
  {\bibinfo {volume} {41}},\ \bibinfo {eid} {101231} (\bibinfo {year}
  {2023})},\ \Eprint {https://arxiv.org/abs/2203.08967} {arXiv:2203.08967
  [hep-ph]} \BibitemShut {NoStop}%
\bibitem [{\citenamefont {Gerosa}\ and\ \citenamefont
  {Fishbach}(2021)}]{Gerosa:2021mno}%
  \BibitemOpen
  \bibfield  {author} {\bibinfo {author} {\bibfnamefont {D.}~\bibnamefont
  {Gerosa}}\ and\ \bibinfo {author} {\bibfnamefont {M.}~\bibnamefont
  {Fishbach}},\ }\bibfield  {title} {\bibinfo {title} {{Hierarchical mergers of
  stellar-mass black holes and their gravitational-wave signatures}},\ }\href
  {https://doi.org/10.1038/s41550-021-01398-w} {\bibfield  {journal} {\bibinfo
  {journal} {Nature Astron.}\ }\textbf {\bibinfo {volume} {5}},\ \bibinfo
  {pages} {749} (\bibinfo {year} {2021})},\ \Eprint
  {https://arxiv.org/abs/2105.03439} {arXiv:2105.03439 [astro-ph.HE]}
  \BibitemShut {NoStop}%
\bibitem [{\citenamefont {{Olejak}}\ \emph {et~al.}(2020)\citenamefont
  {{Olejak}}, \citenamefont {{Fishbach}}, \citenamefont {{Belczynski}},
  \citenamefont {{Holz}}, \citenamefont {{Lasota}}, \citenamefont {{Miller}},\
  and\ \citenamefont {{Bulik}}}]{Olejak:2020}%
  \BibitemOpen
  \bibfield  {author} {\bibinfo {author} {\bibfnamefont {A.}~\bibnamefont
  {{Olejak}}}, \bibinfo {author} {\bibfnamefont {M.}~\bibnamefont
  {{Fishbach}}}, \bibinfo {author} {\bibfnamefont {K.}~\bibnamefont
  {{Belczynski}}}, \bibinfo {author} {\bibfnamefont {D.~E.}\ \bibnamefont
  {{Holz}}}, \bibinfo {author} {\bibfnamefont {J.~P.}\ \bibnamefont
  {{Lasota}}}, \bibinfo {author} {\bibfnamefont {M.~C.}\ \bibnamefont
  {{Miller}}},\ and\ \bibinfo {author} {\bibfnamefont {T.}~\bibnamefont
  {{Bulik}}},\ }\bibfield  {title} {\bibinfo {title} {{The Origin of
  Inequality: Isolated Formation of a 30+10 M$_{{\ensuremath{\odot}}}$ Binary
  Black Hole Merger}},\ }\href {https://doi.org/10.3847/2041-8213/abb5b5}
  {\bibfield  {journal} {\bibinfo  {journal} {\apjl}\ }\textbf {\bibinfo
  {volume} {901}},\ \bibinfo {eid} {L39} (\bibinfo {year} {2020})},\ \Eprint
  {https://arxiv.org/abs/2004.11866} {arXiv:2004.11866 [astro-ph.HE]}
  \BibitemShut {NoStop}%
\bibitem [{\citenamefont {Oriol}\ \emph {et~al.}(2023)\citenamefont {Oriol},
  \citenamefont {Virgile}, \citenamefont {Colin}, \citenamefont {Larry},
  \citenamefont {J.}, \citenamefont {Maxim}, \citenamefont {Ravin},
  \citenamefont {Jupeng}, \citenamefont {C.}, \citenamefont {A.}, \citenamefont
  {Michael}, \citenamefont {Ricardo}, \citenamefont {Thomas},\ and\
  \citenamefont {Robert}}]{pymc:2023}%
  \BibitemOpen
  \bibfield  {author} {\bibinfo {author} {\bibfnamefont {A.-P.}\ \bibnamefont
  {Oriol}}, \bibinfo {author} {\bibfnamefont {A.}~\bibnamefont {Virgile}},
  \bibinfo {author} {\bibfnamefont {C.}~\bibnamefont {Colin}}, \bibinfo
  {author} {\bibfnamefont {D.}~\bibnamefont {Larry}}, \bibinfo {author}
  {\bibfnamefont {F.~C.}\ \bibnamefont {J.}}, \bibinfo {author} {\bibfnamefont
  {K.}~\bibnamefont {Maxim}}, \bibinfo {author} {\bibfnamefont
  {K.}~\bibnamefont {Ravin}}, \bibinfo {author} {\bibfnamefont
  {L.}~\bibnamefont {Jupeng}}, \bibinfo {author} {\bibfnamefont {L.~C.}\
  \bibnamefont {C.}}, \bibinfo {author} {\bibfnamefont {M.~O.}\ \bibnamefont
  {A.}}, \bibinfo {author} {\bibfnamefont {O.}~\bibnamefont {Michael}},
  \bibinfo {author} {\bibfnamefont {V.}~\bibnamefont {Ricardo}}, \bibinfo
  {author} {\bibfnamefont {W.}~\bibnamefont {Thomas}},\ and\ \bibinfo {author}
  {\bibfnamefont {Z.}~\bibnamefont {Robert}},\ }\bibfield  {title} {\bibinfo
  {title} {Pymc: A modern and comprehensive probabilistic programming framework
  in python},\ }\href {https://doi.org/10.7717/peerj-cs.1516} {\bibfield
  {journal} {\bibinfo  {journal} {{PeerJ} Computer Science}\ }\textbf {\bibinfo
  {volume} {9}},\ \bibinfo {pages} {e1516} (\bibinfo {year}
  {2023})}\BibitemShut {NoStop}%
\bibitem [{\citenamefont {Virtanen}\ \emph {et~al.}(2020)\citenamefont
  {Virtanen}, \citenamefont {Gommers}, \citenamefont {Oliphant}, \citenamefont
  {Haberland}, \citenamefont {Reddy}, \citenamefont {Cournapeau}, \citenamefont
  {Burovski}, \citenamefont {Peterson}, \citenamefont {Weckesser},
  \citenamefont {Bright}, \citenamefont {{van der Walt}}, \citenamefont
  {Brett}, \citenamefont {Wilson}, \citenamefont {Millman}, \citenamefont
  {Mayorov}, \citenamefont {Nelson}, \citenamefont {Jones}, \citenamefont
  {Kern}, \citenamefont {Larson}, \citenamefont {Carey}, \citenamefont {Polat},
  \citenamefont {Feng}, \citenamefont {Moore}, \citenamefont {{VanderPlas}},
  \citenamefont {Laxalde}, \citenamefont {Perktold}, \citenamefont {Cimrman},
  \citenamefont {Henriksen}, \citenamefont {Quintero}, \citenamefont {Harris},
  \citenamefont {Archibald}, \citenamefont {Ribeiro}, \citenamefont
  {Pedregosa}, \citenamefont {{van Mulbregt}},\ and\ \citenamefont {{SciPy 1.0
  Contributors}}}]{Virtanen:2020}%
  \BibitemOpen
  \bibfield  {author} {\bibinfo {author} {\bibfnamefont {P.}~\bibnamefont
  {Virtanen}}, \bibinfo {author} {\bibfnamefont {R.}~\bibnamefont {Gommers}},
  \bibinfo {author} {\bibfnamefont {T.~E.}\ \bibnamefont {Oliphant}}, \bibinfo
  {author} {\bibfnamefont {M.}~\bibnamefont {Haberland}}, \bibinfo {author}
  {\bibfnamefont {T.}~\bibnamefont {Reddy}}, \bibinfo {author} {\bibfnamefont
  {D.}~\bibnamefont {Cournapeau}}, \bibinfo {author} {\bibfnamefont
  {E.}~\bibnamefont {Burovski}}, \bibinfo {author} {\bibfnamefont
  {P.}~\bibnamefont {Peterson}}, \bibinfo {author} {\bibfnamefont
  {W.}~\bibnamefont {Weckesser}}, \bibinfo {author} {\bibfnamefont
  {J.}~\bibnamefont {Bright}}, \bibinfo {author} {\bibfnamefont {S.~J.}\
  \bibnamefont {{van der Walt}}}, \bibinfo {author} {\bibfnamefont
  {M.}~\bibnamefont {Brett}}, \bibinfo {author} {\bibfnamefont
  {J.}~\bibnamefont {Wilson}}, \bibinfo {author} {\bibfnamefont {K.~J.}\
  \bibnamefont {Millman}}, \bibinfo {author} {\bibfnamefont {N.}~\bibnamefont
  {Mayorov}}, \bibinfo {author} {\bibfnamefont {A.~R.~J.}\ \bibnamefont
  {Nelson}}, \bibinfo {author} {\bibfnamefont {E.}~\bibnamefont {Jones}},
  \bibinfo {author} {\bibfnamefont {R.}~\bibnamefont {Kern}}, \bibinfo {author}
  {\bibfnamefont {E.}~\bibnamefont {Larson}}, \bibinfo {author} {\bibfnamefont
  {C.~J.}\ \bibnamefont {Carey}}, \bibinfo {author} {\bibfnamefont
  {{\.I}.}~\bibnamefont {Polat}}, \bibinfo {author} {\bibfnamefont
  {Y.}~\bibnamefont {Feng}}, \bibinfo {author} {\bibfnamefont {E.~W.}\
  \bibnamefont {Moore}}, \bibinfo {author} {\bibfnamefont {J.}~\bibnamefont
  {{VanderPlas}}}, \bibinfo {author} {\bibfnamefont {D.}~\bibnamefont
  {Laxalde}}, \bibinfo {author} {\bibfnamefont {J.}~\bibnamefont {Perktold}},
  \bibinfo {author} {\bibfnamefont {R.}~\bibnamefont {Cimrman}}, \bibinfo
  {author} {\bibfnamefont {I.}~\bibnamefont {Henriksen}}, \bibinfo {author}
  {\bibfnamefont {E.~A.}\ \bibnamefont {Quintero}}, \bibinfo {author}
  {\bibfnamefont {C.~R.}\ \bibnamefont {Harris}}, \bibinfo {author}
  {\bibfnamefont {A.~M.}\ \bibnamefont {Archibald}}, \bibinfo {author}
  {\bibfnamefont {A.~H.}\ \bibnamefont {Ribeiro}}, \bibinfo {author}
  {\bibfnamefont {F.}~\bibnamefont {Pedregosa}}, \bibinfo {author}
  {\bibfnamefont {P.}~\bibnamefont {{van Mulbregt}}},\ and\ \bibinfo {author}
  {\bibnamefont {{SciPy 1.0 Contributors}}},\ }\bibfield  {title} {\bibinfo
  {title} {{{SciPy} 1.0: Fundamental Algorithms for Scientific Computing in
  Python}},\ }\href {https://doi.org/10.1038/s41592-019-0686-2} {\bibfield
  {journal} {\bibinfo  {journal} {Nature Methods}\ }\textbf {\bibinfo {volume}
  {17}},\ \bibinfo {pages} {261} (\bibinfo {year} {2020})}\BibitemShut
  {NoStop}%
\bibitem [{\citenamefont {{Ashton}}\ \emph {et~al.}(2019)\citenamefont
  {{Ashton}}, \citenamefont {{H{\"u}bner}}, \citenamefont {{Lasky}},
  \citenamefont {{Talbot}}, \citenamefont {{Ackley}}, \citenamefont
  {{Biscoveanu}}, \citenamefont {{Chu}}, \citenamefont {{Divakarla}},
  \citenamefont {{Easter}}, \citenamefont {{Goncharov}}, \citenamefont
  {{Hernandez Vivanco}}, \citenamefont {{Harms}}, \citenamefont {{Lower}},
  \citenamefont {{Meadors}}, \citenamefont {{Melchor}}, \citenamefont
  {{Payne}}, \citenamefont {{Pitkin}}, \citenamefont {{Powell}}, \citenamefont
  {{Sarin}}, \citenamefont {{Smith}},\ and\ \citenamefont
  {{Thrane}}}]{Ashton:2019}%
  \BibitemOpen
  \bibfield  {author} {\bibinfo {author} {\bibfnamefont {G.}~\bibnamefont
  {{Ashton}}}, \bibinfo {author} {\bibfnamefont {M.}~\bibnamefont
  {{H{\"u}bner}}}, \bibinfo {author} {\bibfnamefont {P.~D.}\ \bibnamefont
  {{Lasky}}}, \bibinfo {author} {\bibfnamefont {C.}~\bibnamefont {{Talbot}}},
  \bibinfo {author} {\bibfnamefont {K.}~\bibnamefont {{Ackley}}}, \bibinfo
  {author} {\bibfnamefont {S.}~\bibnamefont {{Biscoveanu}}}, \bibinfo {author}
  {\bibfnamefont {Q.}~\bibnamefont {{Chu}}}, \bibinfo {author} {\bibfnamefont
  {A.}~\bibnamefont {{Divakarla}}}, \bibinfo {author} {\bibfnamefont {P.~J.}\
  \bibnamefont {{Easter}}}, \bibinfo {author} {\bibfnamefont {B.}~\bibnamefont
  {{Goncharov}}}, \bibinfo {author} {\bibfnamefont {F.}~\bibnamefont
  {{Hernandez Vivanco}}}, \bibinfo {author} {\bibfnamefont {J.}~\bibnamefont
  {{Harms}}}, \bibinfo {author} {\bibfnamefont {M.~E.}\ \bibnamefont
  {{Lower}}}, \bibinfo {author} {\bibfnamefont {G.~D.}\ \bibnamefont
  {{Meadors}}}, \bibinfo {author} {\bibfnamefont {D.}~\bibnamefont
  {{Melchor}}}, \bibinfo {author} {\bibfnamefont {E.}~\bibnamefont {{Payne}}},
  \bibinfo {author} {\bibfnamefont {M.~D.}\ \bibnamefont {{Pitkin}}}, \bibinfo
  {author} {\bibfnamefont {J.}~\bibnamefont {{Powell}}}, \bibinfo {author}
  {\bibfnamefont {N.}~\bibnamefont {{Sarin}}}, \bibinfo {author} {\bibfnamefont
  {R.~J.~E.}\ \bibnamefont {{Smith}}},\ and\ \bibinfo {author} {\bibfnamefont
  {E.}~\bibnamefont {{Thrane}}},\ }\bibfield  {title} {\bibinfo {title}
  {{BILBY: A User-friendly Bayesian Inference Library for Gravitational-wave
  Astronomy}},\ }\href {https://doi.org/10.3847/1538-4365/ab06fc} {\bibfield
  {journal} {\bibinfo  {journal} {\apjs}\ }\textbf {\bibinfo {volume} {241}},\
  \bibinfo {eid} {27} (\bibinfo {year} {2019})},\ \Eprint
  {https://arxiv.org/abs/1811.02042} {arXiv:1811.02042 [astro-ph.IM]}
  \BibitemShut {NoStop}%
\bibitem [{\citenamefont {{Romero-Shaw}}\ \emph {et~al.}(2020)\citenamefont
  {{Romero-Shaw}}, \citenamefont {{Talbot}}, \citenamefont {{Biscoveanu}},
  \citenamefont {{D'Emilio}}, \citenamefont {{Ashton}}, \citenamefont
  {{Berry}}, \citenamefont {{Coughlin}}, \citenamefont {{Galaudage}},
  \citenamefont {{Hoy}}, \citenamefont {{H{\"u}bner}}, \citenamefont
  {{Phukon}}, \citenamefont {{Pitkin}}, \citenamefont {{Rizzo}}, \citenamefont
  {{Sarin}}, \citenamefont {{Smith}}, \citenamefont {{Stevenson}},
  \citenamefont {{Vajpeyi}}, \citenamefont {{Ar{\`e}ne}}, \citenamefont
  {{Athar}}, \citenamefont {{Banagiri}}, \citenamefont {{Bose}}, \citenamefont
  {{Carney}}, \citenamefont {{Chatziioannou}}, \citenamefont {{Clark}},
  \citenamefont {{Colleoni}}, \citenamefont {{Cotesta}}, \citenamefont
  {{Edelman}}, \citenamefont {{Estell{\'e}s}}, \citenamefont
  {{Garc{\'\i}a-Quir{\'o}s}}, \citenamefont {{Ghosh}}, \citenamefont {{Green}},
  \citenamefont {{Haster}}, \citenamefont {{Husa}}, \citenamefont {{Keitel}},
  \citenamefont {{Kim}}, \citenamefont {{Hernandez-Vivanco}}, \citenamefont
  {{Maga{\~n}a Hernandez}}, \citenamefont {{Karathanasis}}, \citenamefont
  {{Lasky}}, \citenamefont {{De Lillo}}, \citenamefont {{Lower}}, \citenamefont
  {{Macleod}}, \citenamefont {{Mateu-Lucena}}, \citenamefont {{Miller}},
  \citenamefont {{Millhouse}}, \citenamefont {{Morisaki}}, \citenamefont
  {{Oh}}, \citenamefont {{Ossokine}}, \citenamefont {{Payne}}, \citenamefont
  {{Powell}}, \citenamefont {{Pratten}}, \citenamefont {{P{\"u}rrer}},
  \citenamefont {{Ramos-Buades}}, \citenamefont {{Raymond}}, \citenamefont
  {{Thrane}}, \citenamefont {{Veitch}}, \citenamefont {{Williams}},
  \citenamefont {{Williams}},\ and\ \citenamefont {{Xiao}}}]{Romero-Shaw:2020}%
  \BibitemOpen
  \bibfield  {author} {\bibinfo {author} {\bibfnamefont {I.~M.}\ \bibnamefont
  {{Romero-Shaw}}}, \bibinfo {author} {\bibfnamefont {C.}~\bibnamefont
  {{Talbot}}}, \bibinfo {author} {\bibfnamefont {S.}~\bibnamefont
  {{Biscoveanu}}}, \bibinfo {author} {\bibfnamefont {V.}~\bibnamefont
  {{D'Emilio}}}, \bibinfo {author} {\bibfnamefont {G.}~\bibnamefont
  {{Ashton}}}, \bibinfo {author} {\bibfnamefont {C.~P.~L.}\ \bibnamefont
  {{Berry}}}, \bibinfo {author} {\bibfnamefont {S.}~\bibnamefont {{Coughlin}}},
  \bibinfo {author} {\bibfnamefont {S.}~\bibnamefont {{Galaudage}}}, \bibinfo
  {author} {\bibfnamefont {C.}~\bibnamefont {{Hoy}}}, \bibinfo {author}
  {\bibfnamefont {M.}~\bibnamefont {{H{\"u}bner}}}, \bibinfo {author}
  {\bibfnamefont {K.~S.}\ \bibnamefont {{Phukon}}}, \bibinfo {author}
  {\bibfnamefont {M.}~\bibnamefont {{Pitkin}}}, \bibinfo {author}
  {\bibfnamefont {M.}~\bibnamefont {{Rizzo}}}, \bibinfo {author} {\bibfnamefont
  {N.}~\bibnamefont {{Sarin}}}, \bibinfo {author} {\bibfnamefont
  {R.}~\bibnamefont {{Smith}}}, \bibinfo {author} {\bibfnamefont
  {S.}~\bibnamefont {{Stevenson}}}, \bibinfo {author} {\bibfnamefont
  {A.}~\bibnamefont {{Vajpeyi}}}, \bibinfo {author} {\bibfnamefont
  {M.}~\bibnamefont {{Ar{\`e}ne}}}, \bibinfo {author} {\bibfnamefont
  {K.}~\bibnamefont {{Athar}}}, \bibinfo {author} {\bibfnamefont
  {S.}~\bibnamefont {{Banagiri}}}, \bibinfo {author} {\bibfnamefont
  {N.}~\bibnamefont {{Bose}}}, \bibinfo {author} {\bibfnamefont
  {M.}~\bibnamefont {{Carney}}}, \bibinfo {author} {\bibfnamefont
  {K.}~\bibnamefont {{Chatziioannou}}}, \bibinfo {author} {\bibfnamefont
  {J.~A.}\ \bibnamefont {{Clark}}}, \bibinfo {author} {\bibfnamefont
  {M.}~\bibnamefont {{Colleoni}}}, \bibinfo {author} {\bibfnamefont
  {R.}~\bibnamefont {{Cotesta}}}, \bibinfo {author} {\bibfnamefont
  {B.}~\bibnamefont {{Edelman}}}, \bibinfo {author} {\bibfnamefont
  {H.}~\bibnamefont {{Estell{\'e}s}}}, \bibinfo {author} {\bibfnamefont
  {C.}~\bibnamefont {{Garc{\'\i}a-Quir{\'o}s}}}, \bibinfo {author}
  {\bibfnamefont {A.}~\bibnamefont {{Ghosh}}}, \bibinfo {author} {\bibfnamefont
  {R.}~\bibnamefont {{Green}}}, \bibinfo {author} {\bibfnamefont {C.~J.}\
  \bibnamefont {{Haster}}}, \bibinfo {author} {\bibfnamefont {S.}~\bibnamefont
  {{Husa}}}, \bibinfo {author} {\bibfnamefont {D.}~\bibnamefont {{Keitel}}},
  \bibinfo {author} {\bibfnamefont {A.~X.}\ \bibnamefont {{Kim}}}, \bibinfo
  {author} {\bibfnamefont {F.}~\bibnamefont {{Hernandez-Vivanco}}}, \bibinfo
  {author} {\bibfnamefont {I.}~\bibnamefont {{Maga{\~n}a Hernandez}}}, \bibinfo
  {author} {\bibfnamefont {C.}~\bibnamefont {{Karathanasis}}}, \bibinfo
  {author} {\bibfnamefont {P.~D.}\ \bibnamefont {{Lasky}}}, \bibinfo {author}
  {\bibfnamefont {N.}~\bibnamefont {{De Lillo}}}, \bibinfo {author}
  {\bibfnamefont {M.~E.}\ \bibnamefont {{Lower}}}, \bibinfo {author}
  {\bibfnamefont {D.}~\bibnamefont {{Macleod}}}, \bibinfo {author}
  {\bibfnamefont {M.}~\bibnamefont {{Mateu-Lucena}}}, \bibinfo {author}
  {\bibfnamefont {A.}~\bibnamefont {{Miller}}}, \bibinfo {author}
  {\bibfnamefont {M.}~\bibnamefont {{Millhouse}}}, \bibinfo {author}
  {\bibfnamefont {S.}~\bibnamefont {{Morisaki}}}, \bibinfo {author}
  {\bibfnamefont {S.~H.}\ \bibnamefont {{Oh}}}, \bibinfo {author}
  {\bibfnamefont {S.}~\bibnamefont {{Ossokine}}}, \bibinfo {author}
  {\bibfnamefont {E.}~\bibnamefont {{Payne}}}, \bibinfo {author} {\bibfnamefont
  {J.}~\bibnamefont {{Powell}}}, \bibinfo {author} {\bibfnamefont
  {G.}~\bibnamefont {{Pratten}}}, \bibinfo {author} {\bibfnamefont
  {M.}~\bibnamefont {{P{\"u}rrer}}}, \bibinfo {author} {\bibfnamefont
  {A.}~\bibnamefont {{Ramos-Buades}}}, \bibinfo {author} {\bibfnamefont
  {V.}~\bibnamefont {{Raymond}}}, \bibinfo {author} {\bibfnamefont
  {E.}~\bibnamefont {{Thrane}}}, \bibinfo {author} {\bibfnamefont
  {J.}~\bibnamefont {{Veitch}}}, \bibinfo {author} {\bibfnamefont
  {D.}~\bibnamefont {{Williams}}}, \bibinfo {author} {\bibfnamefont {M.~J.}\
  \bibnamefont {{Williams}}},\ and\ \bibinfo {author} {\bibfnamefont
  {L.}~\bibnamefont {{Xiao}}},\ }\bibfield  {title} {\bibinfo {title}
  {{Bayesian inference for compact binary coalescences with BILBY: validation
  and application to the first LIGO-Virgo gravitational-wave transient
  catalogue}},\ }\href {https://doi.org/10.1093/mnras/staa2850} {\bibfield
  {journal} {\bibinfo  {journal} {\mnras}\ }\textbf {\bibinfo {volume} {499}},\
  \bibinfo {pages} {3295} (\bibinfo {year} {2020})},\ \Eprint
  {https://arxiv.org/abs/2006.00714} {arXiv:2006.00714 [astro-ph.IM]}
  \BibitemShut {NoStop}%
\bibitem [{\citenamefont {Kumar}\ \emph {et~al.}(2019)\citenamefont {Kumar},
  \citenamefont {Carroll}, \citenamefont {Hartikainen},\ and\ \citenamefont
  {Martin}}]{Kumar:2019}%
  \BibitemOpen
  \bibfield  {author} {\bibinfo {author} {\bibfnamefont {R.}~\bibnamefont
  {Kumar}}, \bibinfo {author} {\bibfnamefont {C.}~\bibnamefont {Carroll}},
  \bibinfo {author} {\bibfnamefont {A.}~\bibnamefont {Hartikainen}},\ and\
  \bibinfo {author} {\bibfnamefont {O.}~\bibnamefont {Martin}},\ }\bibfield
  {title} {\bibinfo {title} {Arviz a unified library for exploratory analysis
  of bayesian models in python},\ }\href {https://doi.org/10.21105/joss.01143}
  {\bibfield  {journal} {\bibinfo  {journal} {Journal of Open Source Software}\
  }\textbf {\bibinfo {volume} {4}},\ \bibinfo {pages} {1143} (\bibinfo {year}
  {2019})}\BibitemShut {NoStop}%
\bibitem [{\citenamefont {Foreman-Mackey}(2016)}]{corner:2016}%
  \BibitemOpen
  \bibfield  {author} {\bibinfo {author} {\bibfnamefont {D.}~\bibnamefont
  {Foreman-Mackey}},\ }\bibfield  {title} {\bibinfo {title} {corner.py:
  Scatterplot matrices in python},\ }\href
  {https://doi.org/10.21105/joss.00024} {\bibfield  {journal} {\bibinfo
  {journal} {The Journal of Open Source Software}\ }\textbf {\bibinfo {volume}
  {1}},\ \bibinfo {pages} {24} (\bibinfo {year} {2016})}\BibitemShut {NoStop}%
\bibitem [{\citenamefont {Hunter}(2007)}]{Hunter:2007}%
  \BibitemOpen
  \bibfield  {author} {\bibinfo {author} {\bibfnamefont {J.~D.}\ \bibnamefont
  {Hunter}},\ }\bibfield  {title} {\bibinfo {title} {Matplotlib: A 2d graphics
  environment},\ }\href {https://doi.org/10.1109/MCSE.2007.55} {\bibfield
  {journal} {\bibinfo  {journal} {Computing in Science \& Engineering}\
  }\textbf {\bibinfo {volume} {9}},\ \bibinfo {pages} {90} (\bibinfo {year}
  {2007})}\BibitemShut {NoStop}%
\end{thebibliography}%

\end{document}